\documentstyle[prd,aps,preprint]{revtex}
\tightenlines
\input epsf.tex
\def\DESepsf(#1 width #2){\epsfxsize=#2 \epsfbox{#1}}
\begin{document}

\draft

\preprint{\vbox{
\hbox{OSU-HEP-98-11}\hbox{UMD-PP-99-063}
\hbox{IASSNS-HEP-98-80}
}}
\title{\large\bf Fermion Masses, Neutrino Oscillations, and Proton Decay\\
in the Light of SuperKamiokande}
\author{{\bf K. S. Babu$^1$, Jogesh C. Pati$^2$} and {\bf Frank Wilczek$^3$} }

\address{$^1$Department of Physics, Oklahoma State University, Stillwater, OK,
74078\\
$^2$Department of
Physics, University of Maryland, College Park, MD, 20742 \\
$^3$School of Natural Sciences, Institute for Advanced Study, Princeton, NJ,
08540}
\date{December, 1998}
\maketitle
\begin{abstract}

Within the framework of unified gauge models,
interactions responsible for neutrino masses can also provide
mechanisms for nucleon instability.
We discuss their implications concretely in the
light of recent results on neutrino oscillation from the
SuperKamiokande collaboration.  We construct
a predictive $SO(10)$-based framework
that describes the masses and mixing of all quarks and leptons.
An overconstrained global fit
is obtained, that makes five successful predictions for quarks and
charged leptons.  The same description provides agreement with the
SuperK results on atmospheric neutrinos and supports a small-angle MSW
mechanism.  We find that
current limits on nucleon stability put
significant stress on the framework.
Further, a distinctive feature of the $SO(10)$ model developed here is
the likely prominence of the $\mu^+ K^0$ mode in addition to the
$\overline{\nu} K^+$ mode of proton decay.
Thus improved searches in these  channels  for proton decay
will either turn up events, or force us outside this circle of ideas.
\end{abstract}

\vskip2pc

%
%

%
\section{Introduction}
Recent SuperKamiokande observations on atmospheric neutrinos \cite{superk}
establish the oscillation of $\nu_\mu$ to $\nu_\tau$  with a mass
splitting $\delta m^2 \sim (10^{-2}~{\rm to}~ 10^{-3})$  eV$^2$ and an
oscillation angle $\sin^2 2\theta_{\mu\tau}^{\rm osc} =
(0.82-1.0)$.  To be more precise, the observations do not directly
exclude oscillation into
some other $\nu_{\rm X}$, as long as it is {\it not\/} to $\nu_e$,
but Occam's razor and the framework adopted in this paper suggest
$X = \tau$, and we shall assume that in what follows, without further
comment.
These observations
clearly require new physics beyond what is usually contemplated in
the Standard Model.

     As we shall presently discuss, a tau neutrino mass consistent
with the  observed oscillations fits extremely naturally into the
framework supplied by unified gauge theories of the strong, weak, and
electromagnetic interactions \cite{unif,gg} that includes the symmetry
structure
$G_{224} = SU(2)_L \times SU(2)_R \times SU(4)_C$ \cite{unif},  the minimal
such
symmetry being $SO(10)$ \cite{so10}.  In this framework, the low-energy
degrees of
freedom need be no larger than the Standard Model.  Neutrino masses
appear as effective nonrenormalizable (dimension 5) operators.

     Unified gauge theories were already very impressive on other
grounds.   They
combine the scattered multiplets of the Standard Model (five per
family) into a significantly smaller number (two for $SU(5)$, one for
$SO(10)$).  They rationalize the otherwise bizarre-looking hypercharge
assignments in the Standard Model \cite{unif,gg}.  Finally, especially in their
supersymmetric version \cite{drw}, they account
quantitatively for the relative values of the strong, weak, and
electromagnetic couplings \cite{gqw,langacker}.

     This last feat is accomplished by renormalizing the separate
couplings down from a single common value at a unification scale,
taking into account the effects of vacuum polarization due to virtual
particles, down to the much lower mass scales at which they are
observed experimentally.  A by-product of this overconstrained, and
singularly successful, calculation, is to identify the mass scale at
which the unified symmetry is broken, to be $M_{\rm U} \sim 2 \times 10^{16}$
GeV \cite{langacker}.

This value is interesting in several respects.
First, from data and concepts purely internal to
gauge theories of particle interactions, it brings us
to the threshold of the fundamental scale of quantum gravity,
namely the Planck
mass $2 \times 10^{18}$ GeV (in rational units).
Reading it the other way, by demanding unification, allowing for both
the classical power-law
running of the gravitational coupling and the quantum logarithmic
running of gauge
couplings,
we obtain a roughly
accurate calculation of the observed strength of gravity.

Second, it
sets the scale for phenomena directly associated with unification but
forbidden in the Standard Model, notably nucleon decay and neutrino
masses.  Prior
to the SuperKamiokande observations,
the main phenomenological virtues of
the large value of the unification mass scale were its negative
implications.  It explained why nucleon decay is rare, and neutrino
masses are small, although both are almost inevitable
consequences of
unification.  Now the scale can also
be {\it positively\/}
identified, at least semi-quantitatively.

Indeed, any unification based on $G_{224}$ \cite{unif}
requires the existence
of right-handed neutrinos $\nu^R$.  When $G_{224}$ is
embedded in $SO(10)$, $\nu^R$
fills out, together with the 15
left-handed quark and lepton fields in each Standard Model family, the
16 dimensional spinor representation of $SO(10)$.
The $\nu^R$ are Standard Model
singlets, so that they can, and generically will, acquire large
Majorana masses at the scale where unified $SO(10)$ symmetry breaks to
the Standard Model $SU(3)\times SU(2)\times U(1)$.  The ordinary,
left-handed neutrinos couple to these
$\nu^R$ much as ordinary quarks and leptons couple to their
right-handed partners, through $SU(2)\times U(1)$ non-singlet Higgs
fields.  For the quarks and leptons, condensation of those Higgs
fields transforms such interactions directly into mass terms.
For neutrinos the effect of this condensation is slightly more
involved.  As mentioned, the $\nu^R$ have an independent, and much
larger, source of mass.  As a result, through the ``see-saw''
mechanism \cite{seesaw},
the effective masses for the
left-handed neutrinos, acquired through their virtual transitions into
$\nu^R$ and back, are predictably tiny.

     In this paper we do two things.  First, we flesh out with
quantitative detail the rough picture just sketched.  Its most
straightforward embodiment leads to the hierarchical pattern
$m_{\nu_e} << m_{\nu_\mu} << m_{\nu_\tau}$, and to a value of
$m_{\nu_\tau}$ very consistent with the SuperK observations,
interpreted as $\nu_\mu - \nu_\tau$ oscillations.  We demonstrate that
the large mixing angle observed does not force us to swerve from this
straightforward direction.  Indeed, it can arise rather
plausibly in the context of
ideas which have been applied successfully
to understanding
quark masses and mixings.  Motivated by
the success of the circle of ideas mentioned above, we insist
that the pattern of neutrino masses and mixings should be discussed
together with those of the quarks and charged leptons, and not in
isolation.   A simple and predictive $SO(10)$--based mass structure that
describes the observed masses and mixings of all fermions including those
of the neutrinos will be presented and analyzed.
We thus demonstrate by example how
the large $\nu_\mu-\nu_\tau$ oscillation angle can be obtained quite
naturally along with a large hierarchy in the $\nu_\mu-\nu_\tau$ masses.
(This is in contrast to several recent attempts
where such a large oscillation angle is explained as a consequence of
the near degeneracy of the
$\nu_\mu-\nu_\tau$ system.  The bulk of the papers in
Ref. \cite{degeneracy,sterile}
belong to this category.)  In this scheme,
$\nu_e-\nu_\mu$ oscillation
drives the small angle MSW explanation \cite{msw} of the solar neutrino
puzzle.

Second, we revisit
a previously noted link between neutrino masses and nucleon
decay, in the framework of supersymmetric $SO(10)$ unified models \cite{bpw}.
Previously, motivated in part by possible cosmological
indications for a hot dark matter component, we used numerical estimates for
$m_{\nu_\tau} \sim$ 1 eV, considerably larger than are now favored
by the SuperK result ($\sim 1/20$ eV).  This amounts to an increase
in the Majorana mass of $\nu^R_\tau$ compared to previous work, and
correspondingly an increase in the strength of the neutrino mass related
$d=5$ proton decay rate.
Another important change is caused by the large $\nu_\mu-\nu_\tau$
oscillation angle suggested by the SuperK result.  With hierarchical
neutrino masses, we will argue, their result
strongly suggests  substantial
mixing in the charged lepton ($\mu-\tau$) sector.  That, in
turn, affects the strength and the flavor structure not only of the neutrino
related, but also of the standard $d=5$ proton decay operators
induced by the exchange of color triplet partners of the electroweak
Higgs doublets \cite{d5}.   These adjustments in our expectations for proton
decay
turn out to be quite
significant quantitatively.
They
considerably heighten the tension around nucleon decay: either it is
accessible, or the framework fails.

As an accompanying major result, we observe that, in contrast to the minimal
supersymmetric $SU(5)$ model for which the charged lepton mode $p \rightarrow
\mu^+ K^0$ has a negligible branching ratio ($\sim 10^{-3}$, see text), for
the $SO(10)$ model developed here the corresponding branching ratio should
typically lie in the range of 20 to even 50\%, if the relevant hadronic
matrix elements have comparable magnitudes.  This becomes possible because
of the presence of the new $d=5$ proton decay operators mentioned above,
which are related to neutrino masses.  Thus the $\mu^+ K^0$ mode, if observed,
would provide an indication in favor of an $SO(10)$ model of the masses
and mixings of all fermions including neutrinos, as developed here.

This paper is organized as follows.  In Sec. II, we discuss the scale of
new physics implied by the SuperK observations.  In Sec. III we describe
a caricature model that accommodates large neutrino oscillation angle
as suggested by SuperK
without assuming neutrino mass degeneracy.  Sec. IV is devoted to a more
ambitious $SO(10)$ model that accounts for the masses of second and
third generation quarks and leptons
including the large neutrino oscillation angle.
In Sec. V we suggest, by
way of example, a predictive way to incorporate the first family fermions
into the $SO(10)$ scheme that retains the success of Sec. IV, leading to
a total of eight successful predictions for the masses and the mixings of the
fermions
including the neutrinos, and supports a small angle MSW mechanism.
In Sec. VI we discuss the issue of proton decay
in the context of neutrino masses.  Four Appendices (A,B,C and D) contain
relevant technical details of our proton decay calculations including
unification scale threshold corrections to $\alpha_3(m_Z)$.
Finally, a summary of our results and some concluding remarks are given in
Sec. VII\footnote{Preliminary results
of this investigation were announced at summer conferences \cite{takayama}.}.


\section{$m_{\nu_\tau}$ and the Unification Scale}

     Using the degrees of freedom of the Standard Model, small
Majorana masses for neutrinos arise from dimension-5 operators in the
form \cite{weinberg}
\begin{equation}
     {\cal L} ~=~ \lambda_{ij} { L_i L_j \phi \phi  \over M} ~+~
     {\rm h.c.}
\end{equation}
where $L_i = (\nu_i, \ell_i)^T$ denote the lepton doublets and
$\phi = (\phi^+, \phi^0)^T$ the Higgs doublet of the Standard Model.
Interpreting the SuperK result as a measure of $m_{\nu_\tau} =
(1/30-1/10)$ eV, momentarily ignoring mixing, and using
$\langle \phi^0 \rangle = 246$ GeV, we find $M/\lambda_{33} =
(6-18) \times 10^{14}$ GeV.

According to the seesaw mechanism, and again putting off the
question of mixing, the tau neutrino mass is given as
\begin{equation}
     m_{\nu_\tau} ~=~ {(m^D_{\nu_\tau})^2\over M^R_\tau}
\end{equation}
where $m^D$ is the Dirac mass that the neutrino would acquire
in the absence of the large Majorana mass of the right-handed
neutrino, and $M^R$ is the value of this Majorana mass.

If one assumes, within $SO(10)$, that the Dirac masses of the
third family are dominated by a contribution from a fundamental Higgs
${\bf 10_H}$ condensate then one obtains the relation $m_\tau (M_{\rm U})
= m_b(M_{\rm U})$ between the masses of the tau lepton and bottom
quark at the unification scale, which is known to be successful \cite{buras}.
This
suggests that the third family fermions get their masses primarily
from a ${\bf 10_H}$ condensate
through a Yukawa coupling ${\bf 16}_3 {\bf 16}_3
{\bf 10_H}$. (${\bf 16}_i$, $i=1-3$ denotes the three generations of
fermions.)  This hypothesis entails the relation
\begin{equation}
m^D_{\nu_\tau} (M_{\rm U}) \approx m_t(M_{\rm U}) \approx
(100-120)~{\rm GeV}~.
\end{equation}
(This numerical estimate is valid for most values of the MSSM parameters, so
long as
$\tan\beta$ is neither too large ($\ge 30$) nor too small ($\le 2$).)
Combining this with the seesaw formula and the SuperK
measurement\footnote{As is well known, for very low values of $\tan\beta < 2$, which
corresponds to a large $\nu_\tau$ Yukawa coupling, renormalization group
effects lead to significant increase in $m_{\nu_\tau}$, by more than
40\%, as the running momentum ($\mu$) is lowered from $M_{\rm U}$ to 1 GeV \cite{parida}.
Such effects however become progressively smaller ($\le 12\%$) for $\tan\beta \ge 3$.
Barring very low values of $\tan\beta \le 2$ which seem to be disfavored (see
remarks later), such renormalization effects on $m_{\nu_\tau}$ are thus small.
Henceforth we neglect them.}, one obtains $M^R_\tau \approx (1-3) \times 10^{14}$ GeV.

     In $SO(10)$, the Majorana masses of right-handed neutrinos
can be generated using either the
five-index self-dual antisymmetric tensor ${\bf
{\overline{126}_H}}$, or a bilinear product of
the spinorial Higgs ${\bf \overline{16}_H}$.  The relevant
interactions are the
renormalizable interaction
$\tilde{f}_{ij} {\bf 16_i 16_j \overline{126}_H}$ or the effective
nonrenormalizable interaction $f_{ij} {\bf 16_i 16_j \overline{16}_H
\overline{16}_H}/M$, respectively.  In terms of the
$SU(2)_L \times SU(2)_R \times SU(4)_C$
subgroup the relevant multiplets
transform as (1,3,10)$_H$ or (1,2,4)$_H$ respectively.

     If the ${\bf { \overline{126}_H}}$ is used to induce the Majorana
mass of $\nu^R_\tau$, then with $\langle {\bf {\overline{126}_H}} \rangle
\approx M_{\rm U}$ we require the Yukawa coupling $f_{33} \sim
10^{-2}$.  Considering that it was constructed by balancing quantities
of vastly different magnitudes, the nearness of this coupling to the
`natural' value unity is encouraging.

     Still more interesting is the situation that arises if we
employ the ${\bf \overline{16}_H}$.  In this case, as mentioned just above,
we require an effective nonrenormalizable interaction.  Such an
interaction could well arise through the exchange of superheavy states
associated with quantum gravity.  Then using $\langle {\bf \overline{16}_H}
\rangle \sim M_{\rm U} \approx 2\times 10^{16}$ GeV and $M \approx
M_{\rm Planck} \approx 2\times 10^{18}$ GeV we find
\begin{equation}
     M^R_\tau \approx f_{33} {\langle {\bf \overline{16}_H} \rangle^2
\over M }\approx (f_{33}) \times 2 \times 10^{14} {\rm ~GeV} ~.
\end{equation}
This is nicely consistent with the required value of $M^R_\tau$, for
$f_{33}$ close to unity!\footnote{The effects of neutrino mixing and of
possible choice of $M = M_{\rm string} \approx 4 \times 10^{17}$ GeV
(instead of $M = M_{\rm Planck}$) on $M_\tau^R$ are discussed later.}

Many of the considerations that follow do not depend on which
of these alternatives, $\overline{\bf 16_H}$ or $\overline{\bf 126}_H$,
is chosen.   Motivated partly
by the foregoing numerology, and partly by some suggestions from
higher symmetry schemes and string theory \cite{keith}, we will mainly
discuss models in which the pair $({\bf 16_H},
\overline{\bf 16_H})$ is used to break $(B-L)$.   Let us
note, however, that the ${\bf \overline{126}}_H$ does have some
advantages.   Specifically,
its couplings $\tilde{f}_{ij}$
are renormalizable, and its vacuum expectation value
violates $(B-L)$ by two units, and
thus conserves an R parity automatically \cite{rnm}.  The latter property
is important for
eliminating catastrophic -- dimension 4 -- sources of proton decay.
With the $SO(10)$ spinor condensates, we must
postulate a suitable $Z_2$ symmetry for this purpose separately.

\section{Large $(\nu_\mu-\nu_\tau)$ oscillation angle with
hierarchical masses}

Based on its measurements of atmospheric cosmic ray
neutrino oscillations,
the SuperK group estimates a large oscillation angle
sin$^22\theta_{\mu\tau}^{\rm osc} = (0.82-1)$ \cite{superk}.
If we compare this to the analogous
angle for quarks, sin$^22\theta_{cb} \approx 4 |V_{cb}|^2
\approx 6 \times 10^{-3}$, a challenge arises.  How are we to
understand the enormous difference between these two mixings, in a
framework where quarks and leptons are unified?

One widely considered possibility is to propose that the two relevant
neutrino flavor eigenstates are nearly degenerate \cite{degeneracy,sterile}.
Then a
small perturbation that lifts this degeneracy will induce
maximal mixing.  Such behavior is familiar from
the $K^0-\overline{K^0}$ system.  But while  there
is a well--established fundamental symmetry (CPT) which guarantees
exact degeneracy of $K^0$ and $\overline{K^0}$, no such
symmetry is known to operate for the different
flavors in the quark--lepton system.  In the
context of the seesaw mechanism, such degeneracy is awkward to
accommodate, at best.
Furthermore, if $\nu_\mu$ and $\nu_\tau$ are
nearly mass degenerate with a common mass of $\ge
1/30~ {\rm eV}$, then
there is no simple way to address the solar neutrino puzzle
through $\nu_e-\nu_\mu$ or $\nu_e-\nu_\tau$ oscillation.  The
possibilities appear bizarre -- either $\nu_e$ is nearly
degenerate with $\nu_\mu$ and $\nu_\tau$, or there is a
fourth ``sterile'' neutrino with the right mass and mixing
parameters.
In light of all this, it seems reasonable to consider
alternatives sympathetically.

We will argue now that large $\nu_\mu-\nu_\tau$ oscillation angle
can in fact arise, without requiring the
neutrinos to be nearly degenerate, along the lines of
some not entirely unsuccessful
attempts to relate mixing angles
and hierarchical masses
in the quark sector.

In the two sections following this one we shall analyze a complete model for
the
quark and lepton masses that is both overconstrained and
phenomenologically successful (though, to be sure, its construction
involves considerable guesswork).  Since that analysis becomes
rather complicated and perhaps intimidating,
we shall first, in this section, highlight some
salient features in a caricature model.

A leading idea in many attempts to understand large
mixing in the quark
sector is to utilize the properties of special matrices, whose form
might be constrained by simple symmetry requirements (e.g., requiring
symmetry or antisymmetry) and selection rules.  Such constraints can
easily arise, as we shall see,
from the group theory of unification,
given specific choices for the Higgs fields whose condensation
generates the masses.

The first attempts along these lines, which remain
quite intriguing and instructive, utilized a
$2 \times 2$ system of the form \cite{zee}
\begin{eqnarray}
M_{f} = \left(\matrix{0 & a \cr a & b}\right)_{f} ~, ~~f=(u,d)
\end{eqnarray}
with $a_f \ll b_f$.  Symmetric matrices arise, for example, in
$SO(10)$ (or left--right symmetric models \cite{lr})
where the matter fields are {\bf 16}s if the relevant
Higgs fields are {\bf 10}s.
The vanishing of the (1,1) entry can be ensured
by a suitable flavor symmetry that distinguishes the two relevant families.
The eigenvalues of this matrix are $(m_1,m_2)_f \simeq (a^2/b,b)_f$.  Thus
the off--diagonal element $a_f$ is nearly the geometrical mean of
the two eigenvalues.  For convenience in discussion,
we will refer to
mass matrices of this form as type A.  For type A mass matrices,
the mixing angles in each sector, up and down, are given by
\begin{equation}
{\rm tan}\theta_{f} = \sqrt{\left|{m_1 \over m_2}\right|}_{f}~.
\end{equation}
To obtain the observable
mixing angle one must of course combine the mixing angles
of the up and the down sectors.  In a world with two flavors, this
leads to the well--known
formula for the Cabibbo angle \cite{zee}:
\begin{equation}
\theta_C \simeq \left|\sqrt{{m_d \over m_s}} - e^{i \phi} \sqrt{{m_u
\over m_c}}\right|~.
\end{equation}
Using $\sqrt{m_d/m_s} \simeq 0.22$ and $\sqrt{m_u/m_c} \simeq 0.06$,
we see that Eq. (7) works within
30\% for any value of the phase $\phi$, and perfectly for a value
of the phase parameter $\phi$ around $\pi/2$.

A notable feature of the type A pattern, which remains valid even if the
requirement of symmetry is relaxed,  is that it
can support a strong hierarchy of eigenvalues by means of a much
weaker hierarchy of matrix elements.
For example, with $(a/b)_f
= 1/10$, one obtains a large hierarchy $(m_1/m_2)_f \sim 1/100$.

Now let us consider the implications of
adopting the type A pattern for
the $\mu-\tau$ sector, including the $2 \times 2$ Dirac
mass matrices of the charged leptons ($\mu-\tau$) and the neutrinos
($\nu_\mu-\nu_\tau$), and  the Majorana mass matrix of the
right--handed neutrinos ($\nu_{\mu}^R-\nu_{\tau}^R$).
Including the first family generally will not much
affect the discussion of the
$\mu-\tau$ sector, as we shall see.

The three matrices in the $\mu-\tau$ leptonic sector have the form
\begin{eqnarray}
M_{\ell,\nu}^D = \left(\matrix{0 & d_{\mu \tau} \cr d_{\mu \tau} &
d_{\tau \tau}}\right)^{\ell,\nu}~;~~ M_\nu^R = \left(\matrix{0 & y \cr y
& 1}\right)M_R
\end{eqnarray}
with the understanding that in each case the $(\tau \tau)$ entry
dominates.
One finds easily that the physical mass matrix for the light
left-handed
neutrinos,
$M_\nu^{\rm light} = -M_\nu^D (M_\nu^M)^{-1} (M_\nu^D)^T$,
takes a similar form, {\it viz}.

\begin{eqnarray}
M_\nu^{\rm light} = {1 \over M_R y^2}\left(\matrix{0 & -d_{\mu \tau}^2
y \cr -d_{\mu \tau}^2 y & \{d_{\mu \tau}^2 - 2 d_{\tau \tau} d_{\mu
\tau} y \}}\right)^\nu ~.
\end{eqnarray}
Denoting the two eigenvalues as $m_{\nu_2}$ and $m_{\nu_3}$, this
yields:
\begin{equation}
{m_{\nu_2} \over m_{\nu_3}} \simeq {-y^2 \over (1-2y {d_{\tau \tau}^\nu
\over d^\nu_{\mu \tau}} )^2 }, ~~
{\rm tan}\theta_{\mu \tau}^\nu = \sqrt{
{m_{\nu_2} \over m_{\nu_3}}},~~
{\rm tan}\theta_{\mu \tau}^\ell = \sqrt{
{m_{\mu} \over m_{\tau}}}~.
\end{equation}
Observe that the square--root formula holds for the mixing angles in all
sectors,
including the light neutrinos.

To illustrate the possibilities, let us consider the
phenomenologically interesting
hierarchy ratio $m_{\nu_\mu}/m_{\nu_\tau} \simeq 1/10$.  With this
ratio, one can accommodate $m_{\nu_\tau} \simeq .03$ eV as suggested
by atmospheric neutrino oscillations and a value of $m_{\nu_\mu}$
consistent with the small-angle MSW solution (with very small
$m_{\nu_e}$) \cite{krastev}.
This ratio is
achieved for
$(d_{\mu \tau}^\nu/d_{\tau \tau}^\nu,~ y)= (1/5,~ 1/13)$.
Combining the contributions from the charged
lepton sector and from the neutrino sector, the physical
oscillation angle
following from Eq. (10) is
\begin{equation}
\theta_{\mu \tau}^{\rm osc} \simeq \left|\sqrt{{m_\mu \over m_\tau}} - e^{i
\eta}\sqrt{{m_{\nu_\mu} \over m_{\nu_\tau}}}\right|~.
\end{equation}
Although the mixing angle  is not large in either sector
($\theta_{\mu \tau}^\ell \simeq \sqrt{m_\mu/m_\tau}
\simeq 0.25 \approx 14^o$ and $\theta_{\mu \tau}^\nu \simeq
\sqrt{m_{\nu_\mu}/m_{\nu_\tau}} \simeq \sqrt{1/10} \simeq 0.31 \approx
18^o$), if $\eta \approx \pi $ one obtains a near-maximal value for
the physical mixing parameter.
Actually the small
angle approximation used in Eq. (11) is too crude; the precise expression
is given by
\begin{equation}
{\rm sin}^22\theta_{\mu \tau} = {4\left(\sqrt{{m_\mu \over m_\tau}}\pm
\sqrt{{m_{\nu_\mu} \over m_{\nu_\tau}}}\right)^2\left(1 \mp\sqrt{{m_\mu
\over m_\tau}}\sqrt{{m_{\nu_\mu} \over m_{\nu_\tau}}}\right)^2
\over
\left(1+|{m_\mu \over m_\tau}|\right)^2\left(1+|{m_{\nu_\mu} \over
m_{\nu_\tau}}|\right)^2 }~,
\end{equation}
where the $\pm$ corresponds to $\eta = (\pi,0)$.
For $\eta = \pi$ and $m_{\nu_\mu}/m_{\nu_\tau} = 1/10$,
sin$^22\theta_{\mu \tau}^{\rm osc} \simeq 0.79$.
This example shows that large
oscillation angles can arise in a simple hierarchical model, without
any extreme adjustment of parameters.  As shown below, even larger oscillation
angles can
arise quite natually in more realistic models which can account also for
$V_{cb}$.

\section{A more ambitious $SO(10)$ model}

The model discussed in the previous section is not adequate
for a unified treatment of the
masses and mixings of the second and the third family
fermions.  Indeed, the square-root formula for $V_{cb}$ reads
\begin{equation}
|V_{cb}| \simeq \left|\sqrt{{m_c \over m_t}} - e^{i \chi} \sqrt{{m_s
\over m_b}} \right|~,
\end{equation}
so that with $\sqrt{m_s/m_b} \simeq 0.17$ and $\sqrt{m_c/m_t} \simeq 0.06$,
one cannot obtain observed value $V_{cb} \simeq 0.04 \pm 0.003$
for any value of $\chi$.
Thus the simplest symmetrical type A
mass matrices (Eq. (5)) cannot adequately describe
the hierarchical masses and the mixings of the
quarks.

We now propose to study in detail a concrete proposal for
asymmetric type A mass matrices that can be obtained within
$SO(10)$, and predicts correlations among the
quark--lepton and up--down mass matrices that are phenomenologically
acceptable.  Before plunging into the analysis, let us briefly
summarize its main conclusions.
In this section, we shall temporarily ignore the first family.

The four $2 \times 2$ Dirac mass matrices
$(U,D,L,$ and $N$) will be generated via four Yukawa couplings.  The
Majorana matrix of the right-handed  neutrinos involves two additional
Yukawa couplings.  Thus there are seven parameters (six
Yukawa couplings plus one ratio of vacuum expectation values)
to describe 10 observables (eight
masses -- $(m_c, m_t, m_s,m_b, m_\mu, m_\tau, m_{\nu_2}, m_{\nu_3})$ --
and two mixing angles -- $\theta_{\mu\tau}^{\rm osc}$ and $V_{cb}$) for the
$\mu$
and the $\tau$ families.
The system is overconstrained, and predicts three relations among
observables.
Two of
these concern the charged fermion sector, and they are reasonably well
satisfied.
The third prediction concerns the neutrino sector: the neutrino
oscillation angle will be predicted as a function of the mass ratio
$m_{\nu_2}/m_{\nu_3}$.  For $m_{\nu_2}/m_{\nu_3}=
(1/10-1/30)$, consistent with mass determinations from
the SuperK atmospheric neutrino
data and the small angle MSW solution of the solar neutrino
puzzle, a large $(\nu_\mu-\nu_\tau)$ oscillation angle is obtained,
as indicated by the SuperK atmospheric data.

In forming hypotheses for the form of the couplings responsible for
the mass matrices, we are guided by several considerations.
We assume an underlying $SO(10)$ unification symmetry, and that
unified symmetry breaking occurs through a minimal system of low
dimensional Higgs multiplets -- specifically,
$\left \langle {\bf 45_H} \right
\rangle$, one pair of $ \left \langle {\bf 16_H} \right \rangle$ and
$\left \langle \overline{\bf 16_H} \right \rangle$.  A
single ${\bf 10_H}$ is employed for electroweak symmetry
breaking.
The hierarchical masses will be assumed to arise from
type A mass matrices, as in Eq. (5), generalized to asymmetric form.
In addition, we must of course satisfy the broad
phenomenological
constraints $m_b(M_{\rm U}) \simeq m_\tau(M_{\rm U})$,
$m_s(M_{\rm U}) \neq m_\mu(M_{\rm U})$, and $V_{CKM} \ne 1$.
Combining these considerations
with a restriction to low dimensional
operators ($d \le 5$), we are led to suggest the following set of
Yukawa couplings \cite{abb}:
\begin{eqnarray}
{\cal L}_{\rm Yukawa} &=& h_{33} {\bf 16_3} {\bf 16_3} {\bf 10_H} +
{a_{23} \over M} {\bf 16_2} {\bf 16_3} {\bf 10_H} {\bf 45_H} +
\nonumber \\
&~& {g_{23} \over M} {\bf 16_2} {\bf 16_3} {\bf 16_H}{\bf 16_H} +
h_{23} {\bf 16_2} {\bf 16_3} {\bf 10_H}~.
\end{eqnarray}
Here $M$, a scale
associated with the effective non--renormalizable interactions,
could plausibly lie somewhere between the unification scale $M_{\rm U}$ and
$M_{\rm Planck}$.  For example,
the $a_{23}$ term might result from
integrating out a ${\bf 16} +{\bf \overline{16}}$ superfield pair with
mass of order $M$; or $M$ might be identified as $M_{\rm Planck}$ itself
if the nonrenormalizable interactions are associated with gravity.
A mass matrix of type A results
if the first term, $h_{33}\left \langle
{\bf 10_H} \right \rangle $,  is dominant. This ensures $m_b(M_{\rm U})
\simeq m_\tau(M_{\rm U})$ and $m_t(M_{\rm U}) \simeq m_{\nu_\tau}^D(M_{\rm
U})$.  The remaining terms, responsible for off-diagonal mixings,
must be smaller by about one order of magnitude.

In more detail,
our rationale for favoring the particular
off--diagonal couplings $a_{23}$
and $g_{23}$ is the following.
If the fermions acquired mass only through {\bf 10}s, one would obtain
too much symmetry between down quark and charged lepton masses, and in
particular the bad relation $m_s(M_{\rm U}) = m_\mu(M_{\rm U})$.
To avoid this, while eschewing proliferation
of extraneous Higgs multiplets,
the simplest possibility is to bring in the vacuum expectation value
of the
(1,1,15) component (under $G_{224}$) in the ${\bf 45_H}$ that is proportional
to
$(B-L)$.  A vacuum expectation value of this form figures prominently
in the Dimpoulos-Wilczek mechanism for doublet-triplet splitting in
$SO(10)$ \cite{dw}, as we shall discuss at length in connection with proton
decay.
Now if one restricts to
$d=4$ and $d=5$ operators in the superpotential, the only relevant
$SO(10)$ invariant effective
coupling that involves the ${\bf 45_H}$
is our $a_{ij} {\bf 16_i} {\bf 16_j}{\bf 10_H}{\bf 45_H}/M$.
In this term, only the ${\bf 120}$
in the decomposition $ {\bf 10_H} \times {\bf 45_H}~ (= {\bf 10} +
{\bf 120} + {\bf 320})$ can contribute to fermion mass matrix, and its
contribution
is {\it antisymmetric} in $(i,j)$.  These two couplings still do not
distinguish
up and down quark mass patterns, and so if they were the whole story one
would
have a trivial, identity, Cabibbo-Kobayashi-Maskawa (CKM) matrix.
The effective
couplings $g_{ij} {\bf 16_i}{\bf 16_j} {\bf 16_H}{\bf 16_H}/M$ remedy
this problem, without requiring addition to our
small set of Higgs multiplets.   When
a vacuum expectation value of order the unification scale
for the $SU(3)\times SU(2) \times U(1)$  singlet component of
${\bf 16_H}$ and a vacuum expectation value
of order the electroweak scale for the $SU(2)\times U(1)$ breaking
component of ${\bf 16_H}$ are
inserted, this term contributes to quark and lepton mass matrices.
To be more precise,
since the electroweak doublet contained in ${\bf 16_H}$ has the quantum
numbers of a down--type Higgs doublet, it contributes to the down
quark and charged lepton mass matrices, but not to the up sector.
The resulting up--down asymmetry generates non--zero CKM mixing angles.

With these four effective Yukawa couplings, the Dirac mass matrices
of quarks and leptons of the second and the third families
at the unification scale take the
form:
\begin{eqnarray}
U &=& \left(\matrix{0 & \epsilon + \sigma \cr -\epsilon+\sigma &
1}\right)m_U,~~~~
D=\left(\matrix{0 & \epsilon+ \eta \cr -\epsilon+\eta &
1}\right)m_D, \nonumber \\
N &=& \left(\matrix{0 & -3\epsilon+ \sigma \cr 3\epsilon+\sigma &
1}\right)m_U,~~~~
L=\left(\matrix{0 & -3\epsilon+ \eta \cr 3\epsilon+\eta &
1}\right)m_D~.
\end{eqnarray}
Here the matrices are multiplied by
left--handed fermion fields
from the left and by anti--fermion fields from
the right.  $(U,D)$ stand for the mass matrices of up and
down quarks, while $(N,L)$ are the Dirac mass matrices
of the neutrinos and the charged leptons.

The entries $(1, \epsilon, \sigma)$ arise respectively from the
$h_{33}, a_{23}$ and $h_{23}$ terms in Eq. (14), while $\eta$
entering into $D$ and $L$ receives contributions from both
$g_{23}$ and $h_{23}$; thus $\eta \neq \sigma$.  Note the
quark--lepton correlations between ($U,N$) as well as ($D,L$),
and the up--down correlation between ($U,D$) as well as
($N,L$).  These correlations arise because of the symmetry
structure of $SO(10)$.  The relative factor of $-3$ between
quarks and leptons involving the $\epsilon$ entry reflects the fact
that $\left \langle {\bf 45_H} \right \rangle \propto (B-L)$, while the
antisymmetry in this entry arises from the $SO(10)$ structure
as explained above.

Assuming $\epsilon, \eta, \sigma \ll 1$, we obtain at the unification scale:
\begin{eqnarray}
\left|{m_c \over m_t}\right| &\simeq& |\epsilon^2-\sigma^2|,~~~|{m_s \over
m_b}| \simeq
|\epsilon^2 -\eta^2| ~,\nonumber \\
\left|{m_{\mu} \over m_\tau}\right| &\simeq& |9 \epsilon^2 - \eta^2|,~~|m_b|
\simeq
|m_\tau| |1-8 \epsilon^2|~, \nonumber \\
|V_{cb}| &\simeq& |\sigma - \eta| ~.
\end{eqnarray}
One virtue of the asymmetric nature of the mass matrices (Eq. (15)) is worth
noting.
The simple expression $|\sigma -\eta|$ for $|V_{cb}|$ actually stands for
$\left| \sqrt{m_s/m_b}\left({\eta + \epsilon \over \eta -\epsilon}
\right)^{1/2}
- \sqrt{m_c/m_t} \left( {\sigma + \epsilon \over \sigma -
\epsilon}\right)^{1/2}\right|$.
Comparing with Eq. (13), which was based on symmetric type A mass matrices and
which
led to too big a value of $|V_{cb}|$, we see that the square--root mass ratios
are
multiplied by the respective asymmetry factors
$({\eta + \epsilon \over \eta - \epsilon})^{1/2}$ and
$({\sigma + \epsilon \over \sigma - \epsilon})^{1/2}$ (see Eq. (15)).  For
$\epsilon$
relatively negative compared to $\eta$ and $\sigma$ (as suggested by
considerations
of the leptonic mixings, see below), these factors provide the desired
lowering of
$V_{cb}$ compared to Eq. (13).

In Eq. (16) all the mass and mixing angle parameters are to be identified with
those at the unification scale.
One can evaluate $\sigma, \eta, \epsilon$ in terms of the
observed masses and mixing (extrapolated to the unification scale):
\begin{eqnarray}
\sigma &\simeq& [V_{cb}^2 + m_s/m_b - m_c/m_t]/(2V_{cb})~, \nonumber \\
\eta &\simeq& [-V_{cb}^2 + m_s/m_b-m_c/m_t]/(2V_{cb})~, \nonumber \\
\epsilon^2 &\simeq& \{ \sigma^2 + m_c/m_t\} \simeq (\eta^2 + m_\mu/m_\tau)/9~.
\end{eqnarray}
This leads to the sum rule:
\begin{equation}
{m_s \over m_b} \simeq {m_c \over m_t} -{5 \over 4} V_{cb}^2
\pm V_{cb}\left[{9 \over 16}
V_{cb}^2 + {1 \over 2} {m_\mu \over m_\tau} -{9 \over 2} {m_c \over
m_t}\right]^{1/2} ~.
\end{equation}

A word of explanation is needed here.  The parameters $\sigma, \eta$ are
in general complex.  In Eqs. (16)-(18)  therefore, one should interpret the
mass ratios and $V_{cb}$ to be also complex.  Let us denote $m_c/m_t =
\eta_{ct} |m_c/m_t|, m_s/m_b = \eta_{sb}|m_s/m_b|, V_{cb} =
\eta_{cb}|V_{cb}|$, with $|\eta_{ij}|= 1$.  Noting that each term on
the RHS of Eq. (18) is small compared to $m_s/m_b \simeq (1/30-1/50)$
(e.g: $m_c/m_t \approx 1/300,~ |V_{cb}||m_\mu/(2m_\tau)|^{1/2} \approx
1/140$), we see that essentially only one choice of the phase factors
can possibly allow the sum rule to work so that the
RHS is maximized -- i.e., $\eta_{ct} = -1,\eta_{\mu \tau} = +1$ and
$\eta_{sb} = -1$
with $\{\pm V_{cb}[~~]^{1/2}\}$ having a negative sign.
Thus, all phases are
constrained to be near the CP conserving limit ($0$ or $\pi$).
For simplicity, we will
take the  $\eta_{ij}$'s to be real.  As long as the phases are small ($\le
10^0$, say),
they would not of course affect our results.

The relatively simple pattern shown in Eq. (15) provides a
reasonable fit to all the masses and the mixing of the quarks and the
charged leptons in the second and the third families.  For example,
if we take as input $m_t^{\rm physical} = 174~{\rm GeV}$,
$m_c(m_c) = 1.37~{\rm GeV}$, $V_{cb} = 0.045$, in agreement
with the
values advocated in Ref. \cite{gasser,pdg}, and the
known $\mu$ and
$\tau$ lepton masses, then we obtain the predictions
\begin{eqnarray}
m_b(m_b) &\simeq& 4.9~{\rm GeV} ~,\nonumber \\
m_s (1~{\rm GeV}) &\simeq& 116~{\rm MeV}~.
\end{eqnarray}
In quoting the numbers in Eq. (19), we have extrapolated the GUT scale values
down to low energies using the beta functions
of the minimal supersymmetric extension
of the Standard Model (MSSM),  assuming
$\alpha_s(M_Z)= 0.118$, an effective SUSY threshold of $500~{\rm GeV}$ and
tan$\beta = 5$.  Our results depend only weakly on these input choices,
so long as $\tan\beta$ is neither too large ($\ge 30$) nor too small
($\le 2)$.

The mass of the strange quark is somewhat low compared to the central value
advocated in
Ref. \cite{gasser}, but is closer in value to the recent lattice determinations
\cite{gupta}. The $b$--quark mass prediction is also in reasonable agreement
with determination from $\Upsilon$ spectroscopy.  In this regard, it should be
mentioned that exact $b-\tau$ unification in  supersymmetric unified models,
without taking finite threshold effects due to the gluino into account, would
yield a $b$--quark mass which is about 10-20\% above the experimental value for
a wide range of the parameter $\tan\beta$ \cite{barger}.
In our case, $b$ and $\tau$ masses are not exactly equal at $M_{\rm U}$,
$m_b$ is about 8\% lower than $m_\tau$ (see Eq. (16)).  This difference, which
has its origin
in the $(B-L)$ generator associated with the off--diagonal entry $\epsilon$ in
Eq. (15), leads to
better agreement with the experimental value of $m_b (m_b)$.

The parameters $\sigma, \eta, \epsilon,m_U$ and $m_D$ are found to
be
\begin{eqnarray}
\sigma &\simeq& -0.110~ \eta_{cb}~,~
\eta \simeq -0.151 ~\eta_{cb}~,~
\epsilon \simeq 0.095 ~\eta_\epsilon~, \nonumber \\
m_U &\simeq& m_t(M_{\rm U}) \simeq (100-120)~{\rm GeV}~,~m_D \simeq
m_b(M_{\rm U}) \simeq 1.5~{\rm GeV}~.
\end{eqnarray}
($\eta_\epsilon=\pm 1$ is the phase of $\epsilon$.)
In addition to the two predictions in Eq. (19), the Dirac masses of the
neutrinos
and the left--handed mixing angles in the $\mu-\tau$ sector are determined to
be
\begin{eqnarray}
m^D_{\nu_\tau}(M_{\rm U}) &\simeq& m_t(M_{\rm U}) \simeq 100-120~ {\rm
GeV}~,\nonumber \\
m^D_{\nu_\mu}(M_{\rm U}) &\simeq& (9 \epsilon^2-\sigma^2)m_U \simeq
8~ {\rm GeV}~, \nonumber \\
\theta^\ell_{\mu \tau} &\simeq& -3\epsilon+\eta \simeq -0.437
\eta_\epsilon~({\rm for~
\eta_{cb}/\eta_\epsilon = +1})~.
\end{eqnarray}
$\eta_{cb}/\eta_\epsilon = +1$ is required to maximize
$\theta^\ell_{\mu \tau}$.
Note that the Dirac mass of $\nu_\mu$ is quite different from
$m_c(M_{\rm U}) \simeq {\rm 300~MeV}$.  This is contrary to what is often
adopted in $SO(10)$.
The difference arises because of the type A pattern of the mass matrices
and the factor of $-3$ associated with the $(B-L)$ generator
that goes into the $\nu_\mu$ Dirac mass \cite{shafi}.

Given the bizarre pattern of quark and lepton masses and mixing,
we regard the overall fit to all of them, good to within 10\%,
using the pattern shown in Eq. (15), as reason to take this pattern
seriously.

Note that owing to the asymmetric mass matrix, the square--root
formula for the mixing angle $\theta_{\mu\tau}^\ell$ receives a
correction given by the factor
$[(-3\epsilon+\eta)/(3\epsilon+\eta)]^{1/2}$.  For the values of
$\eta$ and $\epsilon$ determined in Eq. (21), this factor is about
1.8 for $\eta_\epsilon/\eta_{cb} = +1$.

Although all the entries for the
Dirac mass matrix are now fixed,
to obtain the parameters for the light neutrinos
one needs to specify the Majorana
mass matrix.  Unfortunately, here there is much less information to guide
our hypotheses.
For concreteness, let us
imagine that this too takes the type A form:
\begin{eqnarray}
M_\nu^R = \left(\matrix{0 & y \cr y & 1}\right)M_R~,
\end{eqnarray}
where we allow $y= \eta_y|y|$ to have either sign, i.e., $\eta_y = \pm
1$.  Note that Majorana mass matrices are constrained to be symmetric
by Lorentz invariance.  The seesaw mass matrix $(-N
(M_\nu^R)^{-1}N^T$) for the light $(\nu_\mu-\nu_\tau)$ system is then
\begin{eqnarray}
M_\nu^{\rm light} = \left(\matrix{0 & A \cr A & B}\right){m_U^2 \over M_R}~,
\end{eqnarray}
where $A \simeq (\sigma^2-9 \epsilon^2)/y$ and $B\simeq -(\sigma +
3\epsilon)(\sigma + 3\epsilon-2y)/y^2$.  With $A \ll B$,
this yields
\begin{equation}
m_{\nu_3} \simeq B{m_U^2 \over M_R};~~{m_{\nu_2} \over m_{\nu_3}}
\simeq -{A^2 \over B^2};~~ {\rm tan}\theta_{\mu \tau}^\nu = \sqrt{{m_{\nu_2}
\over m_{\nu_3}}}~.
\end{equation}
Correspondingly,
\begin{equation}
y\eta_{\epsilon} \simeq
{{\pm \sqrt{{m_{\nu_2} \over m_{\nu_3}}}
(3|\epsilon|-|\sigma|)}
\over
3|\epsilon|+|\sigma| \pm 2\sqrt{{m_{\nu_2}\over m_{\nu_3}}}}~,
\end{equation}
where $\eta_\epsilon = \pm1$ and we have used the fact that $\epsilon$
and $\sigma$ are relatively negative for $\eta_{cb}/\eta_{\epsilon}=+1$
(See Eq. (21)).  For a given
choice of the sign of $y$ relative to that of $\epsilon$, and for a
given mass ratio $m_{\nu_2}/m_{\nu_3}$, we can now determine
$y\eta_\epsilon$ using Eq. (25) and the values of $\epsilon$ and
$\sigma$ obtained in Eq. (20).  Corresponding to $m_{\nu_2}/m_{\nu_3} =
(1/10, 1/15,1/20,1/25,1/30)$, we calculate $(y\eta_\epsilon)_+=
(0.0543, 0.0500, 0.0468,0.0444,0.0424)$ and $(y\eta_\epsilon)_- =(0.2359,
0.3781, 0.7681,$~$8.4095, -1.0519)$, where the subscripts $\pm$
correspond to $\eta_y \eta_\epsilon = \pm 1$.  The case of $\eta_y
\eta_\epsilon = +1$ typically requires smaller values of $|y|$ than
for the case of $\eta_y \eta_\epsilon = -1$.  The former ($\eta_y
\eta_\epsilon = +1$)
is more in accord with the idea of the type A matrix with
flavor symmetries being the origin of hierarchical masses,
than the latter ($\eta_y \eta_\epsilon = -1$).

We obtain for the neutrino oscillation angle:
\begin{equation}
\theta_{\mu \tau} \simeq \theta_{\mu \tau}^\ell - \theta_{\mu \tau}^\nu \simeq
0.437 \pm \sqrt{{m_{\nu_2} \over m_{\nu_3}}}~.
\end{equation}
If $\eta_y/ \eta_\epsilon = +1$, the relative $+$ sign should be
chosen, and  for $\eta_y/\eta_\epsilon = -1$
the relative $-$ sign should be chosen.
Taking the
relative $+$ sign, i.e., $\eta_y\eta_\epsilon=+1$ (in accord with the smaller
values of $y \le 1/10$),
and using the more precise
expression given in Eq. (12) (and replacing $\sqrt{m_\mu/m_\tau}$ by
the numerical value of the charged lepton mixing angle $\simeq 0.437$), we
obtain
\begin{eqnarray}
{\rm sin}^22\theta_{\mu \tau} &=& ( 0.96, 0.91, 0.86, 0.83, 0.81)
\nonumber \\
{\rm for}~~m_{\nu_2}/m_{\nu_3} &=& (1/10, 1/15, 1/20, 1/25, 1/30)~.
\end{eqnarray}

As previously advertised, we see that
one can derive rather plausibly a large $\nu_\mu-\nu_\tau$ oscillation
angle sin$^22\theta_{\mu\tau} \ge 0.8$,
together with an understanding of
hierarchical masses and mixing of the quarks and the charged leptons,
while maintaining a
large hierarchy in the seesaw derived masses ($m_{\nu_2}/m_{\nu_3} =
1/10-1/30$) of $\nu_\mu$ and $\nu_\tau$, all within a unified
framework
including both quarks and leptons.  In the example exhibited
here, the mixing angles for the mass eigenstates of neither the neutrinos
nor the charged leptons are really large,
$\theta_{\mu \tau}^\ell \simeq 0.437
\simeq 23^0$ and $\theta_{\mu \tau}^\nu \simeq (0.18-0.31) \approx (10-18)^0$,
{\it yet the oscillation angle obtained by combining the two is near-maximal.}
This contrasts with most previous work, in which a large
oscillation angle is obtained either entirely from the neutrino sector
(with nearly
degenerate neutrinos) or entirely from the charged lepton sector.
In our case, the mass eigenstates of the neutrinos and the charged
leptons are approximately also the respective gauge eigenstates.

\section{Inclusion of the first family: $\boldmath{\nu_{\rm e}-\nu_\mu}$
oscillation}

There are several alternative ways to include the first family, without
upsetting the successful predictions of the 2-3 sector.  In the
absence of a deeper understanding, the theoretical
uncertainties in analyzing the masses and mixings of the first family
are much greater than for the heavier families, simply
because the masses of the
first family are so small, that relatively tiny perturbations
can significantly affect their values.  With this warning,
we will now briefly consider, as
an illustrative example and ``proof of principle'', a
minimal extension to the first family,
inspired by the type A pattern, and also a noteworthy variant.

The $3 \times 3$ Dirac
mass matrices in the four sectors take the form:
\begin{eqnarray}
U &=& \left(\matrix{0 & \epsilon' & 0 \cr -\epsilon' & 0
& \epsilon+ \sigma \cr 0 & -\epsilon + \sigma & 1}\right) m_U,~~
D = \left(\matrix{0 & \epsilon'+\eta' & 0 \cr -\epsilon'+\eta' & 0
& \epsilon+ \eta \cr 0 & -\epsilon + \eta & 1}\right)m_D,  \nonumber \\
N &=& \left(\matrix{0 & -3\epsilon' & 0 \cr 3\epsilon' & 0
& -3\epsilon+ \sigma \cr 0 & 3\epsilon + \sigma & 1}\right)m_U,~~
L = \left(\matrix{0 & -3\epsilon'+\eta' & 0 \cr 3\epsilon'+\eta' & 0
& -3\epsilon+ \eta \cr 0 & 3\epsilon + \eta & 1}\right)m_D.
\end{eqnarray}

At the level of underlying primary couplings,
the $\epsilon'$ term is very similar to the $\epsilon$ term,
arising
through a Yukawa coupling $a_{12} {\bf 16}_1 {\bf 16}_2 {\bf 45_H}
{\bf 10_H}$.
The
$\eta'$ term arises through the coupling $g_{12}{\bf 16}_1
{\bf 16}_2 {\bf 16_H} {\bf 16_H}$.
With $\epsilon, \sigma, \eta, m_U$ and $m_D$ determined essentially by
considerations
of the second and the third families (Eq. (20)), we now have just two new
parameters
in Eq. (28) -- i.e. $\epsilon'$ and $\eta'$ -- which describe five new
observables in the quark and charged lepton sector:
$m_u, m_d, m_e, \theta_C$ and $V_{ub}$.  Thus with $m_u \approx 1.5$ MeV
(at $M_{\rm U}$) and
$m_e/m_\mu$ taken as inputs to fix  $\epsilon'$ and $\eta'$, one can calculate
the other three observables.  In addition, $m^D_{\nu_e}$ will also
be determined.

At the outset, it is worth noting that this specific pattern (Eq. (28))
is consistent with the empirical unification scale relations:
$m_d \simeq 3 m_e, m_s \simeq m_\mu/3$ \cite{gj} and $m_b \simeq
m_\tau$, which in turn imply $m_d m_s m_b \approx m_e m_\mu m_\tau$,
and thus Det($D$) $\simeq$ Det($L$).  Eq. (28) obeys this relation to
a good approximation,
for the
following reason.  From Det($U) =\epsilon'^2 m_U^3$, one obtains
$\epsilon' \simeq \sqrt{m_u/m_c} (m_c/m_t) \approx 2 \times 10^{-4}$.
Using Det$(L)=(9 \epsilon'^2-\eta'^2)m_D^3 = m_e m_\mu m_\tau$, and
neglecting the $\epsilon'^2$ term (justified {\it a posteriori\/}), one obtains
$|\eta'| \simeq \sqrt{m_e/m_\mu} (m_\mu/m_\tau)$ $\simeq 4.4 \times
10^{-3}$.  Since $|\epsilon'| \ll |\eta'|$, one gets Det$(D)\simeq
{\rm Det}(L)$, as described.

Combining the two predictions for the second and the third families given by
Eq. (19),
we are thus led to a total of five predictions for observable parameters of the
quark and charged lepton system of the three families.
\begin{eqnarray}
m_b(m_b) &\simeq& 4.9~{\rm GeV} \nonumber \\
m_s(1~{\rm GeV}) &\simeq& 116~{\rm MeV} \nonumber \\
m_d(1~{\rm GeV}) &\simeq& 8~{\rm MeV} \nonumber \\
\theta_C &\simeq& \left|\sqrt{m_d/m_s} - e^{i\phi}\sqrt{m_u/m_c}\right|
\nonumber \\
|V_{ub}/V_{cb}| &\simeq& \sqrt{m_u/m_c} \simeq 0.07 ~.
\end{eqnarray}
Further, the Dirac masses and mixings of the neutrinos and the mixings of the
charged
leptons also get determined.  Including those for the $\mu-\tau$ families
listed in Eq. (21) we obtain:
\begin{eqnarray}
m_{\nu_\tau}^D(M_{\rm U}) &\approx & 100-120~{\rm
GeV};~m_{\nu_\mu}^D(M_{\rm U}) \simeq 8~{\rm GeV},~
\theta_{\mu \tau}^\ell \simeq -.0.437 \eta_\epsilon \nonumber \\
m_{\nu_e}^D &\simeq& [9 \epsilon'^2/(9 \epsilon^2-\sigma^2)]m_U \simeq
0.4~{\rm MeV} \nonumber \\
\theta_{e\mu}^\ell &\simeq& \left[{\eta'-3\epsilon' \over
\eta'+3\epsilon'}\right]^{1/2}
\sqrt{m_e/m_\mu} \simeq 0.85 \sqrt{m_e/m_\mu} \simeq 0.06 \nonumber \\
\theta_{e\tau}^\ell &\simeq& {1 \over .85} \sqrt{m_e/m_\tau} ({m_\mu/m_\tau})
 \simeq 0.0012~.
\end{eqnarray}
In evaluating $\theta_{e\mu}^\ell$, we have assumed $\epsilon'$ and $\eta'$ to
be
relatively positive.

Note that the first five predictions in Eq. (29) pertaining to observed
parameters in
the quark system are fairly successful.  Considering the bizarre pattern of
the masses and mixings of the fermions in the three families (recall comments
on
$V_{cb}$, $m_b/m_\tau$, $m_s/m_\mu$ and $m_d/m_e$), we feel that the success of
the mass pattern exhibited by Eq. (28) is rather remarkable.  It is worth
stressing that (a) the asymmetric mass pattern induced by the $a_{23}$ coupling
(see Eq. (14)), (b) its contribution being proportional to $B-L$ and (c) the
$g_{23}$
coupling leading to non--trivial CKM matrix have all played essential roles
in leading to a successful mass pattern with a minimal Higgs system -- i.e.,
involving a single ${\bf 45}_H$, ${\bf 16_H}$ and ${\bf 10}_H$ -- two of which
are
needed for unification scale symmetry breaking of $SO(10)$ anyhow.  This is one
reason for taking patterns like Eq. (28) seriously as a guide for
considerations on proton decay.  A particularly
interesting variant is obtained in the limit $\epsilon' \rightarrow 0$, as we
will discuss
at the end of this section.

 Turning to the Majorana mass matrix, one could in general introduce (1,1),
 (1,2) and (1,3) entries involving the first family.  We choose the
 (1,2) element to be zero, both because by itself (i.e., with (1,2)
 $\ne 0$, but (1,1) = (1,3) = 0) it would upset the success of the
 $\nu_\mu-\nu_\tau$ sector obtained above, and also for
 economy in parameters.  We therefore consider the following pattern:

\begin{eqnarray}
M_\nu^R = \left(\matrix{x & 0 & z \cr 0 & 0 & y \cr z & y &
1}\right)M_R~.
\end{eqnarray}
Consistent with our presumption that hierarchical masses have
their origin in the type A pattern,
we will assume that $x \sim z^2$ and $z \le
y < 1$ and that the zero in the (2,2) entry only reflects that this
entry is small compared to $y^2$.

Note that all entries in
the Majorana matrix arise through $f_{ij} {\bf 16}_i {\bf 16}_j
{\bf \overline{16}_H}
{\bf \overline{16}_H}$, which however do not contribute to the Dirac mass
matrices.  As we have stressed before,
they therefore need not have the same flavor structure as the
Dirac mass matrices of Eq. (28).
Given the Dirac matrix $N$, we
can work out the light neutrino mass matrix after the seesaw
diagonalization.  It is given by:
\begin{equation}
M_\nu^{\rm light} = {m_U^2 \over M_R}{1 \over xy^2} \times
\end{equation}
\begin{eqnarray}
\left(\matrix{9 \epsilon'^2(x-z^2) &
3\epsilon'y\{-3\epsilon' z +(-3 \epsilon+\sigma)x\} & 3
\epsilon'\{xy-(3\epsilon+\sigma)(
x-z^2)\} \cr
3\epsilon'y\{-3\epsilon' z +(-3 \epsilon+\sigma)x\} & -9 \epsilon'^2 y^2 &
(3 \epsilon+\sigma)y\{3 \epsilon'z-(-3 \epsilon+\sigma)x\} \cr
 3 \epsilon'\{xy-(3\epsilon+\sigma)(x-z^2)\} &
(3 \epsilon+\sigma)y\{3 \epsilon'z-(-3 \epsilon+\sigma)x\} &
(3 \epsilon+\sigma)\{-2xy+(3\epsilon+\sigma)(x-z^2)\} }\right) \nonumber
\end{eqnarray}

Three features arising from  Eq. (30) are especially
noteworthy.  First, that the parameter $x$ figures
prominently only in the mass of $\nu_e$, which is given by
\begin{equation}
m_{\nu_e} \simeq { 81 \epsilon'^4 \over xy^4(9 \epsilon^2-\sigma^2)^2}
({m_U^2 \over M_R}) \simeq {10^{-10} ~{\rm eV} \over x}~,
\end{equation}
where we have inserted the mass of $\nu_\tau \simeq 1/20~{\rm eV}$.
For $x=(10^{-5}-10^{-4})$, $m_{\nu_e} \approx (2 \times 10^{-5}-
2 \times 10^{-6})~{\rm eV}$.  ($x$ should be less than $10^{-4}$,
otherwise one would run into a conflict with proton lifetime, see later.)
Second, the discussion of the $\mu-\tau$ sector in Sec. IV is corrected by
terms that
are
proportional to $\epsilon'$, which is tiny, or by terms proportional to $z$.
As long as $z^2 \le x/3$ (from the (3,3) entry) and $z < 10^3 x$ (from the
(2,3) entry), the discussion of the $(2,3)$ sector will remain
essentially unaltered.
And third, that the $\nu_e-\nu_\mu$
mixing angle arising from the light neutrino mass matrix itself (as
opposed to the contribution from charged lepton mixing) is
tiny, but still significant for the small angle MSW explanation of
the solar neutrino puzzle. The $\nu_e-\nu_\tau$ mixing is also small, since
$\epsilon' \simeq 2 \times 10^{-4}$ is extremely small.

Note that the $3 \times 3$ Majorana mass matrix $M_\nu^R$ introduces four new
parameters: $x,y,z$ and $M_R$, of which the magnitude of $M_R \sim $ (few to
10) $\times
10^{14}$ GeV can quite plausibly be justified in the context of supersymmetric
unification (see Sec. II).  There are altogether six observables in the
three neutrino system: the three masses and the three oscillation angles.
While
these depend on parameters of both the Majorana and the the Dirac mass
matrices,
since the latter are fully determined from the quark and the charged lepton
system, one can expect at least two predictions for the neutrino system.
These can
be taken to be $\theta_{\mu\tau}^{\rm osc}$ and, for example, $m_{\nu_e}$.
In addition, to the extent that the magnitude of $M_R$ can plausibly be
justified, that of $m_{\nu_\tau} \sim (1/10-1/30)~eV$ can also be anticipated.
We should add that for the limiting case of $\epsilon' = 0$ (a parameter of
the Dirac sector), one would obtain predictions for $\theta_{e\mu}^{\rm osc}$
and
$\theta_{e\tau}^{\rm osc}$ as well, regardless of the values of $x,y,z$ (see
below).

As an example, let us take $x=10^{-4}, y=0.05$ and $z = -0.002$ along with the
``central values" of $\sigma, \eta, \epsilon$ given in
Eq. (21) and $\epsilon' = 2 \times 10^{-4}, \eta'= 4.4 \times 10^{-3}$ (from
a fit to the $u$--quark and electron masses).  Let us also put, in accord with
discussions in Sec. II, $M_R \approx (5-15)\times 10^{14}~{\rm GeV}$.  The mass
eigenvalues and the mixing angles in the neutrino sector are then:
\begin{eqnarray}
m_{\nu_\tau} &\approx& (1/10-1/30)~eV \nonumber \\
m_{\nu_\mu} &\simeq& 6.7 \times 10^{-3}(1-1/2)~eV \nonumber \\
m_{\nu_e} &\simeq& 2 \times 10^{-6}(1-1/2)~eV \nonumber \\
\theta_{\mu\tau}^\nu &\simeq& 0.25,~ \theta_{e\mu}^\nu \simeq
0.009,~\theta_{e\tau}^\nu
\simeq 0.0034 ~.
\end{eqnarray}
Combining with the mixing angles in the charged lepton sector, the oscillation
angles, including that for $\nu_\mu-\nu_\tau$ sector given before, are given by
(setting the relative phases to be 0 or $\pi$):
\begin{eqnarray}
\theta_{\mu\tau}^{\rm osc} &\simeq& 0.437 + \sqrt{m_{\nu_2}/m_{\nu_3}}
\nonumber \\
\theta_{e\mu}^{\rm osc} &\simeq& \theta_{e\mu}^\ell - \theta_{e\mu}^\nu
\simeq 0.06 \pm 0.009 \nonumber \\
\theta_{e\tau}^{\rm osc} &\simeq& \theta_{e\tau}^\ell - \theta_{e\tau}^\nu
\simeq  5.5 \times 10^{-4} \pm 0.0034 ~.
\end{eqnarray}

Though the mixing angles involving the first family are small,
$\theta_{e\mu}^\nu$ does play an important role for solar neutrinos.  Combining
$\theta_{e\mu}^\nu$
with the corresponding mixing angle from the charged lepton
sector, $\theta_{e\mu}^\ell \simeq 0.85 \sqrt{m_e/m_\mu} \simeq 0.06$,
where the
factor $0.85$ is a correction factor from the asymmetric $\epsilon'$ term in
Eq. (28), we find that $\theta_{e \mu}^{\rm osc} \simeq 0.05$ if the two
contributions subtract.  This is just about the right value to explain the
solar neutrino data via the small angle MSW mechanism \cite{krastev}.  Note
that without the (1,3) entry $z$ in Eq. (31), $\theta_{e\mu}^\nu$ would have
been less than $0.002$, which would have resulted in $\theta_{e\mu}^{\rm osc}$
being
larger than the required value for MSW by about 15-20\%.  This is a possible
motivation to introduce the (1,3) entry $z$ into Eq. (31).  Since the
discussion of the (2-3) sector is little affected so long as $z < 10^3 x$
and $z^2 \le x/3$, we
will allow a range $|z| = (0.002-0.006)$, corresponding to $x =
(10^{-5}-10^{-4})$.
This will be significant for
the discussion of proton decay,
especially for final states involving charged leptons.

Another consideration for the choice of the various entries in $M_\nu^R$ is
the generation of cosmological baryon asymmetry.
Note that for $x = 10^{-5}-10^{-4}$, the mass of $\nu_e^R$ is in the range
$5 \times (10^9-10^{11})$ GeV, which is of the right magnitude for
producing $\nu_e^R$ following reheating and inducing lepton
asymmetry
in $\nu_e^R$ decay, that is subsequently
converted to baryon asymmetry via the
electroweak sphalerons \cite{sphaleron}.  For the lepton number violating
decay of $\nu_e^R$ to be out of equilibrium, it is necessary that
the mixing of $\nu_e^R$ with $\nu_\tau^R$ be less than a few times
$10^{-3}$.  Otherwise the decay $\nu_e^R \rightarrow H^0 + \nu_\tau^L$,
which can proceed by utilizing the large (order one) Dirac Yukawa coupling
$h_{33}$
of $\nu_\tau^R$ will bring $\nu_e^R$ into equilibrium, inhibiting the
generation
of lepton asymmetry.  The choice $z \simeq 0.002$ is consistent with this
constraint.

While we cannot claim that our several choices have been
unique,
we have demonstrated
that a rather simple pattern for the four Dirac
mass matrices, motivated and constrained by the group structure of
$SO(10)$,
is
consistent within 10 to 20\% with all observed masses and
mixing of the quarks and the charged leptons.  This fit is
significantly overconstrained, as we have discussed.  The same
pattern, supplemented with a similar structure for the Majorana
mass matrix, quite plausibly accommodates both the SuperKamiokande
result with the large $\nu_\mu-\nu_\tau$ oscillation angle
required for the atmospheric neutrinos and a
small $\nu_e-\nu_\mu$ oscillation angle relevant for
theories of the solar neutrino deficit.

\subsection*{The $\epsilon'=0$  Limit: A special case}

Before turning to proton decay, it is worth noting that much of our discussion
of fermion masses and mixings, including those of the neutrinos, is
essentially unaltered if we go to the limit $\epsilon' \rightarrow 0$ of
Eq. (28).  This limit clearly involves:
\begin{eqnarray}
m_u &=& 0,~ \theta_C \simeq \sqrt{m_d/m_s} \nonumber \\
|V_{ub}| &\simeq& \sqrt{{\eta-\epsilon \over \eta+\epsilon}}\sqrt{m_d/m_b}
(m_s/m_b)
\simeq (2.1)(0.039)(0.023) \simeq 0.0019 \nonumber \\
m_{\nu_e} &=& 0 ,~ \theta_{e\mu}^\nu  = \theta_{e\tau}^\nu = 0 ~.
\end{eqnarray}
All other predictions will remain unaltered.
Now, among the observed quantities in the list above, $\theta_C \simeq
\sqrt{m_d/m_s}$
is indeed a good result.  Considering that $m_u/m_t \approx 10^{-5}$, $m_u =
0$ is
also a pretty good result.  There are of course, plausible small corrections
arising
from higher dimensional operators (for example) involving Planck scale physics
which could
induce a small value for $m_u$.  The question arises: Does such a small
correction
contributing to $m_u$ arise primarily through the (1,2) entry (i.e.,
$\epsilon' \neq 0$)
in $U$ and $N$ or does it arise primarily through the (1,1) entry -- to be
called
$\delta \neq 0$ if $\epsilon' = 0$ -- of $U$ and $N$?  In case of the former
(with
$\delta = 0$), one would get $\epsilon' \approx 2 \times 10^{-4}$, as noted
above;
while for the latter (with $\epsilon' = 0, \delta \neq 0$), one would obtain
$\delta \simeq m_u/m_t \approx 10^{-5}$.  Note that putting $\epsilon' = 0$
does not
matter for $D$ and $L$ because $\eta' \gg \epsilon'$.  The two cases therefore
yield
essentially identical results as regard fermion masses and mixings.
Differences
of about 20\% between their predictions can arise only as regards
$\theta_{e\mu}^{\rm osc}$
for which the former can fare better (depending upon the choice of phases)
than the
latter; while the latter ($\delta \neq 0, \epsilon' = 0$) fares better than
the former as
regards $\theta_C$.  One possible way to distinguish these two closely related
variants for fermion masses and mixings will be a more precise determination of
$V_{ub}$.  The former predicts it to be near $\sqrt{m_u/m_c}V_{cb} \simeq
(0.07)(0.04)
\simeq 0.0028$, while the latter is somewhat smaller ($V_{ub} \simeq 0.0019$).
 We
will refer to these two variants as cases I and II respectively.
\begin{eqnarray}
&{\bf {\rm Case~I}}:&  \epsilon' \approx 2 \times 10^{-4},~\delta = 0
\nonumber \\
&{\bf {\rm Case~II}}:& \delta \approx 10^{-5},~ \epsilon' = 0
\end{eqnarray}

As we will see, however, proton decay amplitudes differ
significantly between the two cases.  Typically they are larger for
case I (by a factor of 2-3) than for case II. Not knowing the
precise origin of the small corrections contributing to $m_u$, we
will consider cases I and II as equally viable guides for forming expectations
for proton decay.  We will in fact derive the proton
decay amplitudes with $\epsilon' \neq 0$ (case I).  Given that
$\delta \ll \epsilon'$, we obtain the results for case II simply
by setting $\epsilon' \simeq 0$.

\section{Neutrino masses and proton decay}

In an earlier paper \cite{bpw} we pointed out that the theory of
neutrino masses can significantly
affect  expectations for proton decay,
as regards both its rate and, especially, its branching ratios.
This happens because in supersymmetric unified theories
a new set of color triplet fields is needed to generate
heavy Majorana masses for the right-handed (RH) neutrinos, as
required for the seesaw mechanism.
Exchange of its superpartners generates new dimension 5 operators,
that appear in addition to the
``standard" $d=5$ operators in the effective Lagrangian.
The standard operators
arise through the exchange of color
triplets that are related to the electroweak doublets, such as those
appearing
in the ${\bf 5}+ {\bf \overline{5}}$ of $SU(5)$ or the
${\bf 10}$ of
$SO(10)$ \cite{d5,ellis,nath,hisano}.

The standard $d=5$ operators are estimated to
dominate
over the gauge boson mediated $d=6$ proton decay operators.
Furthermore, owing to a combination of color antisymmetry,
Bose symmetry of the superfields, and hierarchical
Yukawa couplings of the fermions, they predict that the dominant modes
are
$\overline{\nu}_\mu K^+$ and $\overline{\nu}_\mu \pi^+$ modes, while
$e^+ \pi^0$ and $e^+ K^0$ and even $\mu^+ \pi^0$
and $\mu^+ K^0$ modes are highly suppressed, at least
for small and moderate values
of tan$\beta$ ($\le 15$) \cite{d5,ellis,nath,hisano}.

In Ref. \cite{bpw} we argued that
for neutrino masses in a
plausible range
the new $d=5$ operators could
compete favorably with the standard ones,
and induce
potentially observable, though not yet excluded, proton decay rates.
Within the new contribution
charged lepton decay mode amplitudes,
including those for $\mu^+ K^0, \mu^+ \pi^0$
and possibly $e^+ K^0$ and $e^+ \pi^0$, were generally found to be significant
relative to the
$\overline{\nu} K^+$ and $\overline{\nu} \pi^+$ modes,
even for low tan$\beta$.  Thus
charged lepton modes of nucleon decay potentially
furnish an independent signature
for the interactions that play a central
role in the theory of neutrino masses.

As mentioned earlier, in our previous quantitative work
we had to  hypothesize
values for the
neutrino masses.  Now, after the SuperK observations,  we can
revisit the quantitative aspect with some crucial
experimental
information reliably in hand.
Changes are induced not only
for the new $d=5$ operators that are directly related to neutrino
masses, but also for the standard $d=5$ operators.

To address these issues in detail, we must specify the nature of the
color triplet Higgsino couplings.  This issue is closely tied to the
mechanism of doublet--triplet splitting in $SO(10)$, to which we now turn.

\subsection{Natural doublet-triplet splitting in $SO(10)$}

In supersymmetric $SO(10)$, a
natural doublet--triplet splitting can be
achieved by coupling the adjoint
Higgs ${\bf 45_H}$ to a ${\bf 10_H}$ and a ${\bf 10'_H}$, with
${\bf 45_H}$ acquiring a unification--scale VEV in the $B-L$ direction
\cite{dw}:
$\left \langle {\bf 45_H} \right \rangle = (a, a, a, 0, 0 )\times
\tau_2$ with $a\sim M_{\rm U}$.
As discussed in Section II, to generate
CKM mixing for fermions we require an
$\left \langle {\bf 16_H} \right \rangle_d$ that acquires
an electroweak scale vacuum expectation value.
To insure accurate gauge coupling unification,
the effective low energy theory
should not contain split multiplets beyond those of MSSM.
Thus the MSSM Higgs doublets must
be linear combinations of the $SU(2)_L$ doublets in ${\bf 10_H}$
and ${\bf 16_H}$.  A simple set of superpotential terms that ensures this
and incorporates doublet--triplet splitting is:
\begin{equation}
W_H = \lambda {\bf 10_H} {\bf 45_H}{\bf 10'_H} + M_{10} {\bf 10'_H}^2
+ \lambda' \overline{\bf 16_H} \overline{\bf 16_H}{\bf 10_H} +
M_{16} {\bf 16_H}\overline{\bf 16_H}~.
\end{equation}
A complete superpotential for ${\bf 45_H}, {\bf 16_H},
{\bf \overline{16_H}}, {\bf 10}_H, {\bf 10}'_H$ and possibly other fields,
which
ensure that ${\bf 45_H}$, ${\bf 16_H}$
and ${\bf \overline{16_H}}$ acquire unification
scale VEVs with $\left \langle {\bf 45_H} \right \rangle$ being
along the $(B-L)$ direction, that exactly two Higgs doublets
$(H_u,H_d)$ remain light, with $H_d$ being a linear combination
of $({\bf 10_H})_d$ and $({\bf 16_H})_d$, and that there are
no unwanted pseudoGoldstone bosons, can be constructed \cite{bm,bb,br,ab,cm}.
\footnote{It is intriguing that the Higgs fields used in \cite{br} can be
neatly
incorporated into a single adjoint representation of E$_7$.}
The various
possibilities generate different predictions for threshold corrections in
the unification of gauge couplings, for example.  As we will explain, such
differences will have implications for proton decay rate, and we will allow
for such effects.

The Higgs doublet and the color triplet mass matrices following
from Eq. (38) are, in $SU(5)$ notation,
\begin{eqnarray}
\left ( \matrix{\overline{\bf 5}_{10_H} & \overline{\bf 5}_{10'_H}
& \overline{\bf 5}_{16_H}} \right) \left( \matrix {0 & \lambda
\left \langle {\bf 45_H} \right \rangle & \lambda' \left \langle
\overline{\bf 16_H} \right \rangle \cr -\lambda \left \langle {\bf 45_H}
\right \rangle & M_{10} & 0 \cr 0 & 0 & M_{16} }\right)
\left (\matrix{{\bf 5}_{10_H} \cr {\bf 5}_{10'_H} \cr {\bf 5}_{\overline
{16_H}} }\right)~.
\end{eqnarray}
With the vacuum expectation value $\left \langle{\bf 45_H}\right
\rangle$ in the $B-L$ direction it
does not contribute to the doublet matrix, so one pair of
Higgs doublet remains light, while all triplets acquire unification
scale masses.  The light MSSM Higgs doublets are
\begin{equation}
H_u = {\bf 10}_u,~ H_d = \cos\gamma {\bf 10}_d + \sin\gamma
{\bf 16}_d ~,
\end{equation}
with $\tan\gamma \equiv \lambda' \left \langle {\bf \overline{16}_H}
\right\rangle/M_{16}$.
Consequently, $\left \langle {\bf 10}\right \rangle_d =
{\rm cos}\gamma~ v_d, \left \langle {\bf 16}_d \right \rangle =
{\rm sin}\gamma~ v_d$, with $\left \langle H_d \right \rangle = v_d$
and $\left \langle {\bf 16}_d \right \rangle$ and $\left\langle {\bf
10_d}\right
\rangle$ denoting the electroweak
VEVs of those multiplets.  Note that the $H_u$ is purely in ${\bf 10_H}$
and that $\left \langle {\bf 10}_d \right \rangle^2 + \left \langle
{\bf 16}_d \right \rangle^2 = v_d^2$.

This pattern of gauge symmetry breaking, while motivated on separate
grounds, nicely harmonizes
with the fermion mass pattern advocated in Secs. IV and V.  Specifically,
the $g_{ij}$ terms will contribute to the
down--flavored fermion masses, while
there are no analogous terms in the up--flavor sector.  We also
note the relation tan$\gamma ~{\rm tan} \beta \simeq m_t/m_b
\simeq 60$.  Since $\tan \beta$ is separately observable,
the angle $\gamma$, which will prove relevant
for proton decay, is thereby anchored.

\subsection{Baryon number violation}

By combining Eqs. (38)-(40)
with the Yukawa couplings from Eq. (14),
we can now obtain the effective baryon number violating superpotential.
Denote the
color triplet and anti--triplet in ${\bf 10_H}$ as $(H_C, \overline{H}_C)$,
in ${\bf 10'_H}$ as  $(H_C', \overline{H'}_C)$ and in $(\overline{\bf 16_H},
{\bf 16_H})$ as $(\hat{H}_C, \hat{\overline{H}_C})$.  The relevant
Yukawa couplings of these color triplets, extracted from Eq. (14)
are\footnote{Here we have absorbed the factor $\left\langle {\bf 45}_H \right
\rangle/M$
into $a_{ij}$ and the factor $\left\langle {\bf 16}_H\right\rangle/M$
into $g_{ij}$.  We use the same notation ($a_{ij}$ and $g_{ij}$) for these
redefined
quantities.  As for $f_{ij}$, we define $\hat{f}_{ij} \equiv f_{ij}
\left\langle {\bf \overline{16}}_H
\right\rangle/M$.}
(i) $h_{ij}({1 \over 2}Q_iQ_jH_C+Q_iL_j\overline{H}_C)$, (ii) $-a_{ij}[{1
\over 2}
Q_iQ_jH_C(B-L)_{Q_j} + Q_iL_j\overline{H_C}(B-L)_{L_j}]$,
(iii) $(g_{ij} \left \langle {\bf 16_H} \right \rangle/M) Q_i L_j
\hat{\overline{H_C}}$, and (iv) $(f_{ij} \left \langle
\overline{\bf 16_H} \right\rangle/M) Q_i Q_j \hat{H_C}$.
Note that for the $a_{ij}$ coupling
there are two contractions: ${\bf 10_H} \times {\bf 45_H} \supset
{\bf 10}+{\bf 120}$.  Whereas the antisymmetric ${\bf 120}$ contributes
to fermion masses (for $\left \langle {\bf 45_H}\right \rangle
\propto B-L$), as in  Eq. (15), it is the symmetric ${\bf 10}$ that leads
to the Yukawa couplings of $H_C$ and $\overline{H_C}$.  Although the $a_{ij}$
term can arise through quantum gravity or stringy effects,
for concreteness,
we have assumed that it arises by integrating out
a pair of superheavy ${\bf 16}+{\bf \overline{16}}$ states which
couple through the renormalizable interactions $W \supset ({\bf 16_2}
{\bf 16}) {\bf 10_H}+({\bf 16_3}{\bf \overline{16}}){\bf 45_H}
+ M_V {\bf 16}{\bf \overline{16}}$.\footnote{A variant is to interchange
the indices 2 and 3 in this renormalizable interaction.}
This is why the $(B-L)$ generator
appears in the $a_{ij}$ couplings involving ${ H_C}, { \overline
{H_C}}$.  The strength of this coupling is then fixed in terms of the
corresponding doublet coupling.  Similarly, in the $f_{ij}$ coupling,
there are two $SO(10)$ contractions, and we shall assume both to have
comparable strength.

Baryon number violating operators
of the type $QQQL/M$
are the ones that connect to ordinary quarks and leptons by
wino dressing, and which generally dominate (but see below).
Integrating out the color triplet fields, one arrives at the following
effective superpotential terms involving these operators:
\begin{eqnarray}
W^{(L)}_{\rm eff} &=&
M_{\rm eff}^{-1} [(u^T\hat{H}d')\left\{u^T \hat{H} V' \ell
- d'^T\hat{H}V' \nu' \right\}  + 3(u^T\hat{H}d')\left\{u^T\hat{A}
V' \ell - d'^T \hat{A}V' \nu' \right\} \nonumber \\
&-& (u^T\hat{A}d')\left\{u^T \hat{H}V' \ell - d'^T \hat{H}V' \nu \right
\} - 3 (u^T \hat{A} d')\left\{u^T \hat{A}V'\ell - d'^T\hat{A}V' \nu'
\right\} \nonumber \\
&-&{\rm tan}\gamma(u^T \hat{H} d')\left\{ u^T \hat{G}V'\ell
- d'^T\hat{G}V' \nu'\right\} + {\rm tan}\gamma(u^T\hat{A}d')
\left\{u^T \hat{G}V'\ell - d'^T \hat{G}V'\nu'\right\} ]
\nonumber \\
&+& M_{16}^{-1}(u^T\hat{F}d')\left\{u^T\hat{G}V'\ell - d'^T\hat{G}V'\nu'
\right\}~.
\end{eqnarray}
Here $M_{\rm eff} = (\lambda a)^2/M_{10}$.  $u$ and $\ell$ denote the
column matrices of the physical left--handed up quark and charged
lepton superfields in the supersymmetric basis (i.e., a basis in which
neutral gaugino interactions are flavor diagonal).  The $d'$ and
$\nu'$ fields are related to the physical down quark and light neutrino
fields by the CKM matrices for quarks and leptons: $d' = V_{CKM} d$
and $\nu' = V_{CKM}^\ell \nu$, while $V' = V_u^\dagger V_\ell$, where
$V_u$ and $V_\ell$ diagonalize respectively the left--handed up quark
and charged lepton mass matrices: $u^{(g)} = V_u u^{(m)}$, where $(g)$
and $(m)$ denote the gauge and mass eigenstates.  In writing Eq. (41),
the color indices $(\alpha,\beta,\gamma)$ on quark fields are suppressed
and use is made of the fact that $(u^T_\alpha \hat{H} d'_\beta-d'^T_\alpha
\hat{H} u_\beta) = 2 u_\alpha^T \hat{H} d'_\beta$, which holds because of
antisymmetry under the interchange $\alpha \leftrightarrow \beta$ and because
$\hat{H}$ is symmetric.  The superscript $(L)$
on $W_{\rm eff}^{(L)}$ signifies that all the fields in $W_{\rm eff}^{(L)}$
belong to
$SU(2)_L$ doublets.  We comment later on the contributions from
$W^{(R)}_{\rm eff}$,
involving $RRRR$ operators of the form $u^cu^cd^ce^c$, involving $SU(2)_L$
singlets, which can be important in certain range of supersymmetric
parameter space.

The $3 \times 3$ matrices $(\hat{H}, \hat{A}, \hat{G}, \hat{F})$ operate in
the family space and are related to the Yukawa coupling matrices
given by the elements $(h_{ij}, a_{ij}, g_{ij}, f_{ij})$  respectively,
as follows:
\begin{equation}
(\hat{H}, \hat{A}, \hat{G}, \hat{F}) = V_u^T(h,a,g,f)V_u ~.
\end{equation}
The matrices $(h,a,g,f)$ are related to those appearing in Eqs. (28),(31)
and are given by:
\begin{eqnarray}
h&=&h_{33} \left(\matrix{0 & 0 & 0 \cr 0 & 0 & \sigma \cr 0 & \sigma
& 1}\right),~ a = h_{33} \left(\matrix{ 0 & \epsilon' & 0 \cr
\epsilon' & 0 & \epsilon \cr 0 & \epsilon & 0}\right), \nonumber \\
g &=& {h_{33} \over {\rm tan}\gamma} \left( \matrix{0 & \eta' & 0 \cr
\eta' & 0 & \eta-\sigma \cr 0 & \eta - \sigma & 0 }\right),~
f=\hat{f}_{33} \left(\matrix{x & 0 & z \cr 0 & 0 & y \cr z & y & 1
}\right)~,
\end{eqnarray}
where $\hat{f}_{ij} = f_{ij} v_R/M$.  Values of $\sigma, \eta, \epsilon,
\eta', \epsilon'$ and $y$ have been obtained in Sec. IV-V from considerations
of fermion masses and mixings.  From proton decay constraints and the neutrino
sector we have
estimated $x \sim (10^{-5}-10^{-4})$ and $z \sim (0.002-0.006)$.  $V_u$ can be
worked out using Eq. (28) and (20), and  is found to be close to
an identity matrix.   Its largest off--diagonal entry is the (1,2)
element $\simeq \epsilon'/(\epsilon^2-\sigma^2) \simeq -0.06$.
However
the matrix $V_\ell$ that diagonalizes the charged lepton mass matrix
$L$ is far from trivial, and
so $V' = V_u^\dagger V_\ell$ picks up a substantial
(2,3) element $ \simeq -3 \epsilon + \eta \simeq -0.437$,
reflecting sizable $\mu-\tau$ mixing.
The numerical values of the matrices $V_u, V_\ell$ and $V'$ are given in
Appendix A, as are the matrices $(\hat{H}, \hat{A}, \hat{G}, \hat{F}$).

Even if one is skeptical of our
particular
pattern of fermion mass matrices, it is difficult to avoid the
general conclusion\footnote{Barring near-complete accidental
cancellation between
diagonal and off--diagonal contributions to $m_{\nu_\mu}$.}
that if
neutrino masses are hierarchical, a large $\nu_\mu-\nu_\tau$
oscillation  angle (the SuperK result) requires
a sizable $\mu-\tau$ mixing angle (say $\ge 0.3$).
This  has significant implications for proton decay, as we now discuss.

After wino dressing, which dominates over gluino dressing for
tan$\beta \le 20$, each of the seven terms in Eq. (41) leads to twelve
four fermion operators for proton decay into $\overline{\nu} + X$,
a subset of which was exhibited in Ref. \cite{bpw}.
In Appendix B.1 we give the complete expression for the neutrino as
well as the charged lepton decay modes of the proton.
Representative contributions to
the proton decay amplitudes are analyzed in more detail for the dominant
modes in Appendix B.2. We now briefly summarize the results of a lengthy
investigation
of the net effect of all these terms.

To evaluate the strength of each term in Eq. (41), we need $h_{33}$
and $\hat{f}_{33}$ (see Eq. (43)) at $M_{\rm U}$.  $h_{33}$ is determined using
$h_{33}v_u \simeq m_t(M_{\rm U}) \simeq 100-120$ GeV, which yields
$h_{33} \simeq 1/2$.  $\hat{f}_{33}$ can be determined as follows.
Using Eq. (23) for the mass matrix of the light $\nu_\mu-\nu_\tau$
sector,
we have
\begin{equation}
m_{\nu_3} \simeq B m_U^2/M_R,
\end{equation}
where
\begin{equation}
B = -(\sigma+3 \epsilon)(\sigma+3 \epsilon-2y) \simeq 5.
\end{equation}
Here
we have put $\sigma$ and $\epsilon$ from Eq. (20) and used
$y=0.047$, corresponding to $m_{\nu_2}/m_{\nu_3} = 1/15$.  Putting
$m_{\nu_3} = (1/20~ {\rm eV}) \zeta$, where $\zeta =2$ to $1/2$,
corresponding to SuperK results, and $m_U \simeq m_t(M_{\rm U})$, we
get $M_R \simeq 10^{15}~{\rm GeV}/\zeta$.
Using $M_R = f_{33} v_R^2/M$ (see Eq. (4)), with $v_R = \left\langle {\bf
16_H} \right
\rangle = (2 \times 10^{16})~{\rm GeV} \kappa_R$ and $\kappa_R \approx
1/2$ to $2$, we find
\begin{equation}
\hat{f}_{33} = f_{33} v_R/M
\simeq (1/20)(1/\kappa_R \zeta).
\end{equation}
We will use $\hat{f}_{33} \approx
1/20$ with the understanding that it is uncertain by a factor of
2-3 either way.  Note that $\hat{f}_{33}$ is considerably larger,
by about a factor of 200--700, than the value estimated in Ref. \cite{bpw}.
This results partly from
lowering of $m_{\nu_\tau}$ from a few
eV (used in Ref. \cite{bpw}) to about 1/20 eV (the SuperK value) and in part
from interplay between the mixings in the Dirac and
the Majorana mass matrices via the seesaw mechanism.  The latter has the net
effect of enhancing $M_R \approx B m_U^2/m_{\nu_3}$, for a given $m_{\nu_3}$,
precisely by a factor of $B \approx 5$, compared to what it would be without
mixing. (Compare $M_R \approx 10^{15}$ GeV with its counterpart in Sec. II
where, by ignoring mixing, we got $m_R^\tau = (1-3)\times 10^{14}$ GeV.)  The
two
effects together, -- i.e., lowering of $m_{\nu_3}$ and increase through $B$ --
significantly enhance $\hat{f}_{33} = [B m_U^2/m_{\nu_3}]/v_R$.

It is interesting to note that the larger value of $M_R$ arising because
of mixing has a further
implication.
Using $M_R \approx 10^{15}$ GeV (corresponding to $\zeta \approx 1$), $v_R =
2 \times 10^{16}$ GeV (for $\kappa_R \approx 1$) and $f_{33} \approx 1$, one
obtains: $M = f_{33} v_R^2/M_R \approx 4 \times 10^{17}$ GeV, which is the
perturbative string scale \cite{string}.

Note that the net value of $B$ (see Eq. (45)) depends on the
parameters of the Majorana mass matrix as well as on $\sigma$ and
$\epsilon$ from the Dirac mass matrix of the neutrinos, which in
turn are determined within $SO(10)$ by the masses and mixings of
quarks and charged leptons.  That is
why our expectations for proton decay
are significantly affected by our understanding of the masses and
mixings of quarks and charged leptons.

The full set of contributions to the proton decay amplitudes (from all
the operators in Eq. (41) involving all possible combinations of
family indices) was obtained numerically using {\it Mathematica\/} and is
listed in Appendix B.1.  Calculations of a few representative (dominant)
contributions to the amplitudes are exhibited in detail in Appendix B.2,
where estimates of the amplitudes allowing for uncertainties
in the relative phases of different contributions are presented.
A general discussion of proton decay rate is given in Appendix C.
Based on the result of these appendices, we now discuss the general
constraint on the proton decay amplitude and thereby on the mass scale
$M_{\rm eff}$ and $M_{16}\tan\gamma$ corresponding to the numerical
estimate of the amplitude given in Appendix B.  These constraints
arise from existing lower limits on the proton lifetime.

\subsection{Constraints on proton decay amplitudes from the proton lifetime}

In Appendix C we show that with a certain (apparently)
reasonable choice of supersymmetric
spectrum, the lifetime of the proton decaying into neutrinos is:

\begin{eqnarray}
\Gamma^{-1}(p \rightarrow \overline{\nu}_\tau K^+)
&\approx& (2.2 \times 10^{31})~{\rm yrs} \times \nonumber \\
&~& \left({.67 \over A_S}\right)^2\left[{0.006~ {\rm GeV}^3 \over
\beta_H}\right]^2
\left[{(1/6) \over (m_{\tilde{W}}/m_{\tilde{q}})} \right]^2
\left[{m_{\tilde{q}} \over 1~{\rm TeV}}\right]^2 \left[{2 \times
10^{-24}~ {\rm GeV}^{-1}
\over
\hat{A}(\overline{\nu}) }\right]^2.
\end{eqnarray}
Here $\hat{A}(\overline{\nu}) = A(\overline{\nu})/(2 \overline{f})$, where
$A(\overline{\nu})$ is the strength of the four fermion proton decay amplitude
and $\overline{f}$ is the average wino--dressing function (see Eq. (93) in
Appendix B and Eq. (111)-(114) in Appendix C).  The quantity
$\hat{A}(\overline{\nu})$
is simply the product of all the vertex factors in the wino--dressed Higgsino
exchange diagram divided by the effective mass of the relevant color triplet
Higgsino.
For normalization purpose we can define $\hat{A}(\overline{\nu})^{\rm
SU(5)} = (\lambda_c \lambda_s \theta_C^2)/M_{H_C}$, where $\lambda_i$ stand for
the Yukawa couplings of the quarks at $M_{\rm U}$.
(For clarity of discussions, only the second generation
contribution is kept here.)  The quantity $A(\overline{\nu})$ is the full
amplitude,
including the loop factor associated with the wino dressing.  If we substitute
$\hat{A}(\overline{\nu})^{\rm SU(5)}$ defined above into Eq. (47), we will
reproduce the
results given in Ref. \cite{hisano}.  (We have allowed for a factor of 4
enhancement
in the lifetime relative to \cite{hisano}, corresponding to an apparent slip
by a factor of ${1 \over 2}$ in going from Eq. (3.7) to Eq. (3.8) of that
paper.)

Note that in writing Eq. (47), the short
distance renormalization $A_S$ of the $d=5$ operator in going from $M_{\rm U}$
to
$M_{\rm SUSY}$ (see Ref. \cite{hisano}) as well as the running
factor to go from $M_{\rm SUSY}$ to 1 GeV have been included.  $A_S$ has a
central value of about $0.67$, which we shall adopt even for the $SO(10)$
model.

We have discussed in the appendix that typically proton would decay dominantly
into $\overline{\nu}_\tau K^+$ and $\overline{\nu}_\mu K^+$ modes.\footnote{
$\mu^+ K^0$ is also a likely prominent mode of the proton decay (see Sec. VI H)
with a typical branching ratio  $\sim$ 20-30\%.  This will however not alter
our
discussion here appreciably.}  In some cases
the former supercedes the latter by factors of 2-5 in the rate, while the
converse
is true in some other cases (see Table. 1, Appendix C).  We see from Eq.
(47)
that with $\hat{A}(\overline{\nu}_\tau) \approx 2
\times
10^{-24}~{\rm GeV}^{-1}$, a reasonable ``central value" for the partial
lifetime
$\Gamma^{-1}(p \rightarrow \overline{\nu_\tau}K^+)^{-1}$ is
$2.2 \times 10^{31}~{\rm yrs.}$, which corresponds to
the hadronic matrix element $\beta_H = 0.006~{\rm GeV}^3$
(this is the central value quoted in  lattice
calculations \cite{gavela}),
$(m_{\tilde{W}}/m_{\tilde{q}}) \approx 1/6$ and $m_{\tilde{q}} \approx
1~{\rm TeV}$.  Now, adding the $\overline{\nu}_\mu K^+$ mode with a branching
ratio $R$ ($R_{\rm average} \approx 0.3$, say), this is however $25(1+R)$
times smaller than the empirical lower limit \cite{pdk}
\begin{equation}
\Gamma^{-1}(p \rightarrow
\overline{\nu}K^+)_{\rm expt} \ge 5.6 \times 10^{32}~{\rm yr}.
\end{equation}
Thus, if the parameters have nearly their central values, the amplitude
for the dominant of the two modes
must satisfy the bound: $\hat{A}(\overline{\nu}_\ell K^+) \le
2 \times 10^{-24}~{\rm GeV}^{-1}/\sqrt{1+R} \simeq 4 \times
10^{-24}`{\rm GeV}^{-1}/\sqrt{1+R}$, where
$\ell = \mu$ or $\tau$.
Allowing that both  $\beta_H$ and
the ratio of superpartner masses
$(m_{\tilde{W}}/m_{\tilde{q}})$ might well be smaller by factor of 2
(say) than the value
quoted above, and that $m_{\tilde{q}}$ could be (say) 1.4 TeV
rather than 1 TeV, the theoretical value of the lifetime could
plausibly increase by a factor of
32 compared to the ``central value''  $2.2 \times 10^{31}~{\rm yr}$.
This saturation of all uncertainties in the parameters, all in the
direction so as to extend proton lifetime, strains credulity.
If, nevertheless,  one allows for such a variation in the parameters,
the bound on the amplitude mentioned above will be relaxed by a factor of
5 to 6.  Thus the empirical limit on proton lifetime leads to the rather
conservative bound:
\begin{equation}
\hat{A}(\overline{\nu}_i) \le 2.3 \times 10^{-24}~{\rm GeV}^{-1}/\sqrt{1+R}~.
\end{equation}
This limit
 should  be satisfied for every neutrino flavor $\nu_i$ (assuming
$\overline{\nu}_\mu
 K^+$ and $\overline{\nu}_\tau K^+$ are comparable).  For squark masses
not exceeding about 1.5 TeV, we take Eq. (49) as a strict upper bound.
It should however be noted that the lifetime depends quartically on the
squark mass, so increasing $m_{\tilde{q}}$  by a factor of 2 to about
3 TeV, holding $m_{\tilde{W}}$ fixed,
would lengthen proton lifetime by a factor of 16.  If true, this would relax
the bound on $\hat{A}(\overline{\nu})$ quoted in Eq. (49) by as much as a
factor of
4 to $\hat{A}(\overline{\nu}) \le 9 \times 10^{-24}~{\rm GeV}^{-1}/\sqrt{1+R}$.
Such a heavy spectrum ($m_{\tilde{q}} \sim 3$ TeV)
for all three generations of squarks
would however require severe fine--tuning
of parameters in order to keep the vacuum expectation value of the light Higgs
field at the electroweak scale.
We shall assume
a lighter squark spectrum ($m_{\tilde{q}} \le 1.5$ TeV),
for which there is no need for such an adjustment of parameters.  In this case,
the bound in Eq. (49) will have to be satisfied.

\subsection{Constraints on $M_{\rm eff}$ and proton decay via standard
operators}

In Appendix B, Eqs. (101) and (102) (or Table 1),
we show that within the concrete $SO(10)$ model, the
dominant $\overline{\nu}_\tau K^+$ or $\overline{\nu}_\mu K^+$ decay amplitude
from the standard $d=5$ operator for cases I and II are
given by:
\begin{equation}
A(\overline{\nu} K^+)_{\rm std} \simeq
 \left[{2h_{33}^2\hat{f}(c,d)} \over M_{\rm eff}\right]
\left[\matrix{2.8 \times 10^{-5} \cr
1.2 \times 10^{-5} }\right]({1 \over 2} ~{\rm
to}~{3 \over 2})
\epsilon_{\alpha \beta \gamma}
(d^\alpha u^\beta)(s^\gamma \nu_3)~.
\end{equation}
There is an analogous expression for the new neutrino mass--related $d=5$
operator,
that will be discussed in the next subsection (E). The upper and lower entries
in Eq. (50) correspond to cases I and II respectively.

We now compare the upper limit (Eq. (49)) on the amplitude from proton decay
searches
against theoretical expectations  based on the concrete $SO(10)$ model.
This leads to constraints on
$M_{\rm eff}$ from the standard $d=5$ operator  (and on $M_{16}\tan\gamma$
from the new operator).  The
upper bound (Eq. (49)) on $\hat{A}(\overline{\nu}_i)$ applies to the net
amplitude, which
is given by the sum of the contributions from the standard  (Eq. (50)) and the
new
operators (Eq. (51) below).  For the sake of clarity,
we will derive constraints on $M_{\rm eff}$
$(M_{16}\tan\gamma$) under the assumption that the standard
(respectively, the new) operator dominates.
Indeed, near-complete accidental cancellation between the two
contributions
is unlikely
to occur for both $\overline{\nu}_\tau K^+$ and $\overline{\nu}_\mu K^+$ modes.

Using the net contribution from the standard operators to the amplitude given
by
Eq. (50), and the definition of $\hat{A}(\overline{\nu}_i)$, we obtain
(putting $h_{33} \simeq 1/2$)
\begin{equation}
\hat{A}(\overline{\nu}_\tau)_{\rm std} \approx {1 \over M_{\rm
eff}} \left[\matrix{7 \cr 3 }\right]\times 10^{-6}(1/2 ~{\rm to}~ 3/2)~.
\end{equation}
Comparing with the empirical upper limit on $\hat{A}(\overline{\nu}_i)$ (Eq.
(49))
obtained as above, we get:
\begin{equation}
M_{\rm eff} \ge \left[\matrix{4 \cr 1.7}\right] \times 10^{18}~{\rm
GeV}~(1/2~{\rm to}~3/2)~.
\end{equation}

Thus we see that $M_{\rm eff}$ has to be rather large compared to the MSSM
unification
scale of $2 \times 10^{16}~{\rm GeV}$ in order that the standard operators may
not
run into conflict with the observed limits on proton lifetime.  In effect, this
reflects a net enhancement -- by almost two orders of magnitude -- of the
standard
$d=5$ proton decay operators for realistic $SO(10)$, compared to those in
minimal
$SU(5)$, with low $\tan\beta \le 3$ (see discussion in Appendix C).
In the latter case, one need only
require that the color triplet mass
exceed about $(2-3) \times 10^{16}~{\rm GeV}$.

Now, as mentioned in Appendix C,
there are theoretically attractive mechanisms whereby
the mass of ${\bf 10'}_H$, denoted by $M_{10}$ (see
Eq. (38)), can be suppressed relative to the unification scale $M_{\rm
U}$.
In this case, $M_{\rm eff} \equiv
(\lambda a)^2/M_{10}$ can be larger than $\lambda a \sim M_{\rm U}$.  Very
large
values of $M_{\rm eff} \gg M_{\rm U}$ could however lead to large {\it
positive\/} corrections
to $\alpha_3(M_Z)$, just from the doublet--triplet mechanism, above and beyond
the
value expected on the basis of simple coupling unification.  For the
doublet--triplet splitting mechanism described by Eq. (39) the
shift in $\alpha_3(m_Z)$ from this sector alone is found to be (see Appendix D)
\begin{equation}
\Delta \alpha_3(m_Z)|_{DT} = {[\alpha_3(m_Z)]^2 \over 2 \pi} {9 \over 7} {\rm
ln}
\left({M_{\rm eff}\cos\gamma \over M_{\rm U}}\right)~.
\end{equation}
This generalizes the expression given in Ref. \cite{bpw,urano},
where the MSSM Higgs doublets were assumed to be contained entirely in ${\bf
10}_H$ and
not in ${\bf 16}_H$, corresponding to $\cos\gamma = 1$.  The argument of the
logarithm in
Eq. (53) is simply the ratio: (product of the three color triplet masses)/
(product of the two superheavy doublet masses $\times M_{\rm U}$).  From the
determinant of Eq. (39) we see that the product of the color triplet masses
is equal to $M_{\rm eff} M_{10} M_{16}$, while the two heavy doublets have
masses
given by $M_{10}$ and $M_{16}/\cos\gamma$.  Note that the second  heavy doublet
has a mass larger than $M_{16}$.  This is the reason for the presence of the
$\cos\gamma$
factor in Eq. (53).

To evaluate the RHS of Eq. (53), we use the lower bound of $M_{\rm eff}$ that
is
suggested by proton lifetime constraints, viz., $M_{\rm eff} \ge (2-6) \times
10^{18}~
{\rm GeV}$ or $(0.8-2.5) \times 10^{18}~{\rm GeV}$
(see Eq. (52)), and the MSSM unification scale of $M_{\rm U} \simeq
2 \times 10^{16}~{\rm GeV}$.  We should also specify the value of $\cos\gamma$.
It is obtained in terms of $\tan\beta$ as follows.  From $m_t \simeq h_{33}
\left\langle
{\bf 10_H} \right\rangle_u = h_{33}v_u$ and $m_b \simeq h_{33} \left\langle
{\bf 10_H}
\right\rangle_d = h_{33} \cos\gamma v_d$, we have
$m_t/m_b \simeq (v_u/v_d)(1/\cos\gamma)$.  Inserting $m_t/m_b \simeq 60$,
we thus obtain $\cos\gamma \simeq (\tan\beta/60)$.  (This can also be
expressed as $\tan\beta \tan\gamma \simeq m_t/m_b$, which is valid for
$\tan\gamma \ge 3$.)  We see that $\cos\gamma$ is a small number
($\approx 1/30-1/6)$ for small
and moderate values of $\tan\beta$ ($\approx 2-10$).

Now, let us recall that, in the absence of
unification--scale
threshold and Planck--scale effects, the MSSM value of $\alpha_3(m_Z)$ in
the
$\overline{\rm MS}$ scheme, obtained
by assuming gauge coupling unification, is given by $\alpha_3^0(m_Z)|_{MSSM} =
0.125-0.13$ \cite{langacker,bagger}.  This is about 5-10\% higher than the
observed value:
$\alpha_3(m_Z) = 0.118 \pm 0.003$ \cite{pdg}.  Substituting the central value,
$\alpha_3(M_Z) = 0.118$,
and the MSSM unification scale of $M_{\rm U} = 2 \times  10^{16}~{\rm GeV}$,
one obtains for $\cos\gamma = 1/20$:
$\Delta \alpha_3(m_Z)|_{DT} \simeq 0.0046 - 0.0077$, for $M_{\rm eff} \approx
(2-6)\times 10^{18}~{\rm GeV}$.
Thus the constraint on $M_{\rm eff}$ from
proton lifetime for case I
amounts to having in MSSM a net value $\alpha_3(m_Z)|_{net} =
\alpha_3(M_Z)^{(0)}|_{MSSM}
+ \Delta \alpha_3(M_Z)_{DT}^{MSSM} + \Delta_3' \simeq (0.132 - 0.135) +
\Delta'_3$,
for $\cos\gamma = 1/20$,
where $\Delta_3'$ denotes $other$ unification scale threshold and Planck scale
effects evaluated at the electroweak scale.
Including $\Delta \alpha_3(m_Z)_{DT}$ and
$\Delta_3'$, the net theoretical value
of $\alpha_3(m_Z)$ is given by $\alpha_3(m_Z)_{\rm net} = \alpha_3(m_Z)^0_{\rm
MSSM}
+ \Delta \alpha_3(m_Z)_{DT} + \Delta_3'$.  Since $\alpha_3(m_Z)^0_{\rm MSSM}$
is
higher than the observed value, and $\Delta \alpha_3(m_Z)_{DT}$ is relatively
large (depending on $M_{\rm eff}\cos\gamma$) and is positive (see below), one
would
need a net appropriately large negative value of $\Delta_3'$ so that
$\alpha_3(m_Z)_{\rm net}$ may agree with the observed value.  By varying the
parameter $T \equiv M_{\rm eff}\cos\gamma/(2 \times 10^{18}~{\rm GeV})$
appropriately
so as to include the range $M_{\rm eff} \ge (1-6) \times 10^{18}~{\rm GeV}$,
obtained
from proton lifetime constraint, we can evaluate $\Delta \alpha_3(m_Z)_{\rm
net}$ and
thereby assess the needed value of $\Delta_3' = \alpha_3(m_Z)_{\rm
obs}-\hat{\alpha}_3(m_Z)
$, where $\hat{\alpha}_3(m_Z) \equiv \alpha_3(m_Z)_{\rm MSSM} + \Delta
\alpha_3(m_Z)_{DT}$.
This is shown in the Table below.

\begin{center}
\begin{tabular}{|c|c|c|c|c|c|c|}
\hline
$T$ & 1/60 & 1/40 & 1/30 & 1/20 & 1/10 & 1/3 \\
\hline
$\Delta \alpha_3(m_Z)_{DT}$ & 0.0014 & 0.0026 & 0.0034 & 0.0046 & 0.0066 &
0.0100 \\
$\hat{\alpha}_3(m_Z)$ & 0.1284 & 0.1296 & 0.1304 & 0.1316 & 0.1336 & 0.1370 \\
$\delta_3'$ & -8.8\% & -10\% & -10.5\% & -11.5\% & -13.2\% & -16.1\% \\
\hline
\end{tabular}
\end{center}

\noindent Here $\delta_3' \equiv \Delta_3'/\alpha_3(m_Z)_{\rm obs} =
(\alpha_3(m_Z)_{\rm obs} - \hat{\alpha}_3(m_Z))/\alpha_3(m_Z)_{\rm obs}$.  In
above, we have used a reasonable lower limit on $\alpha_3(m_Z)^0_{\rm MSSM}
=0.127$ for
$m_{\tilde{q}} \le 1$ TeV [42] and have used the central value for
$\alpha_3(m_Z)_{\rm obs}
= 0.118$.  Allowing for $\alpha_3(m_Z)_{\rm obs} = 0.118 \pm 0.003$ would
amount
to adding nearly $\pm 2\%$ change to $\delta_3'$.  For concreteness, we will
quote
results for central value of $\alpha_3(m_Z)_{\rm obs}$, but bear in mind the
possibility of the $\pm 2\%$ change in $\delta_3'$.  Note that the variation of
the parameter $T$ used above include proton lifetime constraints (Eq. (48)) of
$M_{\rm eff} \ge (2-6)\times 10^{18}$ GeV for case I, with $\cos\gamma \approx
1/60$ to $1/9$, and $M_{\rm eff} \ge (0.8 - 2.4) \times 10^{18}$ GeV for case
II, with
$\cos\gamma \approx 1/60$ to $1/4$.

The Table shows that if coupling unification should hold, one must assume, for
the case
of MSSM embedded in $SO(10)$, that other unification scale threshold and
Planck scale
effects (excluding $\Delta \alpha_3(m_Z)_{DT}$), denoted by $\Delta_3'$,
provide
a substantial negative contribution to $\alpha_3(m_Z)$, typically varying
between
$-8\%$ to as much as $-16\%$, as evaluated at the electroweak scale.  Before
discussing the feasibility of such large negative correction, it should be
emphasized that the presence of $\cos\gamma$ inside the logarithm of Eq. (53)
has played a significant role in diminishing the positive threshold correction
to
$\alpha_3(m_Z)$.  In its absence, with $M_{\rm eff} \ge (1-2)\times
10^{18}~{\rm GeV}$,
negative threshold corrections exceeding even (18-20)\% would have been
required from
the other unification scale threshold and Planck scale effects.

We now show that it is reasonable to expect the net threshold correction from
other effects, denoted by
$\Delta_3'$, evaluated at the electroweak scale, to be
negative in sign, and to have
magnitude no more than about 6 to 10\%, if one confines
oneself to low dimensional Higgs multiplets such as ${\bf 45}, {\bf 16},
{\bf \overline{16}}$ and ${\bf 10}$ (as we do).  First, let us note that since
the coupling
and the correction get enhanced by nearly the same factor ($\approx 3$ for
$\alpha_3$) as
they run from $M_{\rm U}$ to $m_Z$, a -10\% correction to $\alpha_3$ at $m_Z$
roughly corresponds to a  -3.3\% correction at the unification scale.
The precise value of the correction depends on the details of the model,
including
the nature of the Higgs system that causes mass splittings within complete
$SU(5)$ multiplets.  Fortunately, as previously observed by several authors
(see
e.g. \cite{br}, \cite{bagger}),
for Higgs multiplets that are not too large (such as ${\bf 45}$ and ${\bf 16}$
of
$SO(10)$ as in our model) and with the split masses $(M^\alpha)$ of the
sub-multiplets
belonging to a given $SO(10)$ multiplet being within a factor of $\le 2-10$ of
$M_{\rm U}$ either way, the net threshold correction to $\alpha_3(m_{\rm U})$,
evaluated at the unification scale\footnote{Note that the threshold
corrections owing
to doublet--triplet splitting in $SO(10)$ (discussed in Sec. VI.A) is a
special case,
because the combination
$M_{\rm eff} \cos\gamma = (\lambda a)^2\cos\gamma/M_{10}$, that enters into
the logarithm in Eq. (53), does not represent the masses of the color triplets
(which is of order $\lambda a$).  This arises because of the $\cos\gamma$
factor
in the logarithm, and also because $M_{10}$ representing the mass of the
complete
multiplet of ${\bf 10'}$ (see VI.A) can be much smaller than $M_{\rm U}$ (as
needed
from proton decay constraint, Eq. (52)).  This is the reason that for
$M_{\rm eff}\cos\gamma \ge 2 \times 10^{18}$ GeV, $\Delta \alpha_3(m_Z)_{DT}$
can
lead to large positive corrections to $\alpha_3$ at $m_Z$.} is
typically no more than about 1 to 2\%, and the corrections from a given
sub-multiplet
can be either positive or negative depending upon whether $M^\alpha$ is
greater or
smaller than $M_{\rm U}$, and also on the $SU(3) \times SU(2) \times U(1)$
quantum numbers of the sub-multiplet.

To get a feel for the magnitude of the correction, consider (for
illustration
only) the case of minimal supersymmetric $SU(5)$ (with ${\bf 24}, {\bf 5}, {\bf
\overline{5}}$ of Higgs).  In this case, owing to the color triplets ($H_3,
H_{\overline{3}}$) in $({\bf 5}, {\bf \overline{5}}$), which are separated from
the doublets (albeit by fine-tuning) and acquire unification scale masses,
$\alpha_3(m_{\rm U})$ receives a correction $\Delta \alpha_3(M_{\rm U})_{H_3}
\equiv \alpha_3(M_{\rm U})(\epsilon_3)_{H_3}$ at the unification scale,
where [42] $(\epsilon_3)_{H_3} = [3 \alpha_{\rm unif} {\rm ln}(M_{H_3}/M_{\rm
U})]/5\pi
\approx +(0.5-1.7)\%$ for $M_{H_3}/M_{\rm U} \approx $2 to 10, and
$\alpha_{\rm unif}
\approx 0.04$.  Note this correction is expected to be {\it positive} because
proton lifetime constraint suggests $M_{H_3} > M_{\rm U}$, and even for the
extreme value of the mass ratio of 10, it is less than 2\%.\footnote{One can
obtain
large negative contribution to $\alpha_3(m_Z)$ in $SU(5)$ by introducing large
Higgs multiplets, such as ${\bf 75}$ and ${\bf 50}$ [42] which are used in the
missing partner mechanism for doublet--triplet splitting in $SU(5)$.  This
however
is a special case associated with large multiplets and does not apply to lower
dimensional multiplets like ${\bf 45}, {\bf 16}, {\bf 10}$ of $SO(10)$.  Large
multiplets such as ${\bf 75}$ of $SU(5)$ do not ``explain" the observed
unification
of couplings, but rather accommodate it.}

In the context of the minimal Higgs system (like ours) that utilizes the VEVs
of
just the ${\bf 45}, {\bf 16}$ and ${\bf \overline{16}}$ of Higgs to break
$SO(10)$
to the standard model gauge symmetry, we can identify two (rather definite)
sources
of negative contribution to $\alpha_3(m_Z)$.  These arise from mass-splittings
within (a) the gauge multiplets, and (b) the Higgs sub--multiplets  ${\bf
45_H}$.

First consider the gauge multiplets.
Representing the VEVs of ${\bf 16}_H$ and ${\bf 45_H}$ by $c$ and $a$
respectively
(see Appendix D),
one finds: $$\Delta \alpha_3(m_Z)_{\rm gauge} \simeq~
(+0.002, -0.00026, -0.0015, -0.0028, -0.0031, -0.0030, -0.0018,
-0.001)$$
for
$$p= (0.1, 0.2, 0.3, 0.5, 0.7, 1, 2,3),$$
where $p\equiv 2c/a$.  Thus the contribution can be
positive for sufficiently small $p$, but for most of its range, $p \ge 0.2$,
the contribution is $negative$, with $(\delta_3')_{\rm gauge} \equiv
\Delta \alpha_3(m_Z)_{\rm gauge}/\alpha_3(m_Z)_{\rm obs}$ varying
from $-1.3$ to about $-2.6\%$, for $p$ varying from 0.3 to 2.0.  Note that the
maximum of $-2.6\%$ evaluated at $m_Z$ corresponds to a correction of about
$-0.8\%$ at the
unification scale, in accord with expectations.

The contribution to $\alpha_3(m_Z)$ from the splitting of ${\bf 45_H}$ has been
evaluated in Appendix D (see Eq. (119)).  As shown there, if one ignores the
coupling
of ${\bf 45_H}$ to other multiplets, the allowed superpotential has a simple
form:
$W = M_1 ({\bf 45_H})^2 + \kappa ({\bf 45_H})^2/M$.  One can then argue that
the
masses of the submultiplets are characterized by $M_1 \approx \kappa (M_{\rm
U}^2/M)
\sim 10^{-2} M_{\rm U}$, if $\kappa \sim 1$ and $M \sim M_{\rm Planck}$.  One
then
obtains $\Delta \alpha_3(m_Z)_{\bf 45_H} \simeq -0.0045$.

Thus,
at least in the simplest approximation, the contribution from ${\bf 45_H}$ to
$\alpha_3(m_Z)$ is negative and is about -4\% at the electroweak scale.  Of
course, the
above superpotential is a simplification because one needs to couple ${\bf
45_H}$
to other fields that acquire unification scale VEVs, like ${\bf 16_H}, {\bf
\overline{16}_H}$,
together with possible additional vector--like pairs (${\bf 16_V}, {\bf
\overline{16}_V}$)
without VEVs to avoid uneaten pseudo-Goldstone bosons.  But such couplings do
not
alter the basic feature that $M_1 \ll M_{\rm U}$ (see \cite{br}).

Thus we see that with the minimal Higgs system (i.e., with just ${\bf 45_H},
{\bf 16_H}$
and ${\bf \overline{16}_H}$ acquiring VEVs), there are good reasons for the
unification scale threshold correction to $\alpha_3(m_Z)$ from our
``other effects" denoted by $\Delta_3'$, to receive negative contributions
from the gauge sector of $-1.3$ to about $-2.6\%$ (depending on $p$), and
plausibly about $-3$ to $-4\%$ from the Higgs multiplet ${\bf 45_H}$ at the
electroweak
scale.  These two together can easily combine to yield about $-4.5$ to $-6.6\%$
correction to $\alpha_3(m_Z)$.

Having identified two rather definite sources of negative threshold corrections
to $\alpha_3(m_Z)$, the remaining other contributions (excluding
$\Delta \alpha_3(m_Z)_{DT}$) arise through splittings within ${\bf 16_H},
{\bf \overline{16}_H}, {\bf 16_V}$ and ${\bf \overline{16}_V}$ (which would be
induced through $<{\bf 45_H}>$).  Contributions to each of these multiplets
can be either positive or negative depending on the parameters in $W$.  Since
the induced splittings within these multiplets will be relatively small, as
also
their sizes (for mass ratios $\le 2$), the magnitudes of these corrections
from each of these multiplets to $\alpha_3(m_Z)$ can be estimated to be
less than about 1\%.  Thus the combined correction from ${\bf 16_H},
{\bf \overline{16}_H}, {\bf 16_V}, {\bf \overline{16}_V}$ is expected to be
less than about 2 to 4\% (positive or negative) at the electroweak scale.

Finally, it is worth noting that with only ${\bf 45_H}, {\bf 16_H}$ and
${\bf \overline{16}_H}$ acquiring VEVs (as in our model), contribution
to $\alpha_3(m_Z)$ from Planck scale physics through effective operators
$F^{\mu \nu} F_{\mu \nu} {\bf 45_H}/M$ will vanish because of antisymmetry
in the $SO(10)$ contraction.

Thus, at least for the minimal Higgs system utilizing the VEVs of only
${\bf 45_H}, {\bf 16_H}, {\bf \overline{16}_H}$, we see that the net
``other'' threshold correction to $\alpha_3(m_Z)$ from unification and Planck
scale effects, excluding $\Delta \alpha_3(m_Z)_{DT}$ but including those
from (a) the gauge sector ($-1$ to $-2.6\%$), (b) from ${\bf 45_H}$ ($-3$ to
$-4\%$),
(c) collectively from ${\bf 16_H}, {\bf \overline{16}_H}, {\bf 16_V},
{\bf \overline{16}_V}$ ($\pm 2$ to 4\%), and (d) from Planck scale effects
($\approx 0\%$), is very likely negative.  But after adding all these, it seems
plausible to assume that the magnitude of this net other contribution
is not more than about 8 to 10\%, at the electroweak scale.

This assumption in turn
implies (see the Table above) that $T \equiv M_{\rm eff} \cos\gamma/(2
\times 10^{18}~{\rm GeV})$ is bounded from above by about 1/30, or
conservatively by 1/20.  If we add a +1.5\% correction to the entries
for $\delta_3'$ in the Table to allow for $\alpha_3(m_Z) = 0.118 + 0.0017$,
then $\delta_3'$ (reduced as above) is $\le 10\%$.  This in turn leads to
the upper bound:
\begin{equation}
{M_{\rm eff} \cos\gamma \over 10^{18}~{\rm GeV}} \le 1/10
\end{equation}
For a given $\cos\gamma \simeq \tan\beta/60$, this upper bound yields an upper
limit on $M_{\rm eff}$ and thereby an upper limit on proton lifetime.  In
particular, if we assume
$\tan\beta \ge 2$, and thus $\cos\gamma \ge 1/30$
\footnote{If $\tan\beta \le 1.5$, the top
Yukawa coupling will blow up before reaching $\mu = M_{\rm U}$.  Within
the standard supergravity spectrum of supersymmetry breaking, there are
indications that $\tan\beta \ge 2$ from LEP constraints.}, we obtain $M_{\rm
eff}
\le 3 \times 10^{18}$ GeV.  Putting $\tan\beta = (2,3,6,8,10)$ and thus
$\cos\gamma = (1/30,1/20, 1/10, 1/7.5,1/6)$, the upper bound given above yields
$(M_{\rm eff}/10^{18}$ GeV) $\le (3,2,1,0.75,0.6)$.  Combining this upper
limit on $M_{\rm eff}$ (which holds for both cases I and II) arising
from the threshold corrections to $\alpha_3(m_Z)$ with the lower limits
on $(M_{\rm eff}/10^{18}~{\rm GeV}) \ge (2- 6)$ for case I and $\ge (0.8-2.4)$
for case II, given by the proton lifetime constraint (Eq. (52)), we see that
the
two limits would be in conflict with each other if $\cos\gamma > (1/20,1/7.5)$
-- i.e., if $\tan\beta > (3,8)$ -- for cases (I, II).  These considerations
based on $\Delta \alpha_3(m_Z)$ and the experimental limit on proton lifetime
suggest that rather small values of
$\tan\beta$ -i.e.  $\tan\beta \le 3$ (case I) and $\tan\beta \le 8$
(case II) are favored (but see below).  In
either case, for the MSSM, assuming $\tan\beta \ge 2$, by demanding accurate
coupling
unification we obtain an upper limit
on $M_{\rm eff}$  given
by

\begin{equation}
M_{\rm eff} \le 3 \times 10^{18}~{\rm GeV}~
\end{equation}

On the other hand, the proton lifetime constraint (Eq. (52)) implies that
$M_{\rm eff}$ must exceed (2 to 6)$ \times 10^{18}$ GeV for case I and
(0.85 to 2.5)$ \times 10^{18}$ GeV for case II, where the range in $M_{\rm
eff}$
corresponds to the uncertainty factor $B = $(1/2 to 3/2) in the amplitude
(see Eq. (51) and Appendix B) in a correlated manner.  For instance, for
$M_{\rm eff} \le 3 \times 10^{18}~{\rm GeV}$ (rather than $6 \times 10^{18}$
GeV) $B$ can vary only between (1/2 to 3/4) for case I, while for case II,
B can still vary between (1/2 to 3/2).  Using this range and $M_{\rm eff} \le
3 \times
10^{18}$ GeV, we can obtain a lower limit for the proton decay amplitude
(given by Eq. (51)):

\begin{eqnarray}
\hat{A}(\overline{\nu}_\tau K^+)_{\rm std} \ge [\matrix{ (7 \times 10^{-24}
{\rm GeV}^{-1})(1/6~{\rm to}~1/4)  \cr  (3 \times 10^{-24} ~{\rm
GeV}^{-1})(1/6~
{\rm to}~1/2) }]
\end{eqnarray}

Substituting into Eq. (47) and adding the contribution form the second
competing
mode $\overline{\nu}_\mu K^+$ with a typical branching ratio $R \approx 0.3$,
we
obtain
\begin{eqnarray}
\Gamma^{-1}(\overline{\nu} K^+)_{\rm std} \le
\left[\matrix{(3 \times 10^{31} {\rm yrs.})(1.6~{\rm to}~0.7)
\cr (6.8 \times 10^{31} {\rm yrs.})(4~{\rm to}~ 0.44)}\right](32~{\rm to}~
1/32)
\end{eqnarray}
Here the uncertainty (32 to 1/32) corresponds to the uncertainty in $\beta_H,
(m_{\tilde{W}}/m_{\tilde{q}})$ and $m_{\tilde{q}}$, by factors of 2,2, and
$\sqrt{2}$ respectively, either way, around the ``central" values reflected in
Eq. (47).  Thus we find that for MSSM the inverse partial proton decay rate
should satisfy:
\begin{eqnarray}
\Gamma^{-1}(p \rightarrow \overline{\nu}K^+)_{\rm std} \le
\left[\matrix{3 \times 10^{31 \pm 1.7}~{\rm yrs.}  \cr 6.8 \times
10^{31^{+2.1}_{-1.5}}~{\rm yrs}}\right]
\nonumber \\
\le \left[\matrix {1.5 \times 10^{33}~{\rm yrs.} \cr 7 \times
10^{33}{\rm yrs.}}\right] ~~~~({\rm MSSM})
\end{eqnarray}
The upper limit in Eq. (58) essentially reflects the upper limit on $M_{\rm
eff}$, while
the remaining uncertainties of matrix elements and spectrum are reflected in
the exponents.  We stress that the upper limit on lifetime exhibited in Eq.
(58) has
arisen by first taking an upper limit of 10\% on the net unification scale
threshold correction to $-\alpha_3$ (excluding the contribution from the
doublet--triplet
splitting), which appears to us generous.    Furthermore, the amplitude
limit exhibited in Eq. (56) is obtained only if the uncertainties in the
amplitude,
$\beta_H, (m_{\tilde{W}}/m_{\tilde{q}})$ and $m_{\tilde{q}}$ all go in the same
direction to about their extreme values to extend proton lifetime, so
it is a generous limit.

Before discussing the contributions of the new operator to proton decay, we
wish to
note an interesting possibility, that can be relevant especially if
$\tan\beta$ is
large.  For large values of $\tan\beta \approx 20-30$, corresponding to
$\cos\gamma
\approx 1/3$ to $1/2$, with $M_{\rm eff} \ge (2-6)\times 10^{18}~{\rm GeV}$
(which
is needed for case I to satisfy the proton lifetime constraint), one would have
$T = M_{\rm eff} \cos\gamma/(2 \times 10^{18}~{\rm GeV}) \ge 1/3$ to $3/2$.
In this
case, in the context of MSSM, a negative threshold correction ($\delta_3'$)
exceeding
16 to 20\% in magnitude would be required from other sources (see the Table
given above for $\delta_3'$ as a function of $T$).  As we said
before, although such a large negative correction,
together with a matching positive one, is in principle possible (see
e.g. Ref. [43]),
it diminishes the luster of the observed agreement of simple coupling
unification,
by making it appear somewhat fortuitous.
In this connection it is noteworthy that extra vector--like matter --
specifically a ${\bf 16} + \overline{\bf 16}$ as proposed in the so-called
ESSM (Extended
Supersymmetric
Standard Model) \cite{essm} -- at the TeV scale could greatly ease
this problem, and thereby allow large values of  $\tan\beta$,
while leaving our discussion of ordinary fermion masses
essentially unaltered.
In this case, $\alpha_{\rm unif}$
is raised to nearly $0.25$ to $0.3$, compared to $0.04$ in the MSSM.  Owing
to increased two--loop effects, the scale of unification $M_{\rm U}$ is raised
to
$(1-2) \times 10^{17}~{\rm GeV}$ \cite{essm,kolda},
while $\alpha_3(M_Z)^0|_{ESSM}$ is lowered
to about $0.112-0.118$.  With increased $M_{\rm U}$ the correction
$\Delta \alpha_3(M_Z)|_{DT}$ is also lowered.  As a result, even for
$\tan\beta \simeq 20-30$, i.e., $\cos\gamma \simeq (1/3-1/2)$, and
$M_{\rm eff} \approx 6 \times 10^{18}~{\rm GeV}$, one obtains for the
case of ESSM, the net value of $\alpha_3(m_Z) = \alpha_3(m_Z)^0|_{ESSM}
+ \Delta \alpha_3(m_Z)_{DT} + \Delta_3' \approx (0.123-0.124) + \Delta_3'$.
Thus for ESSM embedded in $SO(10)$, unification scale threshold corrections
from ``other" sources to $\alpha_3(m_Z)$ (represented by
$\Delta_3'/\alpha_3(m_Z)$),
though negative,
need be no more than 5\% in magnitude,
even in the rather extreme case $\tan\beta = 30$, with
$M_{\rm eff} = 6 \times 10^{18}~{\rm GeV}$.  As shown above, for the minimal
Higgs
system a net negative threshold correction to $\alpha_3(m_Z)$ of this magnitude
is not unexpected.  As an added feature, we note that since in the
ESSM case
$M_{\rm eff}$ can be rather large, $\approx (4-8) \times 10^{18}~{\rm
GeV}$,
without requiring large negative threshold corrections from
other sources, we can be compatible with the
observed limit on proton lifetime for the {\it central values\/} of
uncertainties in the
parameters describing the spectrum of the supersymmetric particles and the
relevant matrix elements.

To be specific, for $M_{\rm eff} \approx 6 \times
10^{18}$ GeV and $\tan\beta=30$, which corresponds to $-\delta_3' = 4$ to
5\%), we obtain
(compare with Eq. (58)):
\begin{equation}
\Gamma^{-1}(\overline{\nu} K^+)_{\rm std} \approx \left[\matrix
{1.2 \times 10^{32}~{\rm yrs.}~(1.6~{\rm to}~0.7) \cr
2.7 \times 10^{32}~{\rm yrs.}~(4~{\rm to}~0.44)}\right]~(32~{\rm
to}~1/32),~~~~~{\rm ESSM}
\end{equation}
where the last factor arises from allowing $(m_{\tilde{W}}/m_{\tilde{q}}),
m_{\tilde{q}},$ and $\beta_H$ to vary by factor of 2,2,$\sqrt{2}$ respectively
about their ``central" values.  The upper and lower entries correspond to
cases I
and II respectively.
Note that allowing for a modest factor of 1 to 4 jointly from the two brackets
for
case I (and 1 to 2 for case II),
which correspond to nearly central values
of the parameters mentioned above, keeps
the theoretical value of proton lifetime within the experimental limit.
On the other hand, allowing for a factor of 20 jointly from the two brackets
(for either
case I or case II), the proton lifetime in the ESSM case
typically lies in the range $(1-5) \times 10^{33}$ yrs, which should be
accessible.

\subsection{Constraint on $M_{16}\tan\gamma$ and proton decay via the new
operator}

The decay amplitude for the new operator for the leading mode (which
in this case is $\overline{\nu}_\mu K^+$) is given by

\begin{eqnarray}
A(\overline{\nu}_\mu K^+)_{\rm new} &\simeq& \left( {\hat{f}_{33} h_{33}
\over
M_{16} \tan\gamma}\right)\left[(3 \times 10^{-6})(1/2~{\rm to}~2)\right]
\left[f(t,d)+f(t,l)\right] \times \nonumber \\
&~&\epsilon_{\alpha \beta \gamma}(d^\alpha u^\beta)(s^\gamma \nu_3)~,
\end{eqnarray}
as shown in Appendix C.  This is true (approximately) for both cases I and II.

{}From the definition of $\hat{A}(\overline{\nu})$  we obtain:
\begin{equation}
\hat{A}(\overline{\nu}_\mu)_{\rm new} \approx \left[{\hat{f}_{33}h_{33}
\over
M_{16} \tan\gamma} \right]\left[(3 \times 10^{-6})(1/2~{\rm
to}~2)\right]~.
\end{equation}
The first bracket $P \equiv [\hat{f}_{33} h_{33}/(M_{16} \tan\gamma)]$ may be
evaluated by
using results of Sec. VI A and B, as follows.  Putting $\hat{f}_{33} = f_{33}
v_R/M$
(see Eq. (46)), $M_{16}\tan\gamma = \lambda' v_R$ (see Eq. (40), and $h_{33}
\simeq 1/2$,
we obtain:
\begin{equation}
P = (f_{33}/M) (1/2\lambda')~.
\end{equation}
Here $M$ stands for the mass scale that characterizes the strength of the
effective non--renormalizable operators (see Sec. III).
Using $M_R \equiv f_{33}
v_R^2/M$
and the result  that $M_R \approx B (m_U^2/m_{\nu_3}) \simeq
5[m_t^2(M_{\rm U})/(1/20~{\rm eV}~ \zeta) \simeq 10^{15}~{\rm GeV}/\zeta$
(see Eqs. (44) and (45)), we find:\footnote{We
note that the $SO(10)$ contraction for the $f_{ij}$ coupling contributing to
neutrino
masses is not the only one that contributes to proton decay.  We assume that
the two
possible contractions are comparable in strengh.}
\begin{equation}
f_{33}/M = M_R/v_R^2 \approx (\kappa_R^2 \zeta)^{-1}[4 \times
10^{17}~{\rm GeV}]^{-1}
\end{equation}
Here $\kappa_R$ and $\zeta$ denote uncertainties in $v_R$ and $m_{\nu_3}$ --
i.e.,
$v_R \equiv \kappa_R (2 \times 10^{16}~{\rm GeV}$) and $m_{\nu_3} \equiv
\zeta(1/20)$ eV.  A priori, we expect $v_R \sim M_{\rm U}$ and thus $\kappa_R
\approx
(1/2-3)$.
Since
$<{\bf 16}_H>$ breaks $SO(10)$ to $SU(5)$, $v_R = <{\bf 16}_H>$
may in fact be somewhat larger than $M_{\rm U}$ -- that is, $\kappa_R > 1$.  From
the SuperK results, we have $\zeta \approx
1.5-1/1.5$.
In Sec. III, we have discussed the appropriateness of the
characteristic mass $M$ being either the Planck or the string scale,
subject to the presumption that the effective coupling $f_{33}$ for the third
family is nearly maximal ($\sim 1$).  Given some uncertainty in this
regard, a good choice appears to be $M = 10^{18}$ GeV, which is intermediate
between $M_{\rm Planck}$ and $M_{\rm string}$.  We expect that such an
intermediate value of $M$ should represent fairly well either choice
$M = M_{\rm Planck}$ or $M= M_{\rm string}$, given that we allow a
range in $\kappa_R^2 \zeta$ and thereby in $f_{33}$.  Setting $M = 10^{18}$
GeV, Eq. (63) yields: $f_{33} \approx 2.5/(\kappa_R^2 \zeta)$.  Noting that
values of $f_{33} \gg 1$ are implausible, we thus expect $\kappa_R^2 \zeta \ge
1$
in accord with the remark mentioned above (rather than $<1$).  Under the
presumption that $f_{33}$ is maximal, it therefore seems quite reasonable
to assume that $\kappa_R^2 \zeta \approx $ (1 to 5) $ \approx 2.5 (1/2$ to 2),
which corresponds to $f_{33} \approx $ (2 to 1/2).
Substituting this range
 for $\kappa_R^2 \zeta$ into Eq. (63) and (62), we obtain
 $$P \approx (5 \times
 10^{-19} ~{\rm GeV}^{-1})(1/2~{\rm to}~ 2)/\lambda'$$.
 Now, on the one hand, the
validity
 of the perturbative treatment requires $\lambda' \leq 1$.  One the other hand,
 given  that $M_{16}\tan\gamma = \lambda' v_R$ and that from considerations of
 standard operators for proton decay, we have the constraint that $\tan\beta <
(3,8)$
 and thus $\tan\gamma > (20, 7)$ for cases (I, II) in MSSM, the choice of
 $\lambda' < 1$ will tend to make $M_{16} \ll v_R$, which is implausible.  We
 therefore take $\lambda' \approx 1$.  Substituting this into $P$ and in turn
into
 Eq. (60), we obtain
 \begin{equation}
 \hat{A}(\overline{\nu}_\mu K^+)_{\rm new} \approx (1.5 \times 10^{-24}~{\rm
GeV}^{-1})(1/4~{\rm to}~1.3)
 \end{equation}
Here the upper end of the uncertainty has been restricted to conform to the
limit on proton lifetime, see Eq. (49).

 Comparing with Eq. (56) we see that the contributions of the new and the
standard
 operators (for $M_{\rm eff} \approx (2-3) \times 10^{18}$ GeV) to the proton
decay
 amplitude are {\it  comparable\/} to one another.  Since there
 is no reason to expect near cancellation between them
 (especially for both $\overline{\nu}_\mu K^+$ and $\overline{\nu} K^+$
modes), we
 expect the net amplitude (standard + new) to be in
 the range exhibited for either one.   Assuming
 that the new operator dominates and substituting Eq. (64) into Eq. (47), we
obtain:
 \begin{equation}
 \Gamma^{-1}(\overline{\nu} K^+)_{\rm new} \approx (3 \times 10^{31}~{\rm
yrs})[16~{\rm to}~
 1/1.7]\{32~{\rm to}~ 1/32\}
 \end{equation}
 In this estimate we have included the contribution of the $\overline{\nu}_\tau
 K^+$ mode with a typical branching ratio $R \approx 0.4$ (see Appendix C).
Here the
 second factor, inside the square bracket, reflects the uncertainties in the
 amplitude (see Eq. (64)), while the last factor corresponds to varying
$\beta_H,
( m_{\tilde{W}}/m_{\tilde{q}})$ and $m_{\tilde{q}}$ around the central values
reflected
 in Eq. (47).

 With a net factor of even 20 to 100
 arising jointly from the square and the curly
 brackets, {\it i.e}. without going to extreme ends of all
 parameters, the new operators related
to neutrino masses lead by
themselves to proton decay lifetimes $\approx (0.6-3)\times 10^{33}$ yrs.
Thus if the new operators were the only source of proton
 decay -- i.e., if the standard $d=5$ operators were somehow absent
-- one could be comfortably compatible with existing limits,
but optimistic regarding future observation.  We now briefly elaborate
on
this possibility.

\subsection{The neutrino mass related operator as the sole source of proton
decay}

As we have seen, straightforward minimal embedding of the MSSM in
$SO(10)$, with informed hypotheses about the fermion-Higgs couplings,
leads to contributions from the standard $d=5$ operators are
disturbingly large, if the relevant parameters have nearly their
central values and/or if $M_{\rm eff} < 2 \times 10^{18}$ GeV.  This is
especially true for case I in the MSSM (see Eq. (57)-(58)).
One is therefore
motivated to wonder
whether the standard operators
might not be present.  This
possibility is realized, if the higher gauge symmetry is
$G_{224}$  (or $G_{2113} = SU(2)_L \times I_{3R} \times
(B-L) \times SU(3)^c$) rather than $SO(10)$.
Such gauge symmetries have been proposed to appear in solutions
of string theory (see Ref. \cite{rizos} for $G_{224}$ and
Ref. \cite{faraggi2} for $G_{2113}$).
Such possibilities retain some attractive features of $SO(10)$,
notably the unification of quark-lepton families into single multiplets
including the right-handed neutrino $\nu^R$ as indicated for the
neutrino seesaw.  One sacrifices a simple group-theoretic explanation
for the observed unification of coupling, but if the models
derive from an underlying string theory, one might still expect such
unification at the string scale.
Plausible mechanisms to reconcile the string and the MSSM
unification scale have been proposed \cite{dienes}.
The
color triplets related to the electroweak doublets, which generate the standard
$d=5$ proton decay operators, need not exist.
The standard source of $d=5$ operators can be absent in
such models \cite{jcp}.

It is possible that after projecting out appropriate fields
the couplings of those that remain reflect the original higher
symmetry\footnote{For example, in a class
of string solutions leading to $G_{2113}$, the cubic level top and $\nu_\tau$
Yukawa
couplings are claimed to be equal at the string scale despite $SU(4)_C$
breaking \cite{faraggi2}.}.  This is what is believed to occur for the gauge
couplings,
as previously mentioned.  If it also occurs for the Higgs superpotential
couplings, our considerations on fermion and neutrino masses in Sec. II-IV and
their
relationships to proton decay will remain valid.

\subsection{Baryon number violation from the $RRRR$ operator}

So far we have focused on the charged wino dressing of the effective
superpotential $W_{\rm eff}^{(L)}$ of Eq. (41).  For small values of
$\tan\beta$ and $\mu$, this gives the dominant contribution to
the proton decay amplitude.  However, if $\mu\tan\beta$ is large,
($\ge 2~{\rm TeV}$), dressing of the effective baryon number violating
operator involving only the right--handed fields, $u^c u^c d^c e^c$,  by the
charged
Higgsino can become important \cite{lucas,goto,bs}.  The
ratio of this amplitude to the usual wino contribution scales
as $\tan\beta(\mu/m_{\tilde{W}})$ (for $m_{\tilde{q}} \gg m_{\tilde{W}})$.
In minimal supersymmetric
$SU(5)$, the Higgsino dressing becomes more important than the wino dressing
when $\tan\beta \ge 9 (m_{\tilde{W}}/\mu)$
\cite{bs}.  This estimate takes into account the differences
in the  renormalization
of the $(RRRR)$ compared to the $(LLLL)$ operator owing to running
from $M_{\rm U}$ to $M_{\rm SUSY}$, and the difference between their
strong matrix elements.  We now show that, by contrast, in the $SO(10)$ model
the $RRRR$
operator is strongly suppressed relative to the $LLLL$ operator, and
that they will be comparable only if ($\mu \tan\beta/m_{\tilde{W}})
\approx 500$.

In the $SO(10)$ model, once the quark and lepton masses and mixings are
fixed, the strength of the $RRRR$ operator is determined by the corresponding
effective superpotential (analogous to $W_{\rm eff}^{(L)}$) which is given by:
\begin{eqnarray}
W_{\rm eff}^{(R)} &=& -M_{\rm eff}^{-1} [ 2 (u^{c~T} \tilde{H} \tilde{V'}
e^c)(u^{c~T} \tilde{H}
K d^c) + (u^{c~T} \tilde{H} \tilde{V'} e^c)(u^{c~T} \tilde{A}K d^c) \nonumber
\\
&-&3(u^{c~T} \tilde{A}\tilde{V'}e^c)(u^{c~T}\tilde{H}Kd^c) - 3(u^{c~T}
\tilde{A}\tilde{V'}
e^c)(u^{c~T}\tilde{A}K d^c) \nonumber \\
&-&\tan\gamma (u^{c~T}\tilde{H}\tilde{V'}e^c)(u^{c~T} \tilde{G}K d^c) +
3(u^{c~T}\tilde{A}\tilde{V'}e^c)(u^{c~T}\tilde{G}Kd^c)] \nonumber \\
&+& M_{16}^{-1} (u^{c~T}\tilde{F}\tilde{V'}e^c)(u^{c~T}\tilde{G}Kd^c)~.
\end{eqnarray}

\noindent Here

\begin{equation}
(\tilde{H}, \tilde{A}, \tilde{G}, \tilde{F}) \equiv  \tilde{V}_u^T (h, a, g,
f)\tilde{V}_u,
\end{equation}
with $\tilde{V}_u$ the unitary matrix that rotates the $u^c$ fields
$u^{c(g)}= \tilde{V}_u u^{c(m)}$ from the gauge basis to the mass basis,
$\tilde{V}_e$ is
the unitary matrix that similarly rotates the right--handed electron
field,
and
$\tilde{V'} \equiv \tilde{V}_u^\dagger \tilde{V}_e$.
$K$ is the right--handed analog of $V_{CKM}$, $K \equiv \tilde{V}_u^\dagger
\tilde{V}_d$.

The contribution to proton decay amplitude from Eq. (66) is estimated as
follows.
Since a charged Higgsino is involved in the dressing, the internal scalars have
to be from the third generation (other diagrams will be suppressed by small
Yukawa couplings).  This uniquely picks out the $\tilde{t}_R$ and
$\tilde{\tau}_R$
as the internal scalars.  The external quark fields are then fixed to be
$u$ and either a $d$ or an $s$.
This suggests that the combination of indices $(ij)(kl)$ in Eq. (66) must be
$(33)(11)$,
$(33)(12)$, $(13)(32)$ or $(13)(31)$.  Among these four, the combination
$(33)(12)$ can proceed without utilizing any of the right handed mixing
angles.  (Though the right--handed mixing angles in the
$(23)$ sector of both $u$ and $d$ are $\sim 0.2$, not terribly small.)
We find the amplitude, after Higgsino dressing, from this dominant
contribution to be
\begin{eqnarray}
\hat{A}[(u^cs^c)^\dagger(d\nu_\tau)] &\simeq& M_{\rm eff}^{-1} h_{33}^2
\eta'V_{td}
+ (M_{16} \tan\gamma)^{-1} h_{33}\hat{f}_{33}\eta'V_{td} \\
 &\approx& 6.6 \times 10^{-6}/M_{\rm eff} + 1.3 \times 10^{-24} {\rm GeV}^{-1}
 (1/2~{\rm to}~2)~.
\end{eqnarray}
Other contributions are not much bigger.
In going from Eq. (68) to (69), we have used $h_{33}^2 \approx 1/4, \eta'
\approx
4.4 \times 10^{-3}$ (see Sec. IV), $V_{td} \simeq 6 \times 10^{-3}$ and
$P \equiv \hat{f}_{33} h_{33}/M_{16}\tan\gamma  \simeq 5 \times 10^{-19}~
{\rm GeV}^{-1}(1/2-2)$ (see the discussion following Eq. (63)).
It is understood that the relative sign (phase) of the
two terms in Eq. (68) and (69) is arbitrary.
The full amplitude $A[(u^cs^c)^\dagger(d\nu_\tau)]$ will be obtained by
multiplying
the above expression by a loop function analogous to the function for
the wino dressing (see Eq. (92)).  There are two differences in this
function however:  (i) The factor $(\alpha_2/4\pi)$ will be replaced by
$(\lambda_t \lambda_\tau)/(16 \pi^2)$, where the Yukawa couplings are
to be evaluated at the momentum scale $M_{\rm SUSY}$.  (ii) The mass
parameter $m_{\tilde{W}}$ is replaced by $\mu$.
The presence of $\lambda_\tau$ brings in a $\tan\beta$ dependence in the
Higgsino
dressing relative to the wino dressing.  (In minimal supersymmetric $SU(5)$,
the
wino dressing has a $\tan\beta$ dependence, but the Higgsino dressing will have
a $(\tan\beta)^2$ dependence.)

Comparing the first term of Eq. (69)
with the standard $LLLL$ amplitude (see Eq. (51)) and making the
replacements as above, we obtain: $\hat{A}_{RRRR}({\rm
std})/\hat{A}_{LLLL}({\rm
std}) \approx [h_th_\tau \mu/(g_2^2 m_{\tilde{W}})]V_{ts} \times \{6.6 \times
10^{-6}\}/\{[7,3]\times 10^{-6}\} (1/2)$ ~$\approx (\mu
\tan\beta/m_{\tilde{W}})V_{ts}
[1/100, 1/48]$, where we have put $h_t(m_t) \approx 1$ and $h_\tau(m_t) \approx
m_\tau/v_D \approx \tan\beta/100$.  Substituting $V_{ts}\simeq 1/25$, we thus
see
that even for relatively large values of $(\mu\tan\beta/m_{\tilde{W}} \approx
100$,
the $RRRR$ amplitude is smaller than the $LLLL$ amplitude by a factor of
$25-12$.
Thus, unlike the case of $SU(5)$\footnote{The main reason why the standard
$RRRR$ operator is suppressed, relative to the $LLLL$ operator, in $SO(10)$
but not in $SU(5)$, is simply that the standard $LLLL$ operator is
effectively enhanced
by about two orders of magnitude in $SO(10)$, for $M_{\rm eff} =
M_C$.  See the remarks following Eq. (114) in Appendix C.}, in the $SO(10)$
model developed
here, the $RRRR$ operator can safely be ignored compared to the $LLLL$
operator, as
long as $(\mu \tan\beta/m_{\tilde{W}}) \le 200$ (say).

The charged Higgsino dressing results in decay modes of the proton containing
final state neutrinos,
not charged leptons.
Apart from the direct
charged Higgsino exchange, there are diagrams involving
$\tilde{t}_L-\tilde{t_R}$ mixing and $\tilde{\tau}_L-\tilde{\tau}_R$
mixing (the latter being proportional to $\tan\beta$) followed by the
exchange of charged wino.  This contribution is
smaller than the direct charged Higgsino exchange by roughly
a factor $v_u/M_{\rm SUSY}$.  Similar arguments apply to mixed
contributions involving charged Higgsino--wino mixing.

\subsection{Charged lepton decay mode}

In minimal supersymmetric $SU(5)$ and many of its variants, charged lepton
decay
of the proton is suppressed as long as $\tan\beta \le 20$ or so.
This is because of a GIM-type cancellation in the wino dressing diagrams,
which brings in a suppression factor in this amplitude, proportional
to the small $u$--quark mass.
Gluino dressing can lead to charged lepton decay of the proton, but
owing to flavor conservation of the primary gluino vertex and the flavor
antisymmetry of the effective superpotential, this contribution is suppressed
for small values of $\tan\beta~ (\le 20$).  For large values of
$\tan\beta~ (\ge 20)$,  flavor mixing
in the up--squark sector becomes significant and the gluino graph begins to
be important \cite{gluino}.  However, for minimal supersymmetric $SU(5)$,
such large values of $\tan\beta$ ~($\ge 20$) are highly disfavored owing to
limits on proton lifetime (see for example Ref. \cite{hisano}).
Even if we ignore this difficulty,
for large
$\tan\beta \ge 20$ dressing of the $RRRR$ operator by charged Higgsino becomes
the dominant source of proton decay in $SU(5)$ (if $\mu/m_{\tilde{W}} \ge 1$),
and the neutrino modes dominate anyhow.
So in minimal supersymmetric $SU(5)$, and many of its variants,
charged
lepton decay mode of the proton is highly suppressed relative to the neutrino
modes, for {\it all\/} values of $\tan\beta$.

The situation is different in our $SO(10)$ model, for two reasons.
First, the contribution to $p \rightarrow \mu^+K^0$ arising from the standard
$d=5$ wino dressing diagram in this model is not small.
Since we have a realistic spectrum of quark and
lepton masses, especially with $m_s \neq m_\mu$ and $m_d \neq m_e$ at $M_{\rm
U}$,
the wino contribution does not experience a GIM cancellation.
Second, the new $d=5$ operator
related to neutrino masses lends comparable strength to the neutrino and
the charged lepton modes.

In evaluating the expected branching ratio for the $\mu^+ K^0$ mode -- i.e.,
$B(\mu^+ K^0) \equiv \Gamma(\mu^+ K^0)/[\Gamma(\overline{\nu}_\tau K^+ +
\Gamma(\overline{\nu}_\mu K^+) + \Gamma(\mu^+ K^0)]$ -- let us
concentrate on case I.
Very similar results hold for case II.  The amplitudes for
$p \rightarrow \mu^+ K^0$ arising from the standard and the new operators for
case I
are given by (see Appendix B, Eq. (107) and (109) or Table 1):
\begin{eqnarray}
\hat{A}(\mu^+ K^0)_{\rm std} &\approx&
 ({h_{33}^2 \over M_{\rm eff}})(3 \times 10^{-6})(1/2 ~{\rm to}~2)
\approx (0.30)(1/2~{\rm to}~2)\times 10^{-24}{\rm GeV}^{-1} \\
\hat{A}(\mu^+ K^0)_{\rm new} &\approx& P(10^{-6})(1/3 ~{\rm to}~2) \approx
(0.5)(1/6~{\rm to}~4) \times 10^{-24} {\rm GeV}^{-1}
\end{eqnarray}
In Eq. (70) we have inserted $h_{33}^2 \simeq 1/4$ and an average value of
$M_{\rm eff} \approx 2.5 \times 10^{18}$ GeV (see discussions following
Eq. (55), showing $M_{\rm eff} \simeq (2-3) \times 10^{18}$ GeV for
case I).  In Eq. (71), we have inserted the value of $P \equiv \hat{f}_{33}
h_{33}/(M_{16}\tan\gamma) \approx (5 \times 10^{-19} {\rm GeV}^{-1})(1/2-2)$.
obtained in Sec. VI E. The standard and the new amplitudes for the
$\overline{\nu}_\tau K^+$ and the $\overline{\nu}_\mu K^+$ modes (for
$M_{\rm eff} \simeq 2.5 \times 10^{18}$ GeV) for case I are given by (see Eq.
(101),
(102), (105) and (106) and Table 1 in Appendix B):
\begin{eqnarray}
\hat{A}\left(\matrix{\overline{\nu}_\tau K^+ \cr \overline{\nu}_\mu
K^+}\right)_{\rm std}
&\approx& \left\{\matrix{(1.8)(0.8-1.2) \cr (1)(0.8-2)}\right\} \times
10^{-24} {\rm GeV}^{-1} \\
\hat{A}\left(\matrix{\overline{\nu}_\tau K^+ \cr \overline{\nu}_\mu
K^+}\right)_{\rm new}
&\approx& \left\{\matrix{(0.75)(1/4-2.8) \cr (1.5)(1/4-1.3)}\right\} \times
10^{-24} {\rm GeV}^{-1}
\end{eqnarray}
Note that the upper ends of the ranges shown for the different amplitudes have
been
restricted to conform with the limit on proton
lifetime (see Eq. (49)).
Assuming that the relevant matrix elements for the $\mu^+ K^0$ and
$\overline{\nu} K^+$
modes are comparable (but see below), it may be inferred from the amplitudes
noted above (or the discussion in Appendix B) that the standard operators by
themselves
lead to a
branching ratio for the $\mu^+ K^0$ mode in the range $B(\mu^+ K^0)_{\rm std}
\approx
$ 1 to 10\%, while the new operators (related to neutrino masses) by themselves
can lead to  $B(\mu^+ K^0)_{\rm new}$ from a few up to
40 or 50\%.\footnote{In estimating the upper and the lower ends of the sum
$\{\Gamma(\overline{\nu}_\tau K^+) + \Gamma(\overline{\nu}_\mu K^+)\}$ and
thereby
the expected range of $B(\mu^+ K^0)$, we stipulate that the maximum (or
minimum)
of $\hat{A}(\overline{\nu}_\tau K^+)$ is accompanied by a median rather than
the
maximum (or minimum) value of $\hat{A}(\nu_{\mu} K^+)$ and vice versa.  This is
because the various terms contributing to the two amplitudes
are unlikely to be entirely constructive (or destructive) simultaneously for
both of them.
Thus the logical maximum and minimum of $B(\mu^+ K^0)$ lie beyond the
stipulated range quoted above.}

With contributions from both the standard and the new operators present, and
comparable in magnitude, one must of course add the two contributions
allowing for interference
between them.   Adding the contributions of the
standard and the new operators to the amplitudes for all three decay modes
($\overline{\nu}_\tau K^+, \overline{\nu}_\mu K^+, \mu^+ K^0$) we estimate
the ``maximum" and the ``minimum" of the sum of the two contributions (standard
and new) to each of these modes, by allowing for the uncertainties in their
amplitudes reflected in Eq. (70)-(73).  We thereby estimate the following
range for $B(\mu^+ K^0)$:
\begin{equation}
B(\mu^+ K^0)_{\rm std + new} \approx \left[1~{\rm to}~(50-60)\%\right] \rho
\end{equation}
where $\rho$ denotes the ratio of the squares of relevant matrix elements
for the $\mu^+ K^0$ and $\overline{\nu} K^+$ modes.

If one uses the chiral
Lagrangian method as a guide \cite{clm,hisano}, one would obtain $\rho =
[1-(m_p/m_B)(D-F)]^2$
$\times [2/3 (m_p/m_B)D q + m_p/(3 m_B)(D+3F)]^{-2}$, where $D\simeq 0.76,
F \simeq 0.48 $ and $m_B \simeq 1150~MeV$, and $q \equiv C(usd\nu)/C(uds\nu)$.
The entity $q$ denotes the ratio of the coefficients of the $d=5$ amplitudes
leading to spinor contractions $(us)(d\nu)$ and $(ud)(s\nu)$ respectively
(see Appendix B).  In the $SO(10)$ model (developed here), we find typically
$q \approx 0.2$ to 0.6.  Thus the chiral Lagrangain method (CLM) would suggest
$\rho \simeq 1/4$ to 1/5.  It should however be noted that CLM leads to
matrix elements that are inconsistent with the results of lattice calculations
\cite{gavela} by factors of 1.5 to 4, for proton decaying either into $(\ell^+
\pi^0$ and $\overline{\nu} \pi^+$) or $\ell^+ K^0$ or both for all values
of $\beta_H \simeq (0.006 ~{\rm GeV}^3)$(1/2 to 3) (see the
discussion following Eq. (116) for the definition
of $\beta_H$.)  For instance, for $\beta_H \simeq 0.003~{\rm GeV}^3$, the $p
\rightarrow
\pi^0$ and $p \rightarrow \pi^+$ amplitudes are larger by a factor $\approx
(1.5-2)$ for
CLM relative to the lattice calculation, while that for $p \rightarrow
K^0$ is $smaller$ by a similar factor for CLM relative to the lattice value
\cite{gavela}.
Lattice calculations of $\overline{\nu} K^+$ are apparently not
available at present.  Evidently, one cannot regard the
estimate of $\rho$ based on the chiral Lagrangian method as entirely reliable.
Even a modest 25 to 30\% correction\footnote{The
chiral Lagrangian method has played a role in Ref. [34] and thus in our
discussion
in  Appendix and Sec. VI on proton decay rate.
We note that
corrections of 20 to 30\% in the amplitude would not however alter the total
proton decay
rate by more than a factor of 1.5 to atmost 2.} to the matrix element of
$\overline{\nu} K^+$
and $\mu^+ K^0$ (compared to CLM values) can alter the
estimate of $\rho$ by more than a factor of 2.  In the
absence -- presumably temporary -- of a reliable calculation,
one should remain open to the possibility
of larger values, say $\rho \approx 1/2 $ to 1.  Clearly, it is only such
larger values of $\rho$, together with the estimate of $B(\mu^+ K^0)$
presented in Eq. (74), that permit optimism for discovery of the
$\mu^+ K^0$ mode.

Returning to the estimate of Eq.(74), we find that typically for a
large range of parameters, which correspond to the uncertainties
in the amplitudes exhibited in Eq. (72)-(73), the branching ratio
$B(\mu^+ K^0)$ can lie in the range of 20 to 30\% (if $\rho
\approx 1$).  Thus we see that the $\mu^+ K^0$ mode is
likely to be prominent in the $SO(10)$ model presented here,
and if $\rho \approx 1$ it can even become a dominant mode.
This contrasts sharply with the minimal $SU(5)$ model, in which
the $\mu^+ K^0$ is expected to have a branching ratio of only about
$10^{-3}$.  In the $SO(10)$ model, the standard
operator by itself gives a branching ratio that is at least an order of
magnitude larger than the $SU(5)$ value, while {\it the potential prominence
of the $\mu^+ K^0$ mode arises only through
the new operator related to neutrino masses}.
Thus the $\mu^+ K^0$ mode of the proton decay serves as a
signature for the sort of $SO(10)$ model of fermion masses and
mixings explored here.

Because of the hierarchical nature of the Yukawa couplings in the Dirac
as well as the Majorana sectors, proton decay into $e^+ K^0$ and $e^+ \pi^0$
are highly suppressed in our scenario compared to the
$\overline{\nu}K^+$ or  the  $\mu^+ K^0$.

\section{Summary and concluding remarks}

One major goal of this paper has been to understand the masses and mixings of
the neutrinos, suggested by the atmospheric and the solar neutrino anomalies,
{\it in conjunction with\/} those of the quarks and charged leptons.  Adopting
familiar ideas of generating eigenvalues through off--diagonal mixings, we find
that the bizarre pattern of masses and mixings observed in the charged fermion
sector can be adequately described (with $\sim$ 10\% accuracy) within
an economical $SO(10)$ framework.  A concrete proposal was
presented that provides five
successful predictions for the masses and mixings in the quark and the
charged lepton systems.  The same description goes extremely well with a
value of $m_{\nu_\tau} \sim 1/20$ eV as well as with a large
$\nu_\mu-\nu_\tau$
oscillation angle ($\rm sin^22\theta_{\mu \tau}^{\rm osc} \simeq 0.82-0.96)$,
despite
highly non--degenerate masses for the light neutrinos. Both these features are
in good agreement with the SuperK results on atmospheric neutrinos.

The other major goal of this paper has been to revisit the previously noted
link between neutrino masses and nucleon decay \cite{bpw} in the light of the
SuperKamiokande result.   We find that the
mass of $\nu_\tau$~ $(\sim 1/20~{\rm eV}$), together with the large
$\nu_\mu-\nu_\tau$
oscillation angle implied by the SuperK  result, suggest a significant
enhancement in the standard as well as in the new (neutrino mass related) $d=5$
proton decay operators, relative to previous estimates, including those of
Ref. \cite{bpw}.  There are many uncertainties in the prediction for
the proton decay rate, including ones arising
from uncertainties in the SUSY spectrum, from the hadronic matrix
elements, from the relative phases of the many different contributions
(see the discussions
in Appendix B), and
from the allowed extent of unification scale threshold corrections to
$\alpha_3(m_Z)$.
Nevertheless, following a detailed analysis of threshold corrections for the
minimal
Higgs system (${\bf 45_H}, {\bf 16_H}, {\overline{\bf 16}_H}$) used in our
work (see
Sec. VI D and Appendix D)
we found that the standard operator contributions
severely constrain the underlying
model.

Specifically, we found that for MSSM embedded in $SO(10)$, the standard
operators, with
generous allowance for the uncertainties,
lead to lifetime estimate
$\Gamma^{-1}(p \rightarrow \overline{\nu} K^+)_{\rm std}^{\rm MSSM} \le [1.5,7]
\times
10^{33}~{\rm yrs}$ corresponding to cases (I,II) (see Sec. V, Eq. (37) for the
origin of these two cases).  In the process, by combining constraints from the
observed limit on proton lifetime together with a reasonable upper limit on
threshold corrections, we found that $\tan\beta$ must be rather small ($\le
3,8)$
for cases (I, II), in MSSM.
For larger values of $\tan\beta \sim 20$,
one must turn to
the Extended Supersymmetric Standard Model (ESSM) embedded in $SO(10)$.
This allows for two extra (vector-like) families at the TeV scale,
and has been motivated on other grounds \cite{essm}.  In this framework, the
standard
$d=5$ operators still represent perfectly viable sources for proton decay.
We find typically  $\Gamma^{-1}(p \rightarrow \overline{\nu}K^+)_{\rm
std}^{\rm
ESSM}
\le (1-5) \times 10^{33}~{\rm yrs}$  for $\tan\beta \ge 20$ (Sec. VI.D).

As observed in our earlier work \cite{bpw}, one very important consequence of
quark--lepton unification is the likely existence of new $d=5$ proton decay
operators that are related to neutrino masses.  Here we have shown that the
``observed" mass $\nu_\tau \sim 1/20$ eV and large $\nu_\mu-\nu_\tau$
oscillation angle (which go well with theoretical expectations in $SO(10)$)
enhance these operators relative to our previous estimates \cite{bpw}.
 As a result, the standard and the
new operators appear to make comparable contributions (see remarks following
Eq. (64)).  We have remarked in Sec. VI.F
that in some string-inspired models leading to $G_{224}$ or $G_{2113}$
symmetries
(rather
than intact $SO(10)$), the color triplets related to electroweak doublets get
projected out of the spectrum altogether, and thus the standard $d=5$ operators
do not contribute.  In that case, the new $d=5$ operators related to
neutrino masses will survive and dominate.  We found (Sec. VI.E) that
(given the SuperK result)
these new operators by themselves lead to proton lifetime
$\Gamma^{-1}(p \rightarrow \overline{\nu} K^+)_{\rm new}$ with a ``central
value"
of about $3 \times 10^{31}~{\rm yrs}$, and a range that is perfectly compatible
with the observed limit.  Assuming rather generous
uncertainties (i.e., those in $(m_{\tilde{W}}/m_{\tilde{q}})$,
$m_{\tilde{q}}$,  the matrix element $\beta_H$), and the amplitude,
the new operators by themselves lead to a proton lifetime
$\leq    (1-6) \times 10^{33}$ yrs.  Thus for the MSSM (or ESSM) embedded in
$SO(10)$, we
expect the proton lifetime to be shorter than about $10^{34}$ yrs.

A distinctive feature of the $SO(10)$ framework discussed here is the
potential prominence of the charged lepton mode $p \rightarrow \mu^+K^0$.
In minimal $SU(5)$, this mode becomes prominent only for $\tan\beta \ge 20$,
when gluino dressing becomes significant.  But such large values of $\tan\beta$
are highly disfavored in $SU(5)$, owing to observed limit on proton lifetime.
Furthermore, for such large values of
$\tan\beta$, the Higgsino dressing of the
baryon number violating $RRRR$ operator becomes dominant, leading
anyhow to dominance of the neutrino mode.  Thus in conventional $SU(5)$--like
supersymmetric unified models, charged lepton decay of the proton is
(even relatively!) scarce.
In the $SO(10)$ framework, on the contrary, the decay $p \rightarrow \mu^+ K^0$
competes favorably with the neutrino mode.  This becomes possible primarily
because
of enhancement of the new neutrino--mass related $d=5$ operators, in which the
amplitudes for $\mu^+ K^0$ and $\overline{\nu}K^+$ modes are comparable.
Thus observation of the $\mu^+ K^0$ decay mode of the proton would be
very encouraging for
the circle of ideas discussed here.
 Owing to the hierarchical Yukawa couplings
of the Majorana neutrinos suggested by the solar and the atmospheric neutrino
data, the modes $p \rightarrow e^+K^0$ and $p \rightarrow e^+ \pi^0$ are
predicted to be
highly suppressed relative the $p \rightarrow \overline{\nu}K^+$.

While we focused on a specific $SO(10)$ example,
several of our results are likely to be more general.  In any scenario with
just
three light neutrinos, a simultaneous resolution of the solar and atmospheric
neutrino anomalies argues for hierarchical neutrino masses.
In a framework that also unifies quarks and leptons (with modest
mixing angles), a simple way to generate
the near-maximal neutrino oscillation angle needed for the atmospheric neutrino
anomaly is to attribute it partly to the $(\mu-\tau)$
sector and partly to the $(\nu_\mu-\nu_\tau)$ sector.  The precise way this
division is made is model dependent.
As one goes beyond minimal supersymmetric
$SU(5)$ to a unified framework
where small neutrino mass is a compelling feature, the predicted
lifetime of the
proton tends to decrease.  In such unified
models there are various factors contributing to  shorten the lifetime (see
Appendix C)
including (i) an enhanced coupling of the muon to the color triplets (relative
to
the coupling of the strange quark), (ii) an enhanced up--quark coupling to the
color triplet which scales as $\sqrt{m_c m_t}$ rather than $m_c$,
and (iii) the presence of new operators related to neutrino mass.

Proton decay has been anticipated for quite some time in the context of
unified theories.  Recent data from SuperK on neutrino mass makes the
case for observable proton decay still
more compelling.  With
improved searches, especially for the $\overline{\nu}K^+$ and $\mu^+ K^0$
modes,
either proton decay will be revealed, or some promising and otherwise
remarkably successful ideas on unification will be called into
question seriously.

{\bf Acknowledgments:} K.S.B is supported by funds from the Oklahoma State
University.  The research of J.C.P. has been supported
in part by NSF Grant No. Phy-9119745 and by
a Distinguished Research Fellowship awarded by the University of
Maryland. F.W is supported by
DOE grant No. DE-FG02-90ER-40542.

\section*{Appendix A}

\subsection*{Numerical values of various matrices}

Here we give the numerical values of the matrices $V_u$, $V_\ell$ and
 $V' = V_u^\dagger V_l$, where $V_u$ diagonalises the up quark matrix $U$ of
 Eq. (28): $u^{(g)} = V_u u^{(m)}$ with $(g)$ and $(m)$ denoting the gauge
 and mass eigenstates, $V_\ell$ diagonalises the charged lepton mass matrix.
 We also give the numerical values of the
 matrices $(\hat{H}, \hat{A}, \hat{G}, \hat{F})$ as well as their matrix
product
 with $V'$, these are relevant for proton decay amplitude calculations.  From
 the fit to the fermion masses discussed in Sec. IV-V, we have determined
 the (approximate) values of the parameters $(\sigma, \eta, \epsilon,
 \epsilon', \eta';)$.  For numerical purposes we shall choose their ``central
values":
 \begin{eqnarray}
 \sigma &=& -0.1096~ \eta_{cb},~~\eta = -0.1507 ~\eta_{cb},~~\epsilon= 0.0954~
\eta_\epsilon,
 \nonumber \\
 \epsilon' &=& 1.76 \times 10^{-4}~ \eta_{\epsilon'},~~\eta' = 4.14 \times
10^{-3} ~
 \eta_{\eta'}~.
 \end{eqnarray}

With these values, the matrix $V_u$ is given by:
\begin{eqnarray}
V_u &\simeq& \left(\matrix{ 1 & {\epsilon' \over \epsilon^2-\sigma^2} &
\epsilon'(-\epsilon+\sigma)
\cr -{\epsilon' \over \epsilon^2-\sigma^2} & 1 & \epsilon+ \sigma \cr
{\epsilon' \over \epsilon-\sigma} & -(\epsilon+ \sigma) & 1 }\right) \nonumber
\end{eqnarray}
\begin{eqnarray}
 \simeq   \left(\matrix{ 1 & -0.061 \eta_{\epsilon'} & -3.61 \times 10^{-5}
\eta_{cb}
 \eta_{\epsilon'} \cr 0.061 \eta_{\epsilon'} & 1 & -0.0142 \eta_{cb} \cr
 8.6 \times 10^{-4} \eta_{cb} \eta_{\epsilon'} & 0.0142 \eta_{cb} & 1}\right)~.
 \end{eqnarray}
 An analogous expression for $V_\ell$ is obtained by the replacement $\sigma
\rightarrow
 \eta, \epsilon \rightarrow -3\epsilon$ and $\epsilon' \rightarrow
 -3\epsilon'$ in
 the first part of Eq. (76).
 The numerical values of $V_{\ell}$ and $V' = V_u^\dagger V_{\ell}$ are:
 \begin{eqnarray}
 V_{\ell} \simeq \left(\matrix{1 & 0.07 \eta_{\eta'} & 5.6 \times 10^{-4}
\eta_{cb} \eta_{\eta'}
 \cr -0.07\eta_{\eta'} & 0.905 & -0.4369 \eta_{cb} \cr
 -0.031 \eta_{cb} \eta_{\eta'} & 0.4369 \eta_{cb} & 0.905}\right)~, \nonumber
\end{eqnarray}
\begin{eqnarray}
 V' \simeq  \left(\matrix{1 & 0.052 \eta_{\epsilon'} + 0.07 \eta_{\eta'}
 & -0.026 \eta_{cb} \eta_{\epsilon'} \cr
 -(0.052 \eta_{\epsilon'} + 0.07 \eta_{\eta'}) & 0.911 & -0.4227 \eta_{cb} \cr
 -0.030 \eta_{cb} \eta_{\eta'} & 0.4227 \eta_{cb} & 0.911}\right).
 \end{eqnarray}

 Numerical values of the matrices ($\hat{H}, \hat{A}, \hat{G}, \hat{F})$ and
their
 products with $V'$ are:
 \begin{eqnarray}
 \hat{H} \simeq h_{33} \left(\matrix {-1.1 \times 10^{-5} & -1.8 \times
10^{-4} \eta_{\epsilon'} &
 -5.8 \times 10^{-3} \eta_{cb}\eta_{\epsilon'} \cr
 -1.8 \times 10^{-4} \eta_{\epsilon'} & -2.9 \times 10^{-3} & -0.0954\eta_{cb}
\cr
  -5.8 \times 10^{-3} \eta_{cb} \eta_{\epsilon'} & -0.0954 \eta_{cb} &
1}\right)~, \nonumber
\end{eqnarray}
\begin{eqnarray}
\hat{H}V' \simeq h_{33}\left(\matrix{1.8 \times 10^{-4} \eta_{\epsilon'}
\eta_{\eta'} &
 -2.6 \times 10^{-3} \eta_{\epsilon'} & -5.8 \times 10^{-3} \eta_{cb}
\eta_{\epsilon'} \cr
3.0 \times 10^{-3} \eta_{\eta'} & -0.043 & -0.0954 \eta_{cb} \cr
2.9 \times 10^{-3} \eta_{cb} \eta_{\epsilon'} - 0.024 \eta_{cb}\eta_{\eta'} &
0.3273 \eta_{cb} & 1}\right)~,
\end{eqnarray}

\begin{eqnarray}
\hat{A} \simeq h_{33} \left(\matrix{ 3.1 \times 10^{-5} & 3.4 \times 10^{-4}
\eta_{\epsilon'}
& 5.8 \times 10^{-3} \eta_{cb} \eta_{\epsilon'} \cr
3.4 \times 10^{-4}\eta_{\epsilon'} & 2.7 \times 10^{-3} & 0.0954 \eta_{cb} \cr
5.8 \times 10^{-3} \eta_{cb} \eta_{\epsilon'} & 0.0954 \eta_{cb} & -2.7 \times
10^{-3}
}\right) ~,\nonumber
\end{eqnarray}
\begin{eqnarray}
\hat{A}V' \simeq  h_{33} \left(\matrix{-1.9 \times 10^{-4} \eta_{\epsilon'}
\eta_{\eta'} + 3.5 \times
10^{-5} & 2.8 \times 10^{-3} \eta_{\epsilon'} & 5.8 \times 10^{-3} \eta_{cb}
\eta_{\epsilon'}
\cr 3.0 \times 10^{-3} \eta_{\eta'} +5.6 \times 10^{-4} \eta_{\epsilon'} &
0.043 &
0.0954 \eta_{cb} \cr
 -6.7 \times 10^{-3} \eta_{cb}\eta_{\eta'} + 8.6 \times 10^{-4} \eta_{cb}
\eta_{\epsilon'}
& 0.0954 \eta_{cb} & -0.043}\right)~,
\end{eqnarray}

\begin{eqnarray}
\tan\gamma \hat{G} \simeq h_{33} \left(\matrix{5.4 \times 10^{-4}
\eta_{\epsilon'}
\eta_{\eta'} & 4.1 \times 10^{-3} \eta_{\eta'} & -2.5 \times 10^{-3} \eta_{cb}
\eta_{\epsilon'}
\cr 4.1 \times 10^{-3} \eta_{\eta'} & -1.2 \times 10^{-3} & -0.0411 \eta_{cb}
\cr
  -2.5 \times 10^{-3} \eta_{cb} \eta_{\epsilon'} & -0.0411 \eta_{cb} & 1.2
\times 10^{-3}
} \right)~,\nonumber
\end{eqnarray}
\begin{eqnarray}
\tan\gamma \hat{G} V' \simeq h_{33} \left(\matrix{\{3.7 \times 10^{-4}
\eta_{\epsilon'}
\eta_{\eta'}
& \{-1.1 \times 10^{-3} \eta_{\epsilon'}
& \{-2.5 \times 10^{-3} \eta_{cb} \eta_{\epsilon'} \cr
 -2.9 \times 10^{-4}\} & +4.1 \times 10^{-3} \eta_{\eta'}\} &
 -1.8 \times 10^{-3} \eta_{cb} \eta_{\eta'}\} \cr
5.4 \times 10^{-3} \eta_{\eta'} & -0.019 & -0.0411\eta_{cb} \cr
  -3.7 \times 10^{-4} \eta_{cb} \eta_{\epsilon'} + 2.9 \times 10^{-3} \eta_{cb}
\eta_{\eta'}
& -0.0411 \eta_{cb} & 0.019 }\right)
\end{eqnarray}

\begin{eqnarray}
\hat{F} \simeq \hat{f}_{33} \left(\matrix{x+7.3 \times 10^{-7} +1.0 \times
10^{-4} \eta_{cb} y &
1.7 \times 10^{-3} \eta_{\epsilon'}\eta_{cb}y -0.06 \eta_{\epsilon'} x & 0.06
\eta_{\epsilon'} y \cr
1.7 \times 10^{-3} \eta_{\epsilon'} \eta_{cb}y -0.06 \eta_{\epsilon'} x &
0.0284 \eta_{cb}
y & y + 0.0142 \eta_{cb} \cr
0.06 \eta_{\epsilon'} y & y+0.0142 \eta_{cb} & 1}\right)
\end{eqnarray}

\section*{Appendix B}

\subsection*{B.1. Numerical evaluation of the full amplitudes for proton decay}

In converting the superpotential given in Eq. (41) into a proton decay
amplitude,
the first step is to dress
two of the sfermions among the four superfields in each term of Eq. (41)
{\it i.e}., to convert them into fermions.  The dominant contribution to the
amplitude
arises from the dressing involving the charged wino.  Here we list the 12
terms that
arise in dressing each one of the 7 terms of Eq. (41) with a final state
neutrino.
The four fermion amplitude after wino dressing of a generic  superpotential
term
\begin{equation}
W = M_{\rm eff}^{-1}
(u^TF d')\{u^T G \ell-d^\prime T G\nu'\}
\end{equation}
is

\begin{eqnarray}
A(p \rightarrow \overline{\nu}X) &=&
M_{\rm eff}^{-1} \epsilon_{\alpha \beta \gamma} \times [
F_{11}G_{2l}(u^\alpha d^{\prime \beta})(s^{\prime \gamma}\nu'_l) \left(f(c,l)+
f(u,d)\right)
\nonumber \\
&-& F_{22}G_{1l}(u^\alpha s^{\prime \beta})(s^{\prime \gamma}\nu'_l)
\left(f(c,l)+ f(c,d)\right)
 - F_{12}G_{1l}(u^\alpha d^{\prime \beta})(s^{\prime \gamma}\nu'_l)
\left(f(c,l)+ f(u,d)\right)
\nonumber \\
&+& F_{12}G_{2l}(u^\alpha s^{\prime \beta})(s^{\prime \gamma}\nu'_l)
\left(f(c,l)+ f(d,c)\right)
 -F_{13}G_{1l}(u^\alpha d^{\prime \beta})(b^{\prime \gamma}\nu'_l)
\left(f(t,l)+ f(u,d)\right)
\nonumber \\
& +& F_{13}G_{2l}(u^\alpha b^{\prime \beta})(s^{\prime \gamma}\nu'_l)
\left(f(c,l)+ f(d,t)\right)
+ F_{13}G_{3l}(u^\alpha b^{\prime \beta})(b^{\prime \gamma}\nu'_l)
\left(f(t,l)+ f(d,t)\right)
\nonumber \\
& -&
F_{23}G_{1l}(u^\alpha b^{\prime \beta})(s^{\prime \gamma}\nu'_l) \left(f(c,l)+
f(t,d)\right)
 - F_{23}G_{1l}(u^\alpha s^{\prime \beta})(b^{\prime \gamma}\nu'_l)
\left(f(c,d)+ f(t,l)\right)
\nonumber \\
 &-&F_{33}G_{1l}(u^\alpha b^{\prime \beta})(b^{\prime \gamma}\nu'_l)
\left(f(t,l)+ f(t,d)\right)
+ F_{11}G_{3l}(u^\alpha d^{\prime \beta})(b^{\prime \gamma}\nu'_l)
\left(f(t,l)+ f(u,d)\right)
\nonumber \\
& +&
F_{12}G_{3l}(u^\alpha s^{\prime \beta})(b^{\prime \gamma}\nu'_l) \left(f(t,l)+
f(d,c)\right)] \left( {\alpha_2 \over 4 \pi} \right)~.
\end{eqnarray}
The dressing functions $f(a,b)$ are defined later in Eq. (93).
In writing Eq. (83), use has been made of the symmetric nature of $F$
~($F_{ij} = F_{ji}$).
As described in the text, the notation $d' = V_{CKM}d$ and $\nu' =
V_{CKM}^\ell \nu$
has been adopted.

A simpler expression can be obtained for proton decay amplitude with
a charged lepton in the final state (for the same
generic superpotential term as in Eq. (82)):

\begin{eqnarray}
\hat{A}(p \rightarrow \ell^+ X) &=&
M_{\rm eff}^{-1}\epsilon_{\alpha \beta \gamma} \times
[(u^\alpha d^\beta)(u^\gamma \ell_l^-) [V_{cd}(G_{1l}F_{21}-G_{2l}F_{11})+
V_{td}(G_{1l}F_{31}-G_{3l}F_{11}) \nonumber \\
&+& (u^\alpha s^\beta)(u^\gamma \ell^-)[V_{cs}(G_{1l}F_{21}-G_{2l}F_{11})+
V_{ts}(G_{1l}F_{31}-G_{3l}F_{11})]~.
\end{eqnarray}
The amplitude with a hat is the full four--Fermion amplitude divided by
the loop function ($2f)$, defined in Eq. (110) and (111) of Appendix C.
Note that the down type quark fields appearing above (without primes)
are the physical ones.

With the numerical values of the Yukawa coupling matrices relevant for the
color
triplet exchange, we can use Eqs. (83) and (84) to compute the decay amplitude
for any given channel.  Since there are seven terms in Eq. (83), for the
neutrino mode, there will be a total of $7 \times 12 = 84$ terms to be summed.
This is most efficiently done numerically using $Mathematica$.  We now present
some
of the dominant amplitudes.  For these estimates we define $V_{cd} = 0.22
\eta_{cd},
V_{td} = 0.006\eta_{td}, V_{ts} = 0.04 \eta_{ts}$ and $V_{ud} = V_{cs}=1$.  The
$\eta_{ij}$ are the unknown phase factors, they can also be used to vary the
central values of the CKM mixing angles adopted. We drop terms which are
smaller
by more than an order of magnitude compared to  the leading term in each set.
\begin{eqnarray}
\hat{A}[(ud)(s\nu_\tau)] &\simeq& M_{\rm eff}^{-1}h_{33}^2 [1.9 \times 10^{-5}
\eta_{cd}
\eta_{ts} \eta_{\epsilon'} -9.0 \times 10^{-6} \eta_{cd}\eta_{cb}\eta_{\epsilon
'}
\nonumber \\
&-&6.1 \times 10^{-6}\eta_{td}\eta_{ts}\eta_{cb}\eta_{\epsilon'}
+4.8 \times 10^{-6} \eta_{ts} - 2.5 \times 10^{-6} \eta_{cb} \nonumber \\
&+& 2.1 \times 10^{-6}
\eta_{td}\eta_{\eta'}
+ 3.0 \times 10^{-6} \eta_{cd}\eta_{ts} \eta_{\eta'}
+2.2 \times 10^{-6} \eta_{cd}\eta_{cb} \eta_{\eta'}]  \nonumber \\
&+& (M_{16}\tan\gamma)^{-1}
h_{33}\hat{f}_{33}[3.1 \times 10^{-7} \eta_{cd} \eta_{ts}\eta_{\epsilon'} +
6.0 \times 10^{-7} \eta_{td}\eta_{ts}\eta_{cb}\eta_{\epsilon'} \nonumber \\
&+& 4.3 \times 10^{-7}
\eta_{td}\eta_{ts}\eta_{cb}\eta_{\eta'}+
1.5 \times 10^{-7}\eta_{td}\eta_{\eta'} \nonumber \\
&+&2.3 \times 10^{-7} \eta_{cd}\eta_{ts} \eta_{\eta'}-0.0411 \eta_{cb}x
+ 3.1 \times
10^{-5}\eta_{cd}\eta_{\epsilon'}y \nonumber \\
&+& 2.2 \times 10^{-5}\eta_{cd}\eta_{ts}\eta_{cb}
\eta_{\epsilon'}y + 1.1 \times 10^{-5} \eta_{td}\eta_{cb}\eta_{\eta'}y
\nonumber \\
&+&1.6 \times 10^{-5} \eta_{cd}\eta_{ts}\eta_{cb}\eta_{\eta'}y
+ 1.1 \times 10^{-5} \eta_{cd}\eta_{\eta'} y \nonumber \\
&+& z\{-2.47 \times 10^{-4}\eta_{td}\eta_{cb}-1.28 \times 10^{-4}\eta_{cd}
\nonumber \\
&+& 9.94 \times 10^{-5}\eta_{ts}\eta_{cb}\eta_{\epsilon'} + 7.23 \times
10^{-5}\eta_{ts}
\eta_{cb} \eta_{\eta'}\}
]~.
\end{eqnarray}

\begin{eqnarray}
\hat{A}[(ud)(d\nu_\tau)] &\simeq& M_{\rm eff}^{-1}h_{33}^2[2.9 \times 10^{-6}
\eta_{cd}\eta_{td}\eta_{\epsilon'} - 2.0 \times 10^{-6}
\eta_{cb}\eta_{\epsilon'}]
\nonumber \\
&+& (M_{16}\tan\gamma)^{-1}\hat{f}_{33}h_{33}[4.7 \times 10^{-8}
\eta_{cd}\eta_{td}
\eta_{\epsilon'}
+ 8.9 \times 10^{-8} \eta_{cb}\eta_{\epsilon'} \nonumber \\
&+&1.8 \times 10^{-8} \eta_{cb}
\eta_{\eta'}
+ 1.3 \times 10^{-8} \eta_{td} + 6.5 \times 10^{-8} \eta_{cb}
\eta_{\eta'} \nonumber \\
&+& 6.8 \times 10^{-8} \eta_{cd}\eta_{td}\eta_{\eta'}
-1.8 \times
10^{-8} \eta_{cb}\eta_{\eta'}
-9.0 \times 10^{-3} \eta_{cd}\eta_{cb}x \nonumber \\
&-&6.8 \times 10^{-6}
\eta_{\epsilon'} y
+ 3.3 \times 10^{-6} \eta_{cd}\eta_{td}\eta_{cb}\eta_{\epsilon'}y \nonumber \\
&+&4.8 \times 10^{-6}\eta_{cd}\eta_{td}\eta_{cb}\eta_{\eta'}y
+ 2.5 \times 10^{-6}
\eta_{\eta'}y \nonumber \\
&+& z\{ -5.42 \times 10^{-5} \eta_{cd}\eta_{td} \eta_{cb} -2.82 \times
10^{-5} + 1.49
\times 10^{-5} \eta_{td} \eta_{cb} \eta_{\epsilon'} \nonumber \\
&+& 1.08 \times \eta_{td}
\eta_{cb}
\eta_{\eta'}\}
]
\end{eqnarray}

\begin{eqnarray}
\hat{A}[(ud)(s\nu_\mu)] &\simeq & M_{\rm eff}^{-1} h_{33}^2[-3.5 \times 10^{-6}
\eta_{td}\eta_{ts}\eta_{\epsilon'} + 8.5 \times 10^{-6} \eta_{cd}\eta_{ts}
\eta_{cb} \eta_{\epsilon'} \nonumber \\
&-& 3.4 \times 10^{-6}\eta_{cd}\eta_{\epsilon'}
+ 2.1 \times 10^{-6}\eta_{ts}\eta_{cb}-4.7 \times 10^{-6}\eta_{td}\eta_{cb}
\eta_{\eta'} \nonumber \\
&-&7.0 \times 10^{-6}\eta_{cd}\eta_{ts}\eta_{cb}\eta_{\eta'}
+ 5.1 \times 10^{-6} \eta_{cd}\eta_{\eta'} -2.1 \times 10^{-6} \eta_{\epsilon'}
\eta_{\eta'} \nonumber \\
&-&1.9 \times 10^{-6} \eta_{ts}\eta_{cb}\eta_{\epsilon'}\eta_{\eta'}]
+ (M_{16}\tan\gamma)^{-1}\hat{f}_{33}h_{33}[2.6 \times 10^{-7}
\eta_{td}\eta_{ts}\eta_{\epsilon'}
\nonumber \\
&+& 1.4 \times 10^{-7} \eta_{cd}\eta_{ts}\eta_{cb}\eta_{\epsilon'}
-9.9 \times 10^{-7} \eta_{td}\eta_{ts}\eta_{\eta'} \nonumber \\
&-& 3.5 \times 10^{-7}
\eta_{td}\eta_{cb} \eta_{\eta'} - 5.2 \times 10^{-7}
\eta_{cd}\eta_{ts}\eta_{cb}
\eta_{\eta'}-1.8 \times 10^{-7}\eta_{cd}\eta_{\eta'} \nonumber \\
&-&1.4 \times 10^{-7} \eta_{ts}
\eta_{cb}\eta_{\epsilon'}\eta_{\eta'}
-1.6 \times 10^{-3} \eta_{ts}\eta_{cb}x
-1.8 \times 10^{-2}x  \nonumber \\
&+&
9.2 \times 10^{-6} \eta_{cd}\eta_{ts}\eta_{\epsilon'}y
-1.4 \times 10^{-5}\eta_{cd}\eta_{cb}\eta_{\epsilon'}y+2.5 \times 10^{-6}
\eta_{ts}y \nonumber \\
&-& 2.5 \times 10^{-5} \eta_{td}\eta_{\eta'}y -3.6 \times 10^{-5}
\eta_{cd}\eta_{ts}\eta_{\eta'}y
+ 2.6 \times 10^{-5}\eta_{cd}\eta_{cb}\eta_{\eta'}y \nonumber \\
&-&1.0 \times 10^{-5}\eta_{ts}\eta_{\epsilon'} \eta_{\eta'}y-7.1 \times 10^{-6}
\eta_{cb}\eta_{\eta'}\eta_{\epsilon'}y \nonumber \\
&+&z\{ -9.86 \times 10^{-6} \eta_{td}\eta_{ts}\eta_{cb} -1.11 \times
10^{-6}\eta_{td}
\nonumber \\
&-& 5.79 \times 10^{-5}\eta_{cd}\eta_{cb} +4.20 \times 10^{-5} \eta_{ts}
\eta_{\epsilon'}
\}
]
\end{eqnarray}

Similarly, the amplitudes $\hat{A}[(us)(d\nu_\tau)],~\hat{A}[(us)(d\nu_\mu)]$
and $\hat{A}[(ud)(d\nu_\mu)]$ can be computed.  We do not display these results
here, since the amplitudes for these operators are always
somewhat smaller than the ones displayed.  Furthermore, the
matrix element for the former two turn out to be suppressed by about a factor
of 3.

Turning now to the charged lepton decay amplitude,
\begin{eqnarray}
\hat{A}[(us)(u\mu^-)]) &\simeq&  M_{\rm eff}^{-1} h_{33}^2
[V_{cs}\{-8.4 \times 10^{-7} -2.1 \times 10^{-6} \eta_{\epsilon'}\eta_{\eta'}\}
\nonumber \\
&+& V_{ts}\{5.2 \times 10^{-5} -4.8 \times 10^{-5}\eta_{cb}\eta_{\epsilon'}
\eta_{\eta'}\} \nonumber \\
&+& (M_{16}\tan\gamma)^{-1}\hat{f}_{33}h_{33} [\{V_{cs}\{-5.0 \times 10^{-8}
\eta_{\epsilon'}\eta_{\eta'}+ 1.8 \times 10^{-2} x \nonumber \\
&-& 7.1 \times 10^{-6} y \eta_{cb}
\eta_{\eta'}\eta_{\epsilon'}\} +
V_{ts}\{9.3 \times 10^{-7} \eta_{cb}-3.6 \times 10^{-6}
\eta_{cb}\eta_{\eta'}\eta_{\epsilon'} \nonumber \\
&-& 0.0411 \eta_{cb}x +6.4 \times 10^{-5} y
+2.5 \times 10^{-4}y \eta_{\eta'}\eta_{\epsilon'}\} \nonumber \\
&+& z\{4.2 \times 10^{-5} \eta_{ts} \eta_{\epsilon'} -1.59 \times 10^{-5}
\eta_\epsilon
\eta_{\epsilon'}  \nonumber \\
&-&1.66 \times 10^{-4} \eta_{ts} \eta_{\eta'} -5.88 \times 10^{-5}
\eta_\epsilon
\eta_{\eta'} \}
]~.
\end{eqnarray}

\begin{eqnarray}
\hat{A}[(us)(ue^-)] &\simeq& h_{33}^2M_{\rm eff}^{-1}[
V_{cs}\{ 8.4 \times 10^{-8} \eta_{\eta'} +1.5 \times 10^{-7}
\eta_{\epsilon'}\} \nonumber \\
&+& V_{ts}\{8.3 \times 10^{-7} \eta_{cb}\eta_{\epsilon'} -6.9 \times
10^{-6}\eta_{cb}
\eta_{\eta'}+3.3 \times 10^{-6}\eta_{cb}\eta_{\epsilon'}\}  \nonumber \\
&+& (M_{16}\tan\gamma)^{-1} \hat{f}_{33}h_{33}[V_{cs}\{3.5 \times 10^{-9}
\eta_{\epsilon'}
-1.7 \times 10^{-4}\eta_{\epsilon'}x \nonumber \\
&+& 5.3 \times 10^{-3} \eta_{\eta'}x +
5.0 \times 10^{-7}\eta_{cb}\eta_{\epsilon'} y\}
+ V_{ts}\{ 8.3 \times 10^{-9}
\eta_{cb}\eta_{\epsilon'} \nonumber \\
&+& 3.1 \times 10^{-7} \eta_{cb}\eta_{\eta'} +2.5 \times
10^{-7} \eta_{cb}\eta_{\epsilon'} +3.7 \times 10^{-4}\eta_{cb}\eta_{\eta'} x
\nonumber \\
&+& 2.9 \times 10^{-3} \eta_{cb}\eta_{\eta'}x+2.2 \times
10^{-5}\eta_{\eta'}y+1.8 \times
10^{-5}\eta_{\epsilon'}y\} \nonumber \\
&+&z\{-1.44 \times 10^{-5} \eta_{ts} \eta_{\epsilon'} \eta_{\eta'}\}
 ]~.
\end{eqnarray}
The decay amplitude for $p \rightarrow \pi^0 \mu^+$ can be obtained from Eq. (88)
by the replacement $V_{cs} \rightarrow V_{cd}$ and $V_{ts} \rightarrow V_{td}$.
The amplitude for $p \rightarrow e^+\pi^0$ may be obtained from Eq. (89)
by the replacement: $V_{cs} \rightarrow V_{cd}, V_{ts} \rightarrow V_{td}$.

\subsection*{B.2 Representative contributions to proton decay amplitudes and
their estimated magnitudes:}

In this subsection, we exhibit in detail the evaluation of a few representative
(dominant) contributions to proton decay amplitudes.  The full set of
contributions
are listed in the preceding subsection.  We also estimate the magnitudes of
the full amplitudes allowing for uncertainty in the relative phase of the
different contributions.

To obtain some of the leading terms in the results exhibited above, consider
the part
containing the $\nu'$ field -- in the first term of $W^{(L)}$ (Eq. (41)),
which we
specify by subscript $I\nu'$.
\begin{equation}
W_{I\nu'}^{(L)}  = -M_{\rm eff} \epsilon_{\alpha \beta \gamma}
\hat{H}_{ij}(\hat{H}V')_{kl}
(u_i^\alpha d_j^{\prime \beta})(d_k^{\prime \gamma} \nu_l')~.
\end{equation}
Here $(\alpha, \beta, \gamma)$ denote color indices. The transpose symbols on
the relevant
fields are dropped henceforth.  We will first consider case I discussed in
Sec. V for which $\epsilon' \neq 0$.
The first and the second terms in Eq. (82)
arise by making the choices (A and B) and (C and D) respectively,
for the indices as given below:
\begin{eqnarray}
(ij)(kl) &=& (21)(23) ~~~~~({\rm Choice~A}) \nonumber \\
         &=& (22)(13) ~~~~~({\rm Choice~B}) \nonumber \\
         &=& (23)(13) ~~~~~({\rm Choice~C}) \nonumber \\
         &=& (21)(33) ~~~~~({\rm Choice~D})
 \end{eqnarray}
 Note that A and B are related by the interchange of $j \leftrightarrow k$,
and similarly C
 and D.

 \noindent{\it Choice A: (ij)(kl) = (21)(23):}  Starting with the operator
 $(u_2^\alpha d_1^{\prime \beta})(d_2^{\prime \gamma} \nu_3)$ in the
superpotential,
 two fields need to be utilized for wino dressing. Which
 ones remain external gets determined as follows.  The field $d_2^{\prime
\gamma}$ must
 remain external, yielding a strange quark $s^\gamma$, accompanied by the CKM
factor
 $V_{cs} \simeq 1$.\footnote{If $d_2'$ is used to yield an external
$d$--quark, and
 $\tilde{u}_2$ (after dressing) an $s$--quark, one would obtain the
four--fermion
 operator $(su)(d\nu_3)$.  It turns out that the matrix element of this
operator is
 suppressed by about a factor of 2-3 compared to that of $(du)(s\nu_3)$. See
 \cite{hisano}.}  (If $\tilde{d}_2'$ were dressed instead, it would yield an
 external charm quark, which is kinematically disallowed.)  Thus both
$u_2^\alpha$
 and $d_1^{\prime \beta}$ must be utilized for dressing, which, after
conversion of
 $\tilde{c}$ and $\tilde{d}_1'$ at the wino vertex, respectively, yield
$V_{cd}d^\alpha
 \equiv (\eta_{cd}\theta_C)d^\alpha$ and $V_{ub}u^\beta$, where $\theta_C
\simeq 0.22$.
 As discussed before, $\eta_{cd} = \pm 1$. $\eta_{ud}$ and $\eta_{us}$ are
chosen to be
 $+1$ by convention.  Thus, the corresponding contribution to the four--fermion
 proton decay operator leading to $\overline{\nu}_3$ emission is given by
 \begin{eqnarray}
 A^{(1)}_{I,\nu'} &\simeq& \left[{-\hat{f}(c,d) \over M_{\rm eff} } \right]
(\theta_C
 \eta_{cd}) \epsilon_{\alpha \beta \gamma}\hat{H}_{21}
(\hat{H}V')_{23}(d^\alpha u^\beta)
 (s^\gamma \nu_3) \nonumber \\
 &\simeq& -\left[h_{33}^2 \over M_{\rm eff}\right]\hat{f}(c,d)(3.8 \times
10^{-6}) \eta_{cd}
 \eta_{\epsilon'}\eta_{cb}\epsilon_{\alpha \beta \gamma}(d^\alpha
u^\beta)(s^\gamma \nu_3)
 \end{eqnarray}
where
\begin{equation}
\hat{f}(a,b) = \left({\alpha_2 \over 4 \pi}\right){m^2_{\tilde{W}} \over
m_a^2-m_b^2}\left(
{m_a^2 \over m_a^2 - m^2_{\tilde{W}}} {\rm ln}{m_a^2 \over m^2_{\tilde{W}}} -
[a \leftrightarrow b]\right) \equiv \left({\alpha_2 \over 4 \pi}\right)
f(a,b)~.
\end{equation}
In getting Eq. (92), we have used the numerical values of the elements of
$\hat{H}$ and
$\hat{H}V'$ given in Appendix A (Eq. (78)).  The phase factors
$\eta_{\epsilon'}$ and
$\eta_{cb}$ are $\pm 1$.  Likewise, confining still to the index combination A,
the contributions from the remaining five standard operators of $W^{(1)}$
labeled by
the subscripts (II-VI) are found to be:
\begin{eqnarray}
\left(A^{(1)}_{II,\nu'},~A^{(1)}_{III,\nu'},~A^{(1)}_{IV \nu'},
A^{(1)}_{V\nu'},~
A^{(1)}_{VI \nu'}\right) \simeq
(-3, ~+2, ~-6,~-0.4,~-0.8) A^{(1)}_{\nu'}
\end{eqnarray}
Combining the contributions (92) and (94), the total contribution of the terms
containing
$\nu'$ in the first six operators (I to VI) in $W^{L}$ is given by
\begin{equation}
A^{(1)}_{I ~{\rm to}~VI, \nu'} \simeq \left[h_{33}^2\hat{f}(c,d) \over M_{\rm
eff}\right]
\eta_{cd}\eta_{\epsilon'}\eta_{cb}(2.74 \times 10^{-5})\epsilon_{\alpha \beta
\gamma}
(d^\alpha u^\beta)(s^\gamma \nu_3)~.
\end{equation}

Next consider choice B.  In this case, since one starts with the operator
$\epsilon_{\alpha \beta \gamma}(u_2^\alpha d_2^{\prime \beta})(d_1^{\prime
\gamma}\nu_3)$,
using the dressing as in case A, one gets an extra minus sign compared to A,
owing to the
color factor.  Following the same procedure as above, one obtains
\begin{equation}
B^{(1)}_{I ~{\rm to}~VI,\nu'} \simeq \left[h_{33}^2\hat{f}(c,d) \over M_{\rm
eff}\right]
\eta_{cd}\eta_{\epsilon'}\eta_{cb}(-1.82\times 10^{-5})\epsilon_{\alpha \beta
\gamma}
(d^\alpha u^\beta)(s^\gamma \nu_3).
\end{equation}

Adding Eq. (95) and (96), the total contribution of terms containing $\nu'$ in
the first six
operators of $W^{(L)}$ for the index combinations A and B, leading to
$\overline{\nu}_3
K^+$ emission, is given by:
\begin{equation}
(A^{(1)}+ B^{(1)})_{I ~{\rm to}~VI,\nu'}
\simeq \left[h_{33}^2\hat{f}(c,d) \over M_{\rm eff}\right]
\eta_{cd}\eta_{\epsilon'}\eta_{cb}(0.9 \times 10^{-5})\epsilon_{\alpha \beta
\gamma}
(d^\alpha u^\beta)(s^\gamma \nu_3)~.
\end{equation}
This checks with the second term in Eq. (85) obtained numerically,
up to an
overall sign, which is due to the difference in the sign of Eq. (96) compared
to
Eq. (41).  Likewise, the
corresponding contributions for the index combinations C and D from the first
six operators in $W^{(L)}$ is found to be:
\begin{equation}
(C^{(1)}+D^{(1)})_{I ~{\rm to}~VI} \simeq \left[h_{33}^2\hat{f}(c,d) \over
M_{\rm eff}\right]
\eta_{cd}\eta_{\epsilon'}\eta_{ts}(-1.9 \times 10^{-5})\epsilon_{\alpha \beta
\gamma}
(d^\alpha u^\beta)(s^\gamma \nu_3)
\end{equation}
This is the first term of Eq. (85) -- up to an overall sign.  Note that Eqs.
(97)-(98) add
constructively because
$\eta_{ts} = -\eta_{cb}$.  We still have to include (i) the contribution that
arises by interchanging $i \leftrightarrow j$ while keeping $(k,l)$ fixed
(see Eq. (83)) and (ii) that from the terms containing charged lepton field
$\ell$ in $W^{(L)}$.  The latter contribute to neutrino emission by
dressing the charged slepton fields ($\tilde{\ell}$).  Including contributions
from
both (i) and (ii), one can verify that  the net
contribution of each kind -- such as A, B, C, D listed above -- is obtained by
simply making the following substitution in Eq. (97)-(98), etc:
\begin{equation}
\hat{f}(c,d) \rightarrow \hat{f}(c,d) + \hat{f}(c,\ell)~.
\end{equation}
In arriving at this result, use has been made of the symmetric nature of the
matrices
$\hat{H}, \hat{A}, \hat{G}, \hat{F}$, as well as color antisymmetry.  Thus,
adding Eq. (97)-(98),
and making the above substitutions, the net contribution to
$\overline{\nu}_3
K^+$ decay mode from the first six terms of $W^{(L)}$, for the index
combination
[A+B+C+D + ($i \leftrightarrow j$)] is given by:
\begin{eqnarray}
[A+B+C+D+(i \leftrightarrow j)]_{I ~{\rm to}~VI} &\simeq& \left[{h_{33}^2\left[
\hat{f}(c,d) + \hat{f}(c,\ell)\right] \over M_{\rm eff}}\right]
\eta_{cd}\eta_{\epsilon'}\eta_{cb} \times \nonumber \\
&~& (2.8 \times 10^{-5})\epsilon_{\alpha \beta \gamma}
(d^\alpha u^\beta)(s^\gamma \nu_3)~.
\end{eqnarray}

Having shown how the first two terms of the first bracket of Eq. (85) arise,
let us now consider
all the other terms in this bracket, which are proportional to $M_{\rm
eff}^{-1}$ and estimate their net magnitude.  It
is clear that the joint contribution of the first two terms (Eq. (85))
dominates over
the other terms in the same bracket by factors of 5 to 10.  Although the
relative
signs (phases) of these subleading terms are uncertain (since we do not know
the
$\eta_{ij}$), we can see that their combined magnitude even if they all add
constructively, is less than about $(1.5 \times 10^{-5}$), which is about half
of
that given by the first two terms of Eq. (85).
Not knowing the relative signs (phases)
we conclude that the net contribution from the standard operator (first six
terms
of $W$) is given by:\footnote{The reader will note that these unknown phases
(or signs)
are just the analogs of those which enter into $\tilde{t}$ versus $\tilde{c}$
contributions in minimal $SU(5)$, which is represented by a parameter $y_{tK}$
etc,
see Ref. \cite{nath,hisano}} $h_{33}^2[2 \hat{f}(c,d)/M_{\rm eff}](2.8 \times
10^{-5})
(1/2 ~{\rm to}~3/2)$.

Turning to case II ($\epsilon' = 0, \delta \neq 0$, see Sec. V), the net
contribution
from the standard operator to the same amplitude is obtained simply by setting
$\eta_{\epsilon'} = 0$ in the first square bracket in Eq. (85).  Since
$\delta \approx 10^{-5} \simeq \epsilon'/20$, contributions proportional to
$\delta$ can safely be neglected compared to the other terms.  Noting that
$\eta_{ts} = -\eta_{cb}$, the said contribution is thus given by
$[(4.8 + 2.5) \times 10^{-6} \eta_{ts} + (3+2.2)\times 10^{-6} \eta_{ts}
\eta_{cd}\eta_{\eta'}] = (0.73 + 0.55 \eta_{cd}\eta_{\eta'})\eta_{ts}
\times 10^{-5}$.  Allowing for unknown relative phase as above, this
contribution will vary in magnitude between (1.28 to 0.18))$ \times 10^{-5}$,
which may be represented approximately by $(0.7 \times 10^{-5})(1 \pm 0.7)$.

Thus we estimate that the contribution of the standard operator to the
$\overline{\nu}_\tau K^+$ amplitude for the cases I and II are given by:

\begin{equation}
A(\overline{\nu}_\tau K^+)_{\rm std} \simeq \left[h_{33}^22\hat{f}(c,d) \over
M_{\rm eff}\right]\epsilon_{\alpha \beta \gamma}
(d^\alpha u^\beta)(s^\gamma \nu_3)
\times \left\{\matrix{
2.8 \times 10^{-5}({1 \over 2} ~{\rm
to}~{3 \over 2}) ~~ - {\rm Case~I} \cr
0.7 \times 10^{-5} (1 \pm 0.7) ~~ - {\rm Case~II} }
\right\}~.
\end{equation}
Note that the standard amplitude for case II ($\epsilon' = 0$) is typically
smaller than that for case I ($\epsilon' \neq 0$) by a factor of 3.5 to 4.
Of course, if there is a near cancellation for case II for the
$\overline{\nu}_\tau
K^+$ mode, so that the curly bracket in Eq. (101)  (for case II) is as small as
$0.2 \times 10^{-5}$, then contributions from the new operator (related to
neutrino masses) to $\overline{\nu}_\tau K^+$ as well as contributions of the
standard and the new operators to the $\overline{\nu}_\mu K^+$ mode
would turn out to be far more important, as discussed below.

\noindent{\bf Contributions to the $\overline{\nu}_\mu K^+$ mode from the
standard operator (First six terms in W)}

Contributions to the $\overline{\nu}_\mu K^+$ mode from the standard operator
for the cases I and II can be estimated by using Eq. (87) in a manner similar
to that noted above.  Collecting terms proportional to $\eta_{\epsilon'}$ and
those independent of it from the first square bracket of Eq. (87), and using
$\eta_{ts} = - \eta_{cb} = \pm 1$, the said contribution is given by:
$[\{ -1.2 \eta_{cd} -0.35 \eta_{td}\eta_{ts} - (0.21-0.19)\eta_{\eta'}\}
\eta_{\epsilon'} + \{ 1.2 \eta_{cd}\eta_{\eta'} -0.21 -0.47 \eta_{td}
\eta_{cb}\eta_{\eta'}\}]\times 10^{-5}$.  Allowing for uncertainty of
relative phases, a reasonable estimate of the magnitude of this contribution
is thus given by $[1.5(1 \pm 0.5) \times 10^{-5}]$ for case I ($\epsilon' \neq
0$)
and by $[1.2 (1 \pm 0.5) \times 10^{-5}]$ for case II ($\epsilon' = 0$).  We
thus obtain:
\begin{eqnarray}
A(\overline{\nu}_\mu K^+)|_{std} \simeq \left[{h_{33}^2 2 \hat{f}(c,d) \over
M_{\rm eff} }\right]\left(\matrix{1.5  \cr  1.2}\right)(1 \pm 0.5) \times
10^{-5}
\epsilon_{\alpha \beta \gamma} (d^\alpha u^\beta)(s^\gamma \nu_\mu)~.
\end{eqnarray}
The upper and lower entries refer to cases I and II respectively.  We observe
that
the amplitudes for $\overline{\nu}_\mu K^+$ and $\overline{\nu}_\tau K^+$ modes
have very similar contributions from the standard operators; the former is
expected to be leading for case II and the latter for case I.

Combining the rates for the two modes, we would expect that the total proton
decay
rate, represented to a good approximation
by the sum $[\Gamma(\overline{\nu}_\tau K^+ + \Gamma(\overline{\nu}_\mu
K^+)]$, would typically be proportional (for central values of the amplitudes)
to $[(2.8)^2 + (1.5)^2] \approx 10$ for case I, and to $[(0.7)^2+(1.2)^2]
\approx 2$
for case II.  In short, in so far as contributions from the standard operator
is concerned, proton decay rate would typically be suppressed by about a
factor of
$5$ for case II compared to case I.
$$
{\Gamma((\overline{\nu}_\tau K^+)_{std} + \Gamma((\overline{\nu}_\mu
K^+)_{std}]_{\rm Case~ II}
\over
\Gamma((\overline{\nu}_\tau K^+)_{std} + \Gamma((\overline{\nu}_\mu
K^+)_{std}]_{\rm Case ~I}}
 \approx {1 \over 5} ~.
$$
This is an interesting and significant effect on proton decay, having its
origin
entirely in the fermion mass matrix.

\noindent{\bf Contribution to $\overline{\nu}_\tau K^+$ and
$\overline{\nu}_\mu K^+$
mode from the new operator
related
to neutrino masses (the seventh term in $W^{(L)}$)}:

Confining first to the part containing $\nu'$ in the seventh term of $W^{(L)}$
(see Eq. (41)), we have:
\begin{equation}
W^{(L)}_{VII,\nu'} = - {\epsilon_{\alpha \beta \gamma} \over
M_{16}}\hat{F}_{ij}
(\hat{G}V')_{kl} (u_i^\alpha d_j^{\prime \beta}(d_k^{\prime \gamma} \nu_l')~.
\end{equation}
Since $\hat{F}_{33}$ is the leading element of $\hat{F}$, first consider the
following combination of indices:

\noindent{\it Choice E: $(ij)(kl) = (33)(13)$}:  Starting with the operator
$(u_3^\alpha d_3^{\prime \beta})(d_1^{\prime \gamma} \nu_3)$, it is clear
that $d_3^{\prime \beta}$ must be external.  This yields $V_{ts}s^\beta$ (if
$\tilde{d}_3$ is dressed, it would lead to external top, which is forbidden);
thus $\tilde{u}_3^\alpha$ and $\tilde{d}_1^\gamma$ must be dressed to
respectively
yield $V_{td}d^\alpha$ and $u^\gamma$.  Thus one finds:
\begin{eqnarray}
E^{(1)}_{I,\nu'} &\simeq& {\hat{F}_{33} (\hat{G}V')_{13} \over M_{16}} V_{td}
V_{ts}
f(t,d) \epsilon_{\alpha \beta \gamma} (d^\alpha u^\beta)(s^\gamma \nu_3)
\nonumber \\
 &\simeq& \left({\hat{f}_{33}h_{33} \over  M_{16}
\tan\gamma}\right)\left[-(6 \eta_{cb}
 \eta_{\epsilon'} + 4.3 \eta_{cb}\eta_{\eta'})\eta_{td} \eta_{ts} \times
10^{-7} \right]
 \times \nonumber \\
 &~& \zeta_{KM} f(t,d) \epsilon_{\alpha \beta \gamma} (d^\alpha
u^\beta)(s^\gamma \nu_3)~.
 \end{eqnarray}

Here, we have used values of $\hat{F}_{ij}$ and $(\hat{G}V')_{kl}\tan\gamma$
listed
in Appendix A (Eq. (80)-(81)), and have put $V_{td} = 0.006 \eta_{td},~V_{ts}
= 0.04
\eta_{ts}$.
The factor $\zeta_{KM} = 1/2$ to $1$ denote the combined uncertainty in the
two CKM
elements.  Note that the two terms in Eq. (104) check with the second and the
fourth terms in Eq. (85) proportional to $(M_{16}\tan\gamma)^{-1}$.
Some of the other leading contributions arise by
choosing
$(ij) = (23)$ or $(32)$, varying $(kl)$.  These are proportional to $y$ and is
given by the
seventh through the eleventh terms in the second bracket of Eq. (85).  Note
that
the
tenth and the eleventh terms subtract (because $\eta_{ts}\eta_{cb} = -1$), and
thus have
a combined magnitude $\approx 0.5 \times 10^{-5}y$, while the seventh and the
eighth terms
subtract to have a net magnitude of $0.9 \times 10^{-5}y$.  Thus, allowing for
unknown phases of each of these contributions and also of the 10th term, we
estimate that the total contribution from the $y$ dependent terms is $\approx
(1.3
\pm 1) \times 10^{-5} y \simeq (0.7 \pm 0.5) \times 10^{-6}$, where we have put
$y \simeq 1/20$ (see Eq. (25)).  Now the $y$--independent terms in the second
square bracket of Eq. (85), including that exhibited in Eq. (104)),
are each individually smaller by
factors of 1.5 to 2.5 compared to $0.7 \times 10^{-6}$ except that the term
proportional to
$x$ is nearly $4 \times 10^{-6} (1/5~{\rm to}~1)$ for $x \simeq 10^{-4}(1/5
~{\rm to}~1)$.
Although the relative
phases (signs) of the various contributions, which depend on $\eta_{ij}$, are
not known, allowing for possible cancellation between some of the smaller
terms,
it seems most plausible (as a conservative estimate) that the second square
bracket
in Eq. (85) has a net magnitude $\approx (1.5 \times 10^{-6})(1/2.5~{\rm
to}~2.5)$ for case I.  Estimating similarly, one obtains a very similar
magnitude for case II as well.
Now, including the contributions from interchange of $i \leftrightarrow j$,
keeping
$(k,l)$ fixed, as appropriate,  and that involving charged
lepton
dressing, as before, the net contribution from the new operator (the seventh
term
in $W^{(L)}$), related to the neutrino masses is thus given by:
\begin{eqnarray}
A(\overline{\nu}_\tau K^+)_{\rm new} &\simeq& \left( {\hat{f}_{33} h_{33}
\over
M_{16} \tan\gamma}\right)\left[(1.5 \times 10^{-6})(1/2.5~{\rm to}~2.5)\right]
\left[f(t,d)+f(t,l)\right] \nonumber \\
&~&\epsilon_{\alpha \beta \gamma}(d^\alpha u^\beta)(s^\gamma \nu_3)~
\end{eqnarray}
for case I as well as case II.

Using Eq. (87), the contribution of the new operator to the $\overline{\nu}_\mu
K^+$ mode for both cases I and II is estimated to be:
\begin{eqnarray}
\hat{A}(\overline{\nu}_\mu K^+)_{\rm new} \approx \left( {\hat{f}_{33} h_{33}
\over
M_{16}\tan\gamma}\right)\left[(3 \times 10^{-6})(1/2 ~{\rm to}~2)\right]
\nonumber \\
\times [f(t,d)+f(t,l)]\epsilon_{\alpha \beta \gamma}(d^\alpha
u^\beta))(s^\gamma \nu_\mu)
\end{eqnarray}
Here we have used $x \approx 10^{-4}(1/5-1)$ and $z \simeq (1/400-1/200)$, in
accord
with discussions in Sec. V.  Thus, for the new operator, the
$\overline{\nu}_\mu K^+$
mode would typically
 supercede  the $\overline{\nu}_\tau K^+$ mode, and the
amplitudes for either mode remain essentially the same for both cases I and II.
It is worth noting that the main source of the relative enhancement of the
$\overline{\nu}_\mu K^+$ amplitude is due to the fact that the dominant
$y$--independent
terms add for $\overline{\nu}_\mu K^+$ (see Eq. 88), and they subtract for
$\overline{\nu}_\tau K^+$ (see Eq. (85)), if one uses $\eta_{ts} = -\eta_{cb}$.

\noindent{\bf Charged lepton decay modes:}

The amplitude for proton decay into charged leptons, $p \rightarrow \ell^+ X$,
is
given in Eq. (84) for a generic superpotential of the form Eq. (82).  The
number of
terms
are fewer compared to the amplitude for neutrino mode, so they can be
summed easily.
The full result is given in Eqs. (88)-(89) for the muonic and
electronic modes.  We will briefly describe
the origin of the dominant terms in the amplitude involving charged muons (Eq.
(88)).

The standard $d=5$ amplitude receives two contributions, one from the exchange
of $\tilde{c}-\tilde{d'}$ squarks (the term proportional to $V_{cs}$ in
Eq. (88)), and one from the exchange of $\tilde{t}\tilde{d'}$ (the $V_{ts}$
term).  They are comparable numerically.  Note that unlike in minimal $SU(5)$
models, where these contributions undergo a unitarity cancellation, here
the amplitude is not much suppressed compared to the neutrino mode.

Following the same procedure as discussed in the other cases, the contribution
of the standard operator to the $\mu^+ K^0$ mode (given in Eq. (88)) is
estimated
to be
\begin{eqnarray}
A(\mu^+ K^0)_{\rm std} \simeq {h_{33}^2 2 f(a,b) \over M_{\rm eff}} \left[
\matrix{3 \times 10^{-6} (1/2 ~{\rm to}~ 2)~~-{\rm Case~I}  \cr
 (2 \pm 0.8) \times 10^{-6}~~-{\rm Case~II}}\right] \epsilon_{\alpha \beta
\gamma}
(u^\alpha s^\beta)(u^\gamma \mu^-)~.
\end{eqnarray}
Thus the standard amplitudes for the $\mu^+ K^0$ mode are nearly the same for
both cases I and II.  They are typically smaller however than those for both
$\overline{\nu}_\tau K^+$ and $\overline{\nu}_\mu K^+$ modes by a factor of
3 to 10 (compare Eq. (107) with Eq. (101) and (102)):
\begin{equation}
A(\mu^+ K^0)_{\rm std}/A(\overline{\nu}_\ell K^+)_{\rm std}
\simeq 1/3 - 1/10,~\ell = \mu, \tau~.
\end{equation}
We see that for the {\it standard operators},
the charged lepton $(\mu^+$) mode, though relatively enhanced compared to
minimal $SU(5)$ is still expected to be suppressed in the rate, even in the
$SO(10)$ model presented here,
by one to two orders of magnitude, compared to the
neutrino modes.

The contributions of the new operator (related to neutrino masses) to $\mu^+
K^0$ may
also be estimated using Eq. (88).  Suppressing relative phases, it is found to
be:
\begin{eqnarray}
A(\mu^+ K^0)_{\rm new} &\approx& \left( {\hat{f}_{33} h_{33} \over M_{16}
\tan\gamma}\right)
\left[1.8 \times 10^{-6} (x/10^{-4}) + (2~{\rm to}~10) \times 10^{-7}\right]
\nonumber \\
&\approx & \left( {\hat{f}_{33} h_{33} \over M_{16}\tan\gamma}\right)(10^{-6})(
1/3~{\rm to}~
2) ~({\rm for~ I~ and II})
\end{eqnarray}
In above, we have used $z\approx 1/300$ (in Step I) and $x\simeq (1/3-1)
\times 10^{-4}$
(in Step 2) in accord with discussions in Sec. V.  Thus, confining to the
contributions
of only the new operator, we would typically expect:
$\hat{A}(\mu^+K^0)_{\rm new}/\hat{A}(\overline{\nu}_\ell K^+)_{\rm new}
\approx 1-1/3$ for $\ell
= \mu, \tau$.  (Compare Eq. (109) with Eq. (105) and (106).)  This means that
if
somehow the standard operators were absent (or highly suppressed) and the new
$d=5$ operators
were the only (dominant) source of proton decay (a possibility discussed in
Sec. VI.F), the charged lepton $\mu^+K^0$ could be comparable to the
neutrino mode, having a branching ratio
of nearly few to even 50\%.  As discussed in the text (Sec. VI), this is also
found
to be the case by including contributions from the standard as well as the new
operators.

The estimates of the proton decay amplitudes arising from the standard and the
new (neutrino--related) $d=5$ operators for the three relevant modes
$(\overline{\nu}_\tau
K^+, \overline{\nu}_\mu K^+, \mu^+ K^0$) are summarized in Table 1.

\begin{center}
\begin{tabular}{|c|c|c|c|c|}
\hline
Mode & (Case I)$_{\rm std}~ (\epsilon' \neq 0)$ & (Case II)$_{std}~
(\epsilon'=0$) &
(Case I)$_{\rm new}~ (\epsilon' \neq 0)$ & (Case II)$_{\rm new}~
(\epsilon'=0)$ \\
\hline
$\overline{\nu}_\tau K^+$ & $2.8 \times 10^{-5} (1 \pm 0.5)$ & $0.7 \times
10^{-5}
(1 \pm 0.7) $ & $1.5 \times 10^{-6} (1/2.5 - 2.5)$ & $2 \times 10^{-6}
(1/2.5-2.5)$
\\
$\overline{\nu}_\mu K^+$ & $1.5 \times 10^{-5}(1 \pm 0.5)$ & $1.2 \times
10^{-5}
(1 \pm 0.5)$ & $ 3 \times 10^{-6} (1/2-2)$ & $ 3 \times 10^{-6} (1/2-2)$ \\
$\mu^+ K^0$ & $3 \times 10^{-6}(1/2 -2)$ & $(2\pm 0.8) \times 10^{-6}$ &
$10^{-6}(1/3-2)$ & $10^{-6}(1/3-2)$ \\
\hline
\end{tabular}
\end{center}

Table Caption: The magnitudes for the standard and the new operators can be
obtained
by multiplying the respective entries shown in the Table by the factor $X
\equiv
h_{33}^2 2 f(c,d)/M_{\rm eff}$ and $Y \equiv \hat{f}_{33} h_{33} 2
f(c,d)/(M_{16} \tan\gamma)
$ respectively (see the text for definitions, in particular Sec. VI E, where
one
derives $P \equiv \hat{f}_{33} h_{33}/(M_{16}\tan\gamma) \approx 5 \times
10^{-19} {\rm GeV}^{-1} (1/2~{\rm to}~2)$,
and Eqs. (101) and (105) as examples).

\section*{Appendix C. Contribution to proton decay rate from the $d=5$
operator}

In this Appendix we give a general discussion of the numerical value for the
$d=5$ proton decay amplitude.

Define
\begin{equation}
\hat{A}(\overline{\nu}_i) = A_5 (p \rightarrow \overline{\nu}
K^+)/(2\overline{f})~.
\end{equation}
$A_5$ denotes the strength of the respective four--fermion proton decay
amplitudes
arising from the $d=5$ operators in $W$.  One representative contribution to
$A_5$
is given by Eq. (97).  The net value of $A_5$ includes contributions from all
the $d=5$ operators, allowing for all possible combinations of the indices
$(ij)(kl)$.
It also includes the wino dressing factor and the relevant CKM mixing
factors at all
vertices.  The net values are listed in Appendix B.
The quantity $\overline{f}$ is the average of the two relevant dressing
functions.  For instance, for the amplitude in Eq. (100):
\begin{equation}
\overline{f} \equiv \left[\hat{f}(c,d) + \hat{f}(c,l)\right]/2~.
\end{equation}
Assuming approximate degeneracy of the sfermions (for simplicity, see however,
remarks later), $\hat{f}$ is the same for all contributions to the amplitude.
Then $\hat{A}$ defined as above is just the net strength of the corresponding
$d=5$ operator in the superpotential, multiplied by the relevant CKM factors
which occur at the vertices involving the color triplet and the wino exchanges.
Since $\overline{f}$ has dimension of (mass)$^{-1}$, $\hat{A}$ has dimension 5.

For comparison purposes, it is useful to note that for minimal SUSY $SU(5)$
involving
exchange of color triplet $H_C$ between $(\tilde{c}s)$ and
$(\tilde{d}\nu_\mu)$ pairs,
the $d=5$ operator has strength $ = (h_{22}^u h_{12}^d/M_{H_C}) \simeq
(m_c m_s \sin\theta_C/(v_uv_d))/M_{H_C} \simeq (m_c m_s \sin\theta_C/v_u^2)
(\tan\beta/M_{H_C}) \approx 9 \times 10^{-8}(\tan\beta/M_{H_C})$, where
$\tan\beta = v_u/v_d$
and we have put $v_u = 174~{\rm GeV}$ and the extrapolated values of the
fermion masses
at the unification scale -- i.e., $m_c \simeq 300~{\rm MeV}$, and $m_s
\simeq 40~{\rm MeV}$.
Multiplying further by an additional factor of $\sin\theta_C$ due to conversion
of $\tilde{c}$ to $d$ at the wino vertex, for the case of $SU(5)$,
\begin{equation}
\hat{A}_{\tilde{c}\tilde{d}}(SU(5)) \simeq (1.9 \times
10^{-8})(\tan\beta/M_{H_C})~,
\end{equation}
corresponding to exchange of the pair $(\tilde{c},\tilde{d}$).  There is a
similar
contribution involving the exchange of the pair $(\tilde{t},\tilde{d})$ with
the substitution $m_c \sin\theta_C \rightarrow m_t V_{td}$.

In calculating the proton decay rate, we will assume the following spectrum of
supersymmetric particles as a guide:  Squarks are nearly degenerate, with
masses $m_{\tilde{q}} \approx 1~$TeV (1 to 1.5) and wino is lighter than
the squarks -- {\it i.e}., $m_{\tilde{W}}/m_{\tilde q} \approx$ 1/6 (1/2 to
2).  Consistent
with light gaugino masses, we will furthermore assume that $m_{1/2}$ (the
common gaugino mass) is small
compared to $m_0$ (the common scalar mass).
In this case, starting with universal masses for the
scalars
at the unification scale, we would expect the slepton masses to be nearly
degenerate with the squark masses at the electroweak scale.\footnote{Allowing
for
sleptons to be lighter than the squarks by a factor of 2 to 3 while keeping the
squark masses to be about 1 TeV would amount to increasing $\overline{f}$ by
about 50\%, compared to its value for the case of nearly degenerate masses
($m_{\tilde{q}} \simeq m_{\tilde{\ell}}$), and thereby enhancing the rate
by a factor of 2.  This possibility is of course not presently excluded.}
The two $f$--functions
that enter into the amplitude are then nearly equal (eg: in Eq. (99)),
$\hat{f}(c,d)
\simeq \hat{f}(c,l)$) and $\hat{f} \approx (m_{\tilde{W}}/m^2_{sq})(\alpha_2/4
\pi)$.
A SUSY spectrum as described, i.e., $m_{\tilde{W}} \ll m_{\tilde{q}} \approx
m_{\tilde{\ell}} \sim 1~{\rm TeV}$ is needed anyway, {\it a posteriori},
because without
it SUSY grand unified models based on $SU(5)$ or $SO(10)$ are likely to run
into
conflict with the experimental limits on proton lifetime.  A spectrum of this
type
is plausible in several scenarios of SUSY breaking.\footnote{For example,
models of SUSY breaking based on contributions from a family universal
anomalous
$U(1)$ $D$--term, superposed with subdominant dilaton $F$--term contributions
\cite{dine}
would lead to such a spectrum.}

With the strength of $(du)(s\nu_\tau)$ operator being given by $A_5 = \hat{A}
(2
\overline{f})$, as
in
Eq. (109), the inverse decay rate for $p \rightarrow \overline{\nu}_\tau K^+$
is given
by
\begin{eqnarray}
&~& \Gamma^{-1}(p \rightarrow \overline{\nu}_\tau K^+)
\approx (2.2 \times 10^{31})~{\rm yrs} \times \nonumber \\
&~& \left({.67 \over A_S}\right)^2\left[{0.006 {\rm GeV}^3 \over
\beta_H}\right]^2
\left[{(1/6) \over (m_{\tilde{W}}/m_{\tilde{q}})} \right]^2
\left[{m_{\tilde{q}} \over 1~{\rm TeV}}\right]^2 \left[{2 \times
10^{-24}~{\rm GeV}^{-1} \over
\hat{A}(\overline{\nu}) }\right]^2~.
\end{eqnarray}
Here $\beta_H$ denotes the hadronic matrix element defined by $\beta_H
u_L(\vec{k})
\equiv \epsilon_{\alpha \beta \gamma}$~$\left\langle0|(d_L^\alpha
u_L^\beta)u_L^\gamma|p,\vec{k}\right\rangle$.
While the range $\beta_H = (0.003 - 0.03)~{\rm GeV}^3$ has been used in the
past
\cite{hisano}, given that one
lattice calculations yield $\beta_H = (5.6 \pm 0.5) \times
10^{-3}~{\rm GeV}^3$ \cite{gavela},
we will take as a plausible range: $\beta_H = (0.006 ~{\rm GeV}^3)(1/2$ to 2).
$A_S$ stands for the short distance renormalization factor of the $d=5$
operator.
In minimal $SU(5)$ it has a central value of 0.67.  Although this factor is
slightly different in the $SO(10)$ model, we shall adopt $A_S = 0.67$ for
the $SO(10)$ model as well.

Note that the familiar factors that appear in the expression for proton
lifetime
-- i.e., $M_{H_C}$, ($1+y_{tK}$) representing the interference between the
$\tilde{t}$ and $\tilde{c}$ contributions and $\tan\beta$ -- are all
effectively
contained in $\hat{A}(\overline{\nu})$.  In fact, the analog of $M_{H_C}$ for
$SU(5)$ is given in our case by two mass scales: $M_{\rm eff}$ and $M_{16}$,
representing contributions of the standard and the neutrino mass--related
operators, respectively.   The analog of $(1+y_{tK}$) is reflected
by the uncertainty in the net value of the many terms in Eq. (85), which
depends on their unknown relative phases (see discussion in Appendix B, where
a reasonable estimate of the uncertainty owing to the phases is given (Eq.
(101) and (105)).

In minimal SUSY $SU(5)$, $\hat{A}$ is proportional to $(\sin2\beta)^{-1} =
1/2(\tan\beta
+ 1/\tan\beta)$, and thus approximately to $\tan\beta$ for $\tan\beta \ge 3$
or so.
Corresponding to a realistic treatment of fermion masses as in Sec. IV and V,
this
approximate proportionality to $\tan\beta$ does not however hold for $SO(10)$.
The reason is this: If the fermions acquire masses only through the
${\bf 10}_H$ in $SO(10)$, as is well known, the up and the down quark Yukawa
couplings will be equal.  This would give the familiar $t-b-\tau-\nu_\tau^D$
unification.  By itself, it would also lead to a large value of $\tan\beta
=m_t/m_b \simeq 60$, and correspondingly to a large enhancement in proton
decay amplitude.  Furthermore, it would also lead to the bad relations:
$m_c/m_s = m_t/m_b$ and $V_{CKM} = 1$.  However, in the presence of additional
Higgs multiplets contributing to fermion masses, such as ${\bf 16}_H$, which
(a) distinguish between the up and the down sectors and (b) correct the bad
relation mentioned above (see Sec. IV and V), $\tan\beta$ can get lowered
considerably -- for instance to values like 10 - 20.  Now, with the
contributions
from $\left \langle {\bf 10}_H \right \rangle$ and $\left \langle {\bf 16}_H
\right \rangle_d$
to fermion masses, as in Sec. IV and V, the simple proportionality of $\hat{A}$
to $\tan\beta$ disappears.  In this case it is more useful to write $A_5$ or
$\hat{A}$ simply in terms of the relevant Yukawa couplings.  That is what we
have done in writing W (see Eq. (40)) in terms of the products of Yukawa
coupling
matrices like $(\hat{H})(\hat{H}V')$ etc, and likewise for the amplitude.

It is instructive to compare a typical amplitude for $SO(10)$ to
the corresponding one for $SU(5)$.  This ratio is given by:
\begin{eqnarray}
{\hat{A}(\overline{\nu}_\mu K^+)_{\rm std}^{SO(10)} \over
\hat{A}(\overline{\nu}_\mu K^+)^{SU(5)}_{\rm std}} &\approx&
{h_{33}^2\over M_{\rm eff}} {2 \times 10^{-5} \over 1.9 \times 10^{-8}
(\tan\beta/M_{H_c})}
\nonumber \\
&\approx& (m_{H_C}/M_{\rm eff})(88)(3/\tan\beta)~.
\end{eqnarray}
We have put $h_{33}^2 \simeq 1/4$ in going from the first line to
the second in Eq. (114).  Thus we see that $M_{\rm eff}$ has to be
$(88-53)$ times larger (for $\tan\beta=3$ to 5) so that the $SO(10)$
amplitude may be comparable to that of $SU(5)$.  In other words, if
$M_{\rm eff}$ were equal to $M_{H_C}$, one would have a net enhancement
by about two orders of magnitude of the $d=5$ amplitude for proton
decay in $SO(10)$ compared to that of $SU(5)$.  This large enhancement
of the amplitude in $SO(10)$ -- apart from the factor of $M_{H_c}/M_{\rm eff}$
--
has come about due to a combination of several factors: (i) the large
off--diagonal
coupling of the up-type quarks with the color triplets which scale as
$\sqrt{m_cm_t}$
in the $SO(10)$ model in contrast to the diagonal $m_c$ in minimal $SU(5)$,
and (ii)
the larger muon Yukawa coupling to color triplets (relative to the strange
quark
Yukawa in $SU(5)$).

$SO(10)$, however, has a possible source of suppression of the $d=5$ amplitude,
because of the nature of the doublet--triplet splitting mechanism in it.
The suppression would arise if the mass $M_{10}$ of ${\bf 10}_H'$
is considerably smaller  than the scale of $\lambda \left\langle {\bf 45}_H
\right \rangle = \lambda a$, as discussed in Section
VI.A.  In this case, $M_{\rm eff} \equiv (\lambda a)^2/M_{10}$
can far exceed $\lambda a$.  Since $\left\langle {\bf 45}_H \right \rangle$
breaks
$SO(10)$ to $SU(3)_C \times SU(2)_L \times I_{3R} \times (B-L)$, one of course
naturally expects $\lambda a$ to be nearly equal to or somewhat larger than
the unification scale $M_U \approx 2 \times 10^{16}~{\rm GeV}$.  But $M_{10}$
can
in general be one or even two orders of magnitude smaller than $M_U$.  For
instance, ${\bf 10}'_H$ may carry a charge that would forbid a mass term
$({\bf 10}'_H)^2$
at the renormalizable level.  Such a mass term could still effectively arise
through  nonrenormalizable operators
by utilizing the VEVs of certain fields $\phi_i$ which do not conserve the
respective charge.
In this case, the mass term $M_{10}$ may be suppressed by relevant powers of
$[\left \langle \phi_i \right \rangle/M]$ where one may expect $\left \langle
\phi_i \right \rangle/M \sim 1/10$.  If $M_{10} \sim (1/10-1/100)\lambda a$,
$M_{\rm eff}$ can exceed $M_U$ and thus $M_{H_C}$ by one or two orders of
magnitude.  The significance of such a large value of $M_{\rm eff}$
for coupling unification is discussed in the text.

More precise constraints on $M_{\rm eff}$ and $M_{16}\tan\gamma$ are obtained
directly
by using limits on proton lifetime.  They are discussed in the text in Sec. VI.

\section*{Appendix D: Threshold corrections to $\alpha_3(m_Z)$ due to low
dimensional
multiplets of $SO(10)$}

Given $\alpha_1$ and $\alpha_2$ at $m_Z$, and the MSSM spectrum, the
hypothesis of
grand unification allows us to predict $\alpha_3(m_Z)$.  This prediction,
however,
receives so--called threshold corrections owing to mass--splittings, at the
unification scale $M_{\rm U}$, between submultiplets belonging to complete
$SO(10)$ multiplets.  Such mass splittings are of course induced when $SO(10)$
breaks spontaneously to lower symmetries using Higgs multiplets such as
${\bf 45_H}$ and ${\bf 16_H}$.

Denoting the one-loop threshold corrections to $\alpha^{-1}_i(m_Z)$ by
$-\Delta_i$, so that $\alpha_i^{-1}(m_Z) = \alpha_U^{-1}$ ~ $-(b_i/2\pi){\rm
ln}(m_Z/M_U)
- \Delta_i$, we obtain:  $\Delta_i = \sum_{\alpha}(-b_i^\alpha)/(2 \pi)$
~${\rm ln}
(M_U/m_\alpha)$ where $b_i = (33/5, 1,-3)$ yield the familiar one-loop $\beta$
functions for the evolution of the three  gauge couplings ($i=1,2,3$) for the
MSSM spectrum, and $b_i^\alpha$ is the contribution to the evolution from
the $\alpha$th sub-multiplet with mass $m_\alpha$.
Using straightforward
algebra, the threshold correction to $\alpha_3(m_Z)$ is thus found to be:
\begin{equation}
\Delta \alpha_3(m_Z) = [\alpha_3(m_Z)]^2\left( {5 \over 7} \Delta_1 -{12 \over
7}
\Delta_2 + \Delta_3\right)
\end{equation}
For illustration only, consider the contributions to $\Delta \alpha_3(m_Z)$
due to a splitting between the doublet and the triplet members in a ${\bf 5}
+ {\bf \overline{5}}$ of $SU(5)$ (or equivalently a ${\bf 10}$ of $SO(10)$)
with
masses $m_2$ and $m_3$ respectively, which are both of order $M_{\rm U}$.
In this case, one obtains:
\begin{eqnarray}
\Delta_1 = (1/2\pi)(3/5){\rm ln}(m_2/M_U) + (1/2\pi)(2/5){\rm ln}(m_3/M_U)
\nonumber \\
\Delta_2 = (1/2\pi)(1){\rm ln}(m_2/M_U),~ \Delta_3 = (1/2\pi)(1){\rm
ln}(m_3/M_U)
\end{eqnarray}
Using Eq. (116), one then obtains:
\begin{equation}
\{\Delta \alpha_3(m_Z)\}_{DT} = [\alpha_3(m_Z)]^2(9/7)(1/2\pi){\rm ln}(m_3/m_2)
\end{equation}
Using the doublet--triplet splitting mechanism exhibited in Eq. (39) and the
fact that
the MSSM evolution includes the two light Higgs doublets, one obtains the
corresponding formula for $\Delta \alpha_3(m_Z)_{DT}$ exhibited in the text
(Eq. (53)),
where $(m_3/m_2)$ is replaced by $(M_{\rm eff}\cos\gamma/M_U)$ as explained
there.

The threshold corrections owing to mass-splittings within the gauge multiplet
and the
Higgs multiplet ${\bf 45_H}$ can be computed in the same manner.  First
consider
the gauge multiplet.

The masses of the gauge particles acquiring superheavy masses are given by:
$M^2(3,1,2/3) = 4g^2(c^2+a^2)$, $M^2(3,2,1/6) = g^2(4 c^2+a^2)$,
$M^2(1,1,\pm1) = 4 g^2 c^2$, $M^2(3,2,-5/6) = g^2 a^2$, where the
numbers in
parentheses denote $SU(3) \times SU(2) \times U(1)$ quantum numbers, and $g$ is
the unified gauge coupling.  Such a splitting leads to a correction in
$\alpha_3$,
which at the electroweak scale is given by:
\begin{equation}
\Delta \alpha_3(m_Z)_{\rm gauge} = - {\alpha_3(m_Z)^2 \over 70 \pi}[75 {\rm
ln}(4+p^2)
-105 {\rm ln}(1+p^2)+30{\rm ln}p^2]
\end{equation}
where $p=2c/a$.  Eq. (118) includes contributions from the gauge particles as
well as from the corresponding gauginos and respective Higgsinos which together
make 4--component Dirac particles.
Numerical values of this correction are presented in the text.

The contribution to $\alpha_3(m_Z)$ from the splitting of ${\bf 45_H}$
is found to be:
\begin{equation}
\Delta \alpha_3(m_Z)_{\bf 45_H} = {\alpha_3(m_Z)^2 \over 14 \pi}[
6 {\rm ln}({M(1,1,\pm1) \over M_{\rm U}}) -24 {\rm ln}({M(1,3,0) \over M_{\rm
U}} +21
{\rm ln}({M(8,1,0) \over M_{\rm U}})]
\end{equation}
Here the numbers (a,b,c) in $M$ denote the quantum numbers of the respective
sub-multiplet  in ${\bf 45_H}$ under $SU(3) \times SU(2) \times U(1)$.  Within
the simplest model consisting of a ${\bf 45_H}$ only (i.e., if we ignore its
couplings to other multiplets), one can argue that the masses of the
sub-multiplets
are about two (to one) order of magnitude lower than $M_{\rm U}$.  This is
because with only a ${\bf 45_H}$, the effective $SO(10)$--invariant
superpotential
including self-couplings has the form $W_{\bf 45_H} = M_1 ({\bf 45_H})^2 +
\kappa ({\bf 45_H})^4/M$  where $M$ characterizes the effect of quantum
gravity.
Note no cubic term $({\bf 45_H})^3$ is allowed.  Setting $F_{\bf 45_H} = M_1
<{\bf 45_H}>
+ \kappa <{\bf 45_H}>^3/M = 0$, we get $M_1 \sim \kappa <{\bf 45_H}>^2/M
\approx
\kappa M_{\rm U}^2/M$, where we put $<{\bf 45_H}> \sim M_{\rm U}$.  Putting
$\kappa
\sim 1$ and $M \simeq M_{\rm Pl} = 2 \times 10^{18}$ GeV, we get $M_1 \sim
10^{-2}
M_{\rm U}$.  With the masses of the sub-multiplets given by $M_1 \approx
10^{-2} M_{\rm U}$, Eq. (119) yields $\Delta \alpha_3(m_Z) \approx -0.0045$.

Note that
if one had introduced larger Higgs multiplets lke ${\bf 126_H}$ or even ${\bf
54}$ of
$SO(10)$, one would have encountered substantially larger threshold correction
to
$\alpha_3(m_Z)$.  For example, with just a ${\bf 54}$, which contains a
$(6,1,4/3)
+ (\overline{6},1,-4/3)$ and a $(8,1,0)$ sub-multiplets, the threshold
correction
from the color (sextets, octet) to $\alpha_3(m_Z)$ is given by
$\Delta \alpha_3(m_Z)_{6+\overline{6},8} = [(51,21)/(14
\pi)]\alpha_3(m_Z)^2{\rm ln}
(M_{6,8}/M_{\rm U})$.  This yields a correction to $\alpha_3(m_Z)$ exceeding
about 14\% (of either sign) if $M_{6,8}/M_{\rm U} \sim 1/2$ or 2.  It
therefore appears
that $SO(10)$ multiplets with dimension larger than 45 are highly
disfavored, if the observed unification of gauge couplings is not to be
regarded  as
a coincidence.

\end{document}